\documentclass{ieeeaccess}
\usepackage{color}
\usepackage{cite}
\usepackage{amsmath,amssymb,amsfonts}
\usepackage{algorithmic}
\usepackage{graphicx}
\usepackage{textcomp}
\def\BibTeX{{\rm B\kern-.05em{\sc i\kern-.025em b}\kern-.08em
    T\kern-.1667em\lower.7ex\hbox{E}\kern-.125emX}}

\usepackage[english]{babel}
\usepackage[T1]{fontenc}
\usepackage[colorlinks=true, linkcolor=blue, citecolor=blue, urlcolor=blue]{hyperref}
\usepackage{braket}
\usepackage{tikz}
 \NewSpotColorSpace{PANTONE}
  \AddSpotColor{PANTONE} {PANTONE3015C} {PANTONE\SpotSpace 3015\SpotSpace C} {1 0.3 0 0.2}
  \SetPageColorSpace{PANTONE}%
\usepackage{lipsum}
\usepackage{geometry}
\usepackage{subfig}
\usepackage{pgfplots}
\usepgfplotslibrary{groupplots}
\usepackage{float}
\usepackage{placeins}
\usepackage{quantikz}
\usepackage[numbers, compress]{natbib}
\usepackage{anyfontsize}

\newcommand{\pmat}[1]{\begin{pmatrix} #1 \end{pmatrix}}
\begin{document}
\history{Received 19 August 2024; revised 10 January 2025; accepted 14 January 2025; date of publication 20 January 2025; date of current version 20 February 2025}
\doi{10.1109/TQE.2025.3532017}

\title{Variational Quantum Algorithms for Differential Equations on a Noisy Quantum Computer}
\author{\uppercase{Niclas Schillo} and
\uppercase{Andreas Sturm}}
\address[]{Fraunhofer IAO, Fraunhofer-Institut für Arbeitswirtschaft und Organisation IAO, Nobelstraße 12, 70569 Stuttgart, Germany}

\markboth
{Schillo \headeretal: Variational Quantum Algorithms on a Noisy Quantum Computer}
{Schillo \headeretal: Variational Quantum Algorithms on a Noisy Quantum Computer}

\corresp{Corresponding author: Niclas Schillo (email: niclas.schillo@iao.fraunhofer.de).}

\begin{abstract}
The role of differential equations (DEs) in science and engineering is of paramount importance, as they provide the mathematical framework for a multitude of natural phenomena.
Since quantum computers promise significant advantages over classical computers, quantum algorithms for
the solution of DEs have received a lot of attention.
Particularly interesting are algorithms that offer advantages in the current noisy intermediate scale quantum (NISQ) era, characterized by small and error-prone systems.
We consider a framework of variational quantum algorithms, quantum circuit learning (QCL), in conjunction with derivation methods, in particular the parameter shift rule, to solve DEs.
As these algorithms were specifically designed for NISQ computers, we analyze their applicability on NISQ devices by implementing QCL on an IBM quantum computer.
Our analysis of QCL without the parameter shift rule shows that we can successfully learn different functions with three-qubit circuits.
However, the hardware errors accumulate with increasing number of qubits and thus only a fraction of the qubits available on the current quantum systems can be effectively used.
We further show that it is possible to determine derivatives of the learned functions using the parameter shift rule on the IBM hardware.
The parameter shift rule results in higher errors which limits its execution to low-order derivatives.
Despite these limitations, we solve a first-order DE on the IBM quantum computer.
We further explore the advantages of using multiple qubits in QCL by learning different functions simultaneously and demonstrate the solution of a coupled differential equation on a simulator.
\end{abstract}


\titlepgskip=-15pt

\maketitle
\section{Introduction}
The efficient and accurate solution of differential equations is of great importance in numerous scientific fields and in various branches of industry.
However, many differential equations are very difficult to simulate, e.g. due to their high degree of non-linearities or numerical instability.
It is well known that quantum algorithms can significantly speed up the computation of certain problems.
Because of this potential advantage, much research is devoted to solving differential equations using quantum algorithms.
For example, several quantum algorithms for solving differential equations have been proposed that utilize quantum algorithms only as a subroutine \cite{Berry.2014,Montanaro.2016,Berry.2017,Gaitan.2020}. 
A common feature of these algorithms is the use of quantum phase estimation \cite{AlexeiY.Kitaev.1995} as the fundamental component in their quantum subroutines.
Furthermore, they employ oracle-based approaches for data access and rely on amplitude-encoded states. This approach comes with several challenges, including the input and output problem \cite{Biamonte.2017}, as well as substantial computational overhead in constructing quantum oracles. In particular, for nonlinear differential equations, quantum algorithms are scarce, with only a few notable examples \cite{Lloyd.2020}.\\
Additionally, these algorithms require a large-scale, fault-tolerant quantum computer.
However, the current noisy intermediate scale quantum (NISQ) \cite{Preskill.2018}
era is characterized by quantum computers that are error-prone and limited in size. 
Hence, quantum algorithms that work on NISQ computers are of central interest.
The most prominent candidates in this context are variational quantum algorithms. Due to their hybrid quantum-classical architecture they require fewer qubits and quantum gates so that they can cope with 
the limitations of NISQ systems.
\\
In this work, we consider variational quantum algorithms based on the quantum circuit learning (QCL) framework \cite{Mitarai.2018}.
Quantum circuit learning is a loosely defined term that is used differently in the literature.
In our consideration, QCL circuits are multi-qubit circuits that have one data encoding layer at the beginning where a variable is encoded into the quantum state using quantum feature map encoding \cite{Schuld.2018}.
This is followed by a variational layer consisting of parameterized quantum gates.
Finally, an expectation value is measured. 
In this way, we obtain a parameterized function of the encoded variable.
The types of functions that can be represented with this method depend on the data encoding layer and the number of qubits.\\
One application of QCL circuits is to learn one dimensional functions by training the parameters with a classical optimizer.
Building upon this idea, it is possible to use QCL in
combination with the parameter shift rule to solve differential equations.
The parameter shift rule is an approach to obtain gradients of a parameterized quantum circuit \cite{Mitarai.2018, Schuld.2019}.\\
The previously described circuits have already been extensively studied in the literature and their effectiveness has been demonstrated by classical simulations \cite{Mitarai.2018, HatakeyamaSato.2022}.
Using these circuits to solve differential equations has also been analysed and classically simulated \cite{Kyriienko.2021, Heim.2021, Knudsen.2020, Paine2.2023}.
Since these algorithms were proposed specifically for NISQ systems, it is of great interest whether these algorithms can be executed on such systems in practice. So far, however, the full algorithm has never been implemented on NISQ hardware.\\
In this work, we focus on the executability of QCL circuits on current quantum computers and investigate possible error sources.
It is shown that different functions can be learned with a three qubit QCL circuit on a superconducting IBM quantum computer with under one hundred optimization steps.
However, if we increase the number of qubits the hardware errors increase significantly and an execution becomes infeasible.
Furthermore, we show that the parameter shift rule, which is necessary for the solution of differential equations, can be executed on the NISQ hardware but leads to considerable errors because it is very prone to noise.
Despite these difficulties, we successfully solve a simple differential equation with QCL circuits and the parameter shift rule on the IBM quantum computer.\\
Another open question regarding QCL algorithms is how to take advantage of the multi-qubit nature of these circuits.
We present methods to use the multi-qubit character of QCL circuits to efficiently learn multiple functions simultaneously or to solve coupled differential equations.\\
This work is organized as follows: Section \ref{chap:theory_simple} gives an introduction to QCL and the circuits used in this work.
Following this, Section \ref{hardware_introduction} introduces the IBM hardware on which the QCL algorithm is to be tested and provides additional implementation details.
Subsequently, in Section \ref{chap:hardware_noise}, we perform prior simulations with different noise models and experiments on the IBM hardware to determine suitable settings for our subsequent execution of QCL.
In Section \ref{chap:QCL_sim} we learn exemplary functions with a statevector simulator and subsequently, in Section \ref{chap:ehningen}, on the IBM quantum computer.
Following this, the possibility of solving differential equations in combination with the parameter shift rule is explored.
In order to investigate how NISQ-friendly it is to solve differential equations with QCL circuits,
the parameter shift rule is tested on the IBM quantum computer in Section \ref{chap:parameter_shift_rule}.
Additionally, in Section \ref{chap_DGL_real_qc}, a simple differential equation is solved on the quantum computer.
Finally, Section \ref{chap:Multi-Qubit} examines whether it is possible to learn several
functions with a single QCL circuit by measuring multiple qubits.
With this concept, a coupled differential equation is solved with a single QCL circuit on a simulator in Section \ref{chap:coupled_HO}.
\section{Quantum Circuit Learning}
\label{chap:theory_simple}
The general structure of QCL circuits is shown in Figure \ref{circ_qcl_example}.
\def\myvdots{\ \vdots\ }
\begin{figure}[H]
    \centering
    \begin{tikzpicture}
    \node[scale=0.65] {
    \begin{quantikz}
    \lstick{$\ket{0}$}&  \gate[5, style={fill=blue!10}]{\begin{array}{c} \text{Data Encoding} \\ \text{Layer} \\ U(x) \end{array}}\slice{$\alpha(x)$} & \gate[5, style={fill=green!10}]{\begin{array}{c} \text{Variational} \\ \text{Layer} \\ V(\boldsymbol{\theta}) \end{array}}\slice{$\beta(x,\boldsymbol{\theta})$} &  \gate[5]{\begin{array}{c} \text{Expectation} \\ \text{Value} \\ \text{tr}\bigl(\beta(x,\boldsymbol{\theta}) A\bigr) \end{array}}\\[-0.77cm]
    \lstick{$\ket{0}$} & & &\\
    \lstick{$\ket{0}$} & & &\\
    \lstick{$\ket{0}$} & & &\\[-0.17cm]
    \lstick{\vdots \\ $\ket{0}$ \\ \\ }& & & 
    \end{quantikz}
    };
    \end{tikzpicture}
    \caption{Structure of QCL circuits starting with the data encoding layer $U(x)$ followed by the parameterized variational layer $V(\boldsymbol{\theta})$ and the calculation of the expectation value of an observable $A$.}
    \label{circ_qcl_example}
\end{figure}
In the data encoding layer $U(x)$ the variable $x$ is encoded into the quantum state with quantum feature map encoding \cite{Schuld.2018}.
After the data encoding layer follows the variational layer $V(\boldsymbol{\theta})$, which consists of parameterized quantum gates with the parameters $\boldsymbol{\theta} = (\theta_0, \theta_1, ...)$.
At the end, an expectation value of an observable $A$ is calculated.\\
The data encoding layer in this work consists of 
\begin{equation}
    U(x) = \bigotimes_{n = 1}^{N} R_Y(\varphi(x))\,,
    \label{data_encoding}
\end{equation}
where $N$ is the number of qubits and $\varphi(x)$ is an inner function.\\
For example, a data encoding layer with $\varphi(x)\,=\,x$ results in the density matrix
\begin{equation}
    \alpha(x) = \frac{1}{2^N} \bigotimes_{n = 1}^{N} \left( I + \sin(x)X + \cos(x)Z\right).
    \label{rho_trigo_multi_qubit}
\end{equation}
By multiplying out and using addition theorems we can see that the expression contains trigonometric functions of the form $\sin(n x)$ and $\cos(n x)$ with ${n = 1,2,...,N}$.
In this case, the expectation value without the variational layer, ${\braket{A}_\alpha(x) = \text{tr}\bigl(\alpha(x) A\bigr)}$, has the form
\begin{equation}
    \braket{A}_\alpha(x) = c_0 + \sum^{N}_{n=1} \bigl( c_n \sin(n x) + c_{n\texttt{+}N} \cos(n x) \bigr),
\end{equation}
where $c_n$ are scalar coefficients that depend on the observable $A$. To adjust the coefficients $c_n$, the variational layer is introduced.
The new coefficients, which we again denote with $c_n$, now depend on the variational parameters $\boldsymbol{\theta}$ which allows us to control their value.
Hence, the expectation value after the variational layer, ${\braket{A}_\beta(x,\boldsymbol{\theta})\,=\,\text{tr}\bigl(\beta(x,\boldsymbol{\theta}) A\bigr)}$, has the form
\begin{equation}
    \begin{split}
        \braket{A}_\beta(x,\boldsymbol{\theta}) = & c_0(\boldsymbol{\theta})+ \sum^{N}_{n=1} \biggl( c_n(\boldsymbol{\theta}) \sin(n x)\\
        &+ c_{n\texttt{+}N}(\boldsymbol{\theta}) \cos(n x) \biggr).
    \end{split}
    \label{arcsin_function}
\end{equation}
In \cite{Mitarai.2018} a different data encoding layer
\begin{equation}
    U(x) = \bigotimes_{n = 1}^{N} R_Y(\arcsin(x))
\end{equation}
was introduced. This data encoding scheme results in the density matrix
\begin{equation}
    \alpha(x) = \frac{1}{2^N} \bigotimes_{n = 1}^{N} \left( I + x X + \sqrt{1-x^2}Z\right)
\end{equation}
and gives a set of polynomials up to the order of $x^N$ with additional $\sqrt{1-x^2}$-terms.\\
One application of QCL is to learn arbitrary functions $f(x)$.
Here, the parameters are chosen such that the expectation value $\braket{A}_\beta$ matches the function $f(x)$.
For this purpose, a cost function on several training points $x_i$ is minimized with a classical optimizer.
\subsubsection*{Cost Function}
An example of a simple cost function is
\begin{equation}
L(\boldsymbol{\theta}) = \sum_i|f(x_i) - f_{QC}(x_i, \boldsymbol{\theta})|^2
\end{equation}
where $f_{QC}(x, \boldsymbol{\theta}) = \text{tr}\bigl(\beta(x,\boldsymbol{\theta}) A\bigr)$ is the expectation value of the observable $A$ and will be referred to as the quantum model function, $\sum_i$ sums over a number of training points $x_i$ and $f(x)$ is the function to be learned.
While this approach can learn the qualitative behavior of functions, it may show strong deviations in the final result as discussed in Appendix \ref{AppendixA}. To increase the accuracy of function learning, an improved cost function is introduced that includes a post-processing parameter $\theta_{\text{post}}$
\begin{equation}
L(\boldsymbol{\theta}) = \sum_i|f(x_i) - f^{\text{post}}_{QC}(x_i, \boldsymbol{\theta})|^2\,,
\label{cost_function_post}
\end{equation}
where $f^{\text{post}}_{QC}(x, \boldsymbol{\theta}) = f_{QC}(x, \boldsymbol{\theta})\cdot \theta_{\text{post}}$. The post-processing parameter $\theta_{\text{post}}$ is optimized along with the other parameters. This additional degree of freedom significantly increases the expressivity of the model and leads to faster and more accurate learning. It also extends the value range of the quantum model functions to $f^{\text{post}}_{QC}(x, \boldsymbol{\theta}) \in \mathbb{R}$ for $\theta_{\text{post}} \in \mathbb{R}$, overcoming the limitation of the $Z$ expectation value range $[-1,1]$.

\subsubsection*{Quantum Circuit}
In this work we analyze QCL circuits with the following structure: First, the input data is encoded using $R_Y(x)$ or $R_Y(\arcsin(x))$ gates applied to each qubit. Next, an entanglement layer is created using CNOT gates in a linear chain with an additional gate between the first and last qubit, forming a circular entanglement pattern. This entanglement is used to generate functions up to the order $N$ on the first qubit.
Afterwards, $\boldsymbol{\theta}$-parameterized x-, y- and z-rotations are applied to each qubit, allowing for any unitary single-qubit operation, excluding global phase. The entanglement and parameterized rotation layers are repeated $D$ times, using the same parameters $\boldsymbol{\theta}$ in each block. These repetitions increase the expressivity of the circuit without increasing the number of parameters \cite{Schillo.2023}.
From now on, we use $A=Z_0$, i.e. we use the $Z$ expectation value of the first qubit to read out our quantum model function.
The circuit structure for the example of $N=3$ is illustrated in Figure \ref{circ_simple_example_depth}.
\begin{figure}[H]
    \centering
    \begin{tikzpicture}
    \node[scale=0.5] {
    \begin{quantikz}
    \lstick{$\ket{0}$}&  \gate{R_Y(\varphi(x))}  & \ctrl{1} \gategroup[3,steps=6,style={dashed,rounded corners, inner xsep=2pt},background,label style={label position=below,anchor=north,yshift=-0.2cm}]{{$D\times$}}&  \qw & \targ{} & \gate{R_X(\theta_0 )}& \gate{R_Y(\theta_1 )}  & \gate{R_Z(\theta_2)}  &   \meter[]{\braket{Z}} \\
    \lstick{$\ket{0}$} &  \gate{R_Y(\varphi(x))}  & \targ{} &  \ctrl{1} & \qw  & \gate{R_X(\theta_3 )}& \gate{R_Y(\theta_4)}  & \gate{R_Z(\theta_5)}  &   \qw\\
    \lstick{$\ket{0}$} &  \gate{R_Y(\varphi(x))}  & \qw  &  \targ{} & \ctrl{-2}  & \gate{R_X(\theta_6 )}& \gate{R_Y(\theta_7)}  & \gate{R_Z(\theta_8)}  &     \qw
    \end{quantikz}
    };
    \end{tikzpicture}
    \caption{Three qubit QCL circuit with $R_Y(\varphi(x))$ data encoding followed by a variational block consisting of three CNOT gates to achieve circular entanglement and $\boldsymbol{\theta}$-parameterized x-, y- and z-rotations. The variational block is repeated $D$ times. The $Z$ expectation value of the first qubit is measured.}
    \label{circ_simple_example_depth}
\end{figure}
\section{Hardware Implementation Details}
\label{hardware_introduction}
The QCL circuits will be executed on the ibmq\_ehningen quantum computer.
This system is a 27-qubit superconducting IBM quantum computer based on transmon qubits of the Falcon chip class.
One important characteristic of real quantum computers is that their operations are subject to errors. For ibmq\_ehningen we show exemplary error rates in Appendix \ref{AppendixB}. Furthermore, they only have a certain set of native gates that can be executed directly (for ibmq\_ehningen: CNOT, I, RZ, $\sqrt{\text{X}}$, X) and they have limited connectivity between the different qubits as seen in the coupling map in Figure \ref{fig:coupling_map} for ibmq\_ehningen.
\begin{figure}[H]
    \centering
    \scalebox{0.5}{\input{figures/Tikz/coupling_map}}
	\caption{Coupling map of ibmq\_ehningen.}
	\label{fig:coupling_map}
\end{figure}
Despite the limited connectivity, arbitrary circuits can be realized by applying additional SWAP gates. Since a SWAP gate between connected qubits consists of three CNOT gates, this increases the circuit depth, which makes an optimized transpilation to the specific hardware especially important.\\
Our QCL circuits in the form given in Figure \ref{circ_simple_example_depth} cannot be executed directly on ibmq\_ehningen as they do not meet the hardware constraints with regard to the coupling map and the native gate set of this backend.
To transform the circuits into a compatible form we use the qiskit transpiler \cite{QiskitTranspiler.2024} with the \texttt{optimization\_level\,=\,2} setting. This medium optimization level is designed to strike a balance between improving the circuit performance and maintaining reasonable transpilation times. At this level, the transpiler applies the following key transformations to optimize the circuit for execution on the IBM system:
First, the transpiler searches for an initial qubit layout that minimizes the need for SWAP gates when mapped to the coupling map of the device.
The circuit is then unrolled to the basis gates supported by the hardware. Finally, optimizations are performed in the form of commutative gate cancellation and redundant reset removal to minimize the circuit depth.\\
Additionally, all experiments on ibmq\_ehningen are performed with Twirled Readout Error eXtinction (TREX) \cite{vanBerg.2020}.
TREX is a technique that mitigates readout errors in quantum computations by randomly applying Pauli X gates to qubits just before measurement, effectively diagonalizing the noise channel of the measurement. This randomization allows for easier characterization and correction of readout errors without requiring knowledge about the error model.
Apart from this, no error mitigation techniques are applied.
\section{Prior Investigations}
\label{chap:hardware_noise}
Before we run the resource intensive quantum-classical variational workflow to learn functions by optimizing the parameters $\boldsymbol{\theta}$, we first select fixed parameters to make statements about the scalability of the circuits and the number of shots required.
For this purpose, all parameters $\theta_i$ are set to $\frac{\pi}{2}$, as shown in Figure \ref{circ_simple_example_2} for a qubit number of $N=3$, a depth of $D = 3$ and $R_Y(\arcsin(x))$ data encoding.
\begin{figure}[H]
    \centering
    \begin{tikzpicture}
    \node[scale=0.5] {
    \begin{quantikz}
    \lstick{$\ket{0}$} & \gate{R_Y(\arcsin(x))}  & \ctrl{1} \gategroup[3,steps=6,style={dashed,rounded corners, inner xsep=2pt},background,label style={label position=below,anchor=north,yshift=-0.2cm}]{{$3\times$}}&  \qw & \targ{} & \gate{R_X(\tfrac{\pi}{2} )}& \gate{R_Y(\tfrac{\pi}{2} )}  & \gate{R_Z(\tfrac{\pi}{2})}  &   \meter[]{ \braket{Z}} \\
    \lstick{$\ket{0}$} &  \gate{R_Y(\arcsin(x))}  & \targ{} &  \ctrl{1} & \qw  & \gate{R_X(\tfrac{\pi}{2})}& \gate{R_Y(\tfrac{\pi}{2})}  & \gate{R_Z(\tfrac{\pi}{2})}  &   \qw\\
    \lstick{$\ket{0}$} &  \gate{R_Y(\arcsin(x))}  & \qw  &  \targ{} & \ctrl{-2}  & \gate{R_X(\tfrac{\pi}{2})}& \gate{R_Y(\tfrac{\pi}{2})}  & \gate{R_Z(\tfrac{\pi}{2})}  &     \qw
    \end{quantikz}
    };
    \end{tikzpicture}
    \caption{Three qubit QCL circuit with $R_Y(\arcsin(x))$ data encoding and a variational layer with $\theta_i=\frac{\pi}{2}$ for all $i$.}
    \label{circ_simple_example_2}
\end{figure}
It is easy to check that this circuit yields
\begin{equation}
	f_{QC}(x) = (-1)^N \cdot x\,,
\end{equation}
where $N$ is the number of qubits.\\
To transpile the circuit we use the qiskit transpiler endowed with the coupling map and basis gates of ibmq\_ehningen.
As an initial layout, we choose the qubits $0,...,N-1$.
The stochastic components of the optimization process are made reproducible by choosing a fixed transpilation seed (that we chose as 123).
In order to determine a suitable setting for our later experiments on the real IBM quantum backend, we run the transpiled circuit on a simulator with three different noise models.
Our first noise model (noise model 1) has perfect quantum operations and the only error source stems from finite sampling, i.e. from shot noise.
For the second one (noise model 2) we randomly and independently insert Pauli operators after gates and before measurements with probabilities determined by the gate \cite{Knill.2005}.
Single qubit gates are modified with probability $p_1/3$ by one of the non-trivial Pauli operators $\{X, Y, Z\}$ and two qubit gates are followed by a non-identity Pauli product, i.e. one of
$\{P_1 \otimes P_2 \ | \ P_1, P_2 \in \{I, X, Y, Z\}\} \setminus \{I \otimes I\}$, with probability $p_2/15$.
Moreover, a binary measurement in the $Z$ basis has the wrong outcome with probability $p_r$ in this noise model.
The error rates are taken as $p_1 = 0.00024$, $p_2 = 0.0075$ and $p_r = 0.012$.
These error rates correspond to the median error rates as reported in the calibration data from the ibmq\_ehningen quantum computer on January 30, 2024, see Appendix~\ref{AppendixB}.
At last, our third noise model (noise model 3) is derived from this ibmq\_ehningen calibration data via the method {\small{\texttt{NoiseMode.from\_backend\_properties}}} of the Qiskit Aer package \cite{IbmAer.2024} and includes besides single gate, two qubit gate and readout errors also a depolarizing error and a thermal relaxation error.
\subsection{Number of shots}
First, we investigate how the number of shots affects the result in order to determine a suitable number for the subsequent experiments.
To do this, we use the circuit in Figure \ref{circ_simple_example_2} with $N=3$ and the $x$ values $x \in \{-1, -0.5, 0, 0.5, 1\}$. We simulate these circuits to obtain $f_\text{QC}$ with the noise models introduced before and a range of different shot numbers. For each noise model and shot number we repeat the simulation 20 times. Figure \ref{fig:shot_noise} shows the different values of $f_\text{QC}(x)$ for the selected $x$ values as x markers and the exact values as horizontal lines.
\begin{figure*}[!tp]
    \centering
    \scalebox{0.5}{\input{figures/shots_investigation/experiment_1_plot}}
    \caption{$f_{QC}(x)$ of QCL circuit structured as given in Figure \ref{circ_simple_example_2} evaluated at the points $x \in \{-1, -0.5, 0, 0.5, 1\}$, plotted against the number of shots. Three noise models are compared: noise model 1 (only shot noise), noise model 2 (Pauli noise), noise model 3 (derived from calibration data).
    }
    \label{fig:shot_noise}
\end{figure*}
For noise model 1 (only shot noise), the results fluctuate around the exact values and diminish as the number of shots increases.
For noise models 2 and 3, there is a noticeable drift of $f_{QC}(x)$ towards zero, independent of the number of shots.
This drift is more pronounced in noise model 3 (derived from calibration data) compared to noise model 2 (Pauli noise).
Importantly, even at high shot numbers, this deviation persists, indicating that it is an effect of the hardware errors and cannot be compensated with a higher shot number.
Based on these results, we will use 2000 shots for our subsequent experiments.
This number of shots provides a good balance between accuracy and computational efficiency.
At 2000 shots, the results for all noise models show low fluctuations, while still maintaining reasonable execution times for the subsequent hardware experiments.
\subsection{Number of Qubits}
We now analyze how the error scales with increasing qubit count, providing insights into the practical limitations of the circuit size for ibmq\_ehningen.
For this purpose, circuits with the same structure but different qubit numbers are simulated.
In addition, the circuits are also executed on the real ibmq\_ehningen backend.
Figure \ref{fig:qml_depth_ehningen1} shows $(-1)^N f_{QC}(x)$ for different circuit sizes on a simulator with noise model 2 (Pauli noise) and noise model 3 (derived from calibration data) as well as on ibmq\_ehningen. Moreover, the plot shows $f(x) = x$ which corresponds to $(-1)^N f_{QC}(x)$ for an exact statevector simulation.
\begin{figure*}[!tp]
    \centering
    \scalebox{0.52}{\input{figures/ehningen/functions_qml_depth_3}}
    \caption{$(-1)^N f_{QC}(x)$ of QCL circuits structured as given in Figure \ref{circ_simple_example_2} for different numbers of qubits evaluated at the points $x \in \{-1, -0.5, 0, 0.5, 1\}$ on a simulator with noise model 2 (Pauli noise) and noise model 3 (derived from calibration data) as well as on the ibmq\_ehningen.}
    \label{fig:qml_depth_ehningen1}
\end{figure*}
Due to the increasing number of CNOT gates, the circuit depth also increases significantly with the number of qubits.
We observe that the increasing depth and therefore increasing hardware errors lead to the measured expectation values approaching
\begin{equation}
	f_{QC}(x) \equiv 0\,.
    \label{is_zero}
\end{equation}
Among the simulated error models, the noise model 2 (Pauli noise) shows a slower convergence towards zero, which is to be expected since this model assumes fewer sources of error.
On the real ibmq\_ehningen backend, the expectation values approach $f_{QC}(x) \equiv 0$ more rapidly than in the simulations, indicating a faster accumulation of errors.
The discrepancy between the real backend and the error models illustrates the limitations of simplified noise models in representing the error dynamics in the actual quantum hardware.
The error models chosen here can reflect the geneal behavior, but are not detailed enough for a more accurate representation. 
We will discuss possible reasons for this discrepancy after examining the scaling of the error in more detail.\\
For a more comprehensive picture, we analyze the mean absolute error
\begin{equation}
    \text{MAE} = \frac{1}{T}\sum^T_{i=1}|f(x_i) - f_{QC}(x_i, \boldsymbol{\theta})|,
\end{equation}
where $T$ is the number of training points, for different number of qubits $N$.
On real quantum devices every qubit and every gate has an individual error rate and we expect that these error rates have a strong influence on MAE. Thus, we now consider two transpilations of our underlying circuit (Figure \ref{circ_simple_example_2}) that use different qubits of ibmq\_ehningen.
We use the qubits $0, \dots, N-1$ (first qubits) and additionally the qubits $26-N+1, \dots, 26$ (last qubits) as the initial layouts for the transpiler (see Figure~\ref{fig:coupling_map}).
Figure \ref{fig:qml_depth_ehningen2} shows the MAE for qubit numbers from $N=2$ to $N=20$ for the three different noise models and the two inital layouts on a simulator. Moreover, we add the MAE for ibmq\_ehningen when the transpiler selected the best qubits.\\
Clearly, the MAE for noise model 1 is not affected by the qubit number since it uses perfect qubits and gates.
For noise model 2, we see a linear increase of MAE with the number of qubits.
We will discuss this in more detail in the next section.
For our most advanced noise model 3, we observe a higher MAE compared to the second noise model.
Moreover, we can see that the transpilation with the first initial qubit layout suffers from a much higher MAE than the transpilation with the second layout for $N \ge 7$ qubits.
This can be explained since beginning with this number of qubits the first transpilation has to use the CNOT gate between qubits 4 and 7 which has by far the highest error rate, see Table \ref{table-cx-errors-ibmq-ehningen} in Appendix \ref{AppendixB}, whereas the second transpilation can use CNOT gates with lower errors.
The MAE from the real quantum computer ibmq\_ehningen increases even more strongly and already plateaus for $N=5$ qubits on $MAE = 0.6$.
This can be expected since this is the MAE of $f_{QC}(x) \equiv 0$ and, as we have seen in Figure~\ref{fig:qml_depth_ehningen1}, ibmq\_ehningen returns approximately this quantum function for this qubit number regime.
\begin{figure*}[!tp]
    \centering
    \scalebox{0.8}{
\begin{tikzpicture}

\definecolor{darkslategray38}{RGB}{38,38,38}
\definecolor{green01270}{RGB}{0,127,0}
\definecolor{lightgray204}{RGB}{204,204,204}
\definecolor{crimson2143940}{RGB}{214,39,40}
\definecolor{darkorange25512714}{RGB}{255,127,14}
\definecolor{darkslategray38}{RGB}{38,38,38}
\definecolor{forestgreen4416044}{RGB}{44,160,44}
\definecolor{lightgray204}{RGB}{204,204,204}

\definecolor{mediumpurple148103189}{RGB}{148,103,189}
\definecolor{steelblue31119180}{RGB}{31,119,180}
\definecolor{darkgray176}{RGB}{176,176,176}
\definecolor{darkturquoise23190207}{RGB}{23,190,207}
\definecolor{forestgreen4416044}{RGB}{44,160,44}
\definecolor{gray127}{RGB}{127,127,127}
\definecolor{lightgray204}{RGB}{204,204,204}
\definecolor{sienna1408675}{RGB}{140,86,75}
\definecolor{steelblue31119180}{RGB}{31,119,180}

\definecolor{crimson2143940}{RGB}{214,39,40}
\definecolor{darkorange25512714}{RGB}{255,127,14}
\definecolor{darkslategray38}{RGB}{38,38,38}
\definecolor{forestgreen4416044}{RGB}{44,160,44}
\definecolor{lightgray204}{RGB}{204,204,204}
\definecolor{mediumpurple148103189}{RGB}{148,103,189}
\definecolor{steelblue31119180}{RGB}{31,119,180}

\begin{axis}[
axis line style={lightgray204},
legend cell align={left},
legend style={fill opacity=0.8, draw opacity=1, text opacity=1, draw=none, at={(1.05,0.5)}, anchor=west}, 
tick align=outside,
tick pos=left,
x grid style={lightgray204},
xlabel=\textcolor{darkslategray38}{Number of Qubits $N$},
xmajorgrids,
xmin=2, xmax=20,
xtick={2,5,10,15,20,25},
xtick style={color=darkslategray38},
y grid style={lightgray204},
ylabel=\textcolor{darkslategray38}{MAE},
ymajorgrids,
ymin=0, ymax=0.8,
ytick style={color=darkslategray38}
]
\addplot [semithick, black]
table {%
2 0.6
3.04 0.6
4.08 0.6
5.12 0.6
6.16 0.6
7.2 0.6
8.24 0.6
9.28 0.6
10.32 0.6
11.36 0.6
12.4 0.6
13.44 0.6
14.48 0.6
15.52 0.6
16.56 0.6
17.6 0.6
18.64 0.6
19.68 0.6
20.72 0.6
21.76 0.6
22.8 0.6
23.84 0.6
24.88 0.6
25.92 0.6
26.96 0.6
28 0.6
};
\addlegendentry{MAE for $f_{QC}(x)\equiv0$}
\addplot [semithick, red, mark=x, mark size=3, mark options={solid}]
table {%
2 0.035151407761488
3 0.179934785045416
4 0.446987857661168
5 0.658077248807982
6 0.656981131136185
7 0.625304887500365
8 0.62109777743415
9 0.649242000090157
10 0.663779452835733
11 0.627880300222112
12 0.603551873092356
13 0.620426697792023
14 0.614533214269019
15 0.642057731770039
16 0.623312968433214
17 0.606977918720442
18 0.607764655677665
19 0.623723390419195
20 0.638115248478417
21 0.619311699504786
22 0.63776684769035
23 0.5955082614697
24 0.610199779983865
25 0.611582319004709
26 0.598188163580596
27 0.590149944036883
};
\addlegendentry{MAE on ibmq\_ehningen}

\addplot [semithick, forestgreen4416044, mark=x, mark size=3, mark options={solid}]
table {%
2 0.012200000000000
3 0.007600000000000
4 0.014000000000000
5 0.010000000000000
6 0.009000000000000
7 0.004400000000000
8 0.013400000000000
9 0.005200000000000
10 0.006400000000000
11 0.005200000000000
12 0.012400000000000
13 0.008200000000000
14 0.008200000000000
15 0.006000000000000
16 0.013400000000000
17 0.009000000000000
18 0.010800000000000
19 0.004600000000000
20 0.014400000000000
};
\addlegendentry{MAE Noise model 1}
\addplot [semithick, darkorange25512714, mark=x, mark size=3, mark options={solid}]
table {%
2 0.032200000000000
3 0.082000000000000
4 0.114200000000000
5 0.148000000000000
6 0.181000000000000
7 0.205600000000000
8 0.235600000000000
9 0.292000000000000
10 0.286600000000000
11 0.324800000000000
12 0.312000000000000
13 0.418600000000000
14 0.377200000000000
15 0.370400000000000
16 0.388400000000000
17 0.421200000000000
18 0.442400000000000
19 0.440000000000000
20 0.446600000000000
};
\addlegendentry{MAE Noise model 2, first qubits}
\addplot [semithick, darkorange25512714, mark=triangle, mark size=3, mark options={solid}]
table {%
2 0.027000000000000
3 0.073000000000000
4 0.125600000000000
5 0.193800000000000
6 0.188000000000000
7 0.232600000000000
8 0.266000000000000
9 0.253200000000000
10 0.261400000000000
11 0.309200000000000
12 0.330400000000000
13 0.418200000000000
14 0.391000000000000
15 0.396400000000000
16 0.396400000000000
17 0.449400000000000
18 0.423800000000000
19 0.455400000000000
20 0.448600000000000
};
\addlegendentry{MAE Noise model 2, last qubits}
\addplot [semithick, steelblue31119180, mark=x, mark size=3, mark options={solid}]
table {%
2 0.025200000000000
3 0.117800000000000
4 0.166600000000000
5 0.240600000000000
6 0.236800000000000
7 0.463800000000000
8 0.458600000000000
9 0.456000000000000
10 0.472800000000000
11 0.529600000000000
12 0.475600000000000
13 0.605800000000000
14 0.611000000000000
15 0.581000000000000
16 0.596600000000000
17 0.603800000000000
18 0.597600000000000
19 0.591200000000000
20 0.581000000000000
};
\addlegendentry{MAE Noise model 3, first qubits}
\addplot [semithick, steelblue31119180, mark=triangle, mark size=3, mark options={solid}]
table {%
2 0.040800000000000
3 0.108200000000000
4 0.135200000000000
5 0.193000000000000
6 0.200800000000000
7 0.243800000000000
8 0.314000000000000
9 0.291400000000000
10 0.323200000000000
11 0.398200000000000
12 0.437200000000000
13 0.464400000000000
14 0.468200000000000
15 0.490400000000000
16 0.528600000000000
17 0.564800000000000
18 0.546800000000000
19 0.553800000000000
20 0.555200000000000
};
\addlegendentry{MAE Noise model 3, last qubits}

\end{axis}

\end{tikzpicture}}
    \caption{Mean absolute error (MAE) of QCL circuits structured as given in Figure \ref{circ_simple_example_2} for different numbers of qubits evaluated at the points $x \in \{-1, -0.5, 0, 0.5, 1\}$ on a simulator with noise model 1 (only shot noise), noise model 2 (Pauli noise) and noise model 3 (derived from calibration data) and on ibmq\_ehningen. Additionally, two different sets of qubits of ibmq\_ehningen are used (first and last qubits). The black line indicates the error of $MAE = 0.6$ which corresponds to the error in the case $f_{QC}(x) \equiv 0$.}
    \label{fig:qml_depth_ehningen2}
\end{figure*}
We see that that the error models provide better results than the real experiments, which indicates that they cannot capture all error sources present in ibmq\_ehningen.
For example, a study on the ibmq\_ehningen processor revealed significant impacts of crosstalk on gate fidelities, leading to correlated errors between simultaneously executed quantum gates on neighboring qubits \cite{Wellens.2023}.
This research highlights the importance of considering crosstalk in quantum system modeling.
Another study focusing on NISQ systems for quantum optimization uncovered time-dependent errors in IBM quantum computers, showing varying results at different times \cite{Sturm.10.04.2024}.
These findings demonstrate some of the error sources in superconducting quantum hardware. The discrepancy between the error models used in this work and real experiments can be attributed, among other things, to these factors.
\subsubsection*{Role of Circular Entanglement}
\label{sec_circ_ent}
One additional factor, apart from the pure error rates of the quantum operations, is the resilience in the design of the circuit against the spread of errors.
To investigate this, we compare the results from our previous circuit design with circular entanglement, see Figure~\ref{circ_simple_example_2}, with a design that uses only linear entanglement as given in Figure~\ref{circ_simple_example_3}. The two circuits agree in all gates except that the circuit with linear entanglement lacks the last CNOT in the variational layer connecting the first and the last qubit.
\begin{figure}[H]
    \centering
    \begin{tikzpicture}
    \node[scale=0.5] {
    \begin{quantikz}
    \lstick{$\ket{0}$} & \gate{R_Y(\arcsin(x))}  & \ctrl{1} \gategroup[3,steps=5,style={dashed,rounded corners, inner xsep=2pt},background,label style={label position=below,anchor=north,yshift=-0.2cm}]{{$3\times$}}&  \qw & \gate{R_X(\tfrac{\pi}{2} )}& \gate{R_Y(\tfrac{\pi}{2} )}  & \gate{R_Z(\tfrac{\pi}{2})}  &   \meter[]{ \braket{Z}} \\
    \lstick{$\ket{0}$} &  \gate{R_Y(\arcsin(x))}  & \targ{} &  \ctrl{1} & \gate{R_X(\tfrac{\pi}{2})}& \gate{R_Y(\tfrac{\pi}{2})}  & \gate{R_Z(\tfrac{\pi}{2})}  &   \qw\\
    \lstick{$\ket{0}$} &  \gate{R_Y(\arcsin(x))}  & \qw  &  \targ{}   & \gate{R_X(\tfrac{\pi}{2})}& \gate{R_Y(\tfrac{\pi}{2})}  & \gate{R_Z(\tfrac{\pi}{2})}  &     \qw
    \end{quantikz}
    };
    \end{tikzpicture}
    \caption{Three qubit QCL circuit with linear entanglement with $R_Y(\arcsin(x))$ data encoding. The $Z$ expectation value of the first qubit is measured.}
    \label{circ_simple_example_3}
\end{figure}
In order to study the spread of errors we use our second noise model without single gate and readout errors, $p_1 = 0$ and $p_r = 0$, and 1 percent two qubit error rate $p_2 = 0.01$.
We set $x=1$ so that the exact values of $f_{QC}$ are given by $f_{QC}(1) = (-1)^N$ and $f_{QC}(1) = 1$ for the case of circular and linear entanglement, respectively.
Figure~\ref{fig:circ_ent} shows the errors for both circuit designs.
We see that the linear entanglement design is resilient against the spread of errors whereas in the circular entanglement design we observe a linear increase of the error with the qubit number.
This phenomenon arises from the additional pathways for error propagation that circular entanglement introduces.
While both circular and linear entanglement schemes allow errors to propagate through the chain of qubits, circular entanglement creates an extra connection that increases the error measured at the first qubit.
This is the same behavior that we observed in Figure~\ref{fig:qml_depth_ehningen2} for the MAE of our Pauli noise model. 
At this point we emphasize that the more resilient design of the linear entanglement circuit comes with severe disadvantages in solving our original problem of learning functions and solving differential equations.
As we explain in Appendix \ref{AppendixC}, we need circular entanglement to be able to learn arbitrary functions of degree $N$.
Therefore, in the following experiments we will use circular entanglement but limit ourselves to a qubit number of $N=3$.
\begin{figure*}[!tp]
    \begin{minipage}[c]{0.5\textwidth}
        \scalebox{0.6}{\begin{tikzpicture}

\definecolor{darkgray176}{RGB}{176,176,176}
\definecolor{darkorange25512714}{RGB}{255,127,14}
\definecolor{lightgray204}{RGB}{204,204,204}
\definecolor{steelblue31119180}{RGB}{31,119,180}

\begin{axis}[
height=1\textwidth,
axis line style={lightgray204},
legend cell align={left},
legend style={
  fill opacity=1.0,
  text opacity=1,
  at={(0.03,0.97)},
  anchor=north west,
  draw=none
},
minor xtick={},
minor ytick={},
tick align=outside,
tick pos=left,
width=1.5\textwidth,
x grid style={darkgray176},
xlabel={$N$},
xmajorgrids,
xmin=1.1, xmax=20.9,
xtick style={color=black},
ytick={0,2.5,5,7.5,10,12.5,15,17.5,20,22.5},
y grid style={darkgray176},
ylabel={error in \%},
ymajorgrids,
ymin=0.2125, ymax=21.9375,
ytick style={color=black},
xtick={0,2,5,10,15,20}
]
\path [draw=steelblue31119180, semithick]
(axis cs:2,1.2)
--(axis cs:2,2.15);

\path [draw=steelblue31119180, semithick]
(axis cs:3,2.95)
--(axis cs:3,4.35);

\path [draw=steelblue31119180, semithick]
(axis cs:4,3.5)
--(axis cs:4,5.65);

\path [draw=steelblue31119180, semithick]
(axis cs:5,4.65)
--(axis cs:5,6.2);

\path [draw=steelblue31119180, semithick]
(axis cs:6,5.25)
--(axis cs:6,7.3);

\path [draw=steelblue31119180, semithick]
(axis cs:7,5.85)
--(axis cs:7,8.4);

\path [draw=steelblue31119180, semithick]
(axis cs:8,7.35)
--(axis cs:8,9);

\path [draw=steelblue31119180, semithick]
(axis cs:9,8.1)
--(axis cs:9,10.45);

\path [draw=steelblue31119180, semithick]
(axis cs:10,9)
--(axis cs:10,11.2);

\path [draw=steelblue31119180, semithick]
(axis cs:11,9.5)
--(axis cs:11,12.25);

\path [draw=steelblue31119180, semithick]
(axis cs:12,10.7)
--(axis cs:12,12.75);

\path [draw=steelblue31119180, semithick]
(axis cs:13,10.6)
--(axis cs:13,14.3);

\path [draw=steelblue31119180, semithick]
(axis cs:14,11.6)
--(axis cs:14,15.2);

\path [draw=steelblue31119180, semithick]
(axis cs:15,12.9)
--(axis cs:15,15.25);

\path [draw=steelblue31119180, semithick]
(axis cs:16,13.8)
--(axis cs:16,16.45);

\path [draw=steelblue31119180, semithick]
(axis cs:17,14.35)
--(axis cs:17,17.45);

\path [draw=steelblue31119180, semithick]
(axis cs:18,14.95)
--(axis cs:18,18.15);

\path [draw=steelblue31119180, semithick]
(axis cs:19,15.65)
--(axis cs:19,18.2);

\path [draw=steelblue31119180, semithick]
(axis cs:20,16.45)
--(axis cs:20,20.95);

\addplot [thick, steelblue31119180, mark=-, mark size=5, mark options={solid}, only marks]
table {%
2 1.2
3 2.95
4 3.5
5 4.65
6 5.25
7 5.85
8 7.35
9 8.1
10 9
11 9.5
12 10.7
13 10.6
14 11.6
15 12.9
16 13.8
17 14.35
18 14.95
19 15.65
20 16.45
};
\addlegendentry{circular entanglement}

\addplot [thick, darkorange25512714, mark=-, mark size=5, mark options={solid}, only marks]
table {%
2 1.2
3 1.6
4 1.6
5 1.7
6 1.55
7 2.05
8 1.95
9 1.95
10 2.15
11 2.05
12 1.8
13 2.05
14 1.4
15 2.15
16 1.8
17 2.2
18 2
19 2.1
20 2
};
\addlegendentry{linear entanglement}

\addplot [thick, steelblue31119180, mark=-, mark size=5, mark options={solid}, only marks]
table {%
2 2.15
3 4.35
4 5.65
5 6.2
6 7.3
7 8.4
8 9
9 10.45
10 11.2
11 12.25
12 12.75
13 14.3
14 15.2
15 15.25
16 16.45
17 17.45
18 18.15
19 18.2
20 20.95
};
\path [draw=darkorange25512714, semithick]
(axis cs:2,1.2)
--(axis cs:2,2.05);

\path [draw=darkorange25512714, semithick]
(axis cs:3,1.6)
--(axis cs:3,3.2);

\path [draw=darkorange25512714, semithick]
(axis cs:4,1.6)
--(axis cs:4,3.4);

\path [draw=darkorange25512714, semithick]
(axis cs:5,1.7)
--(axis cs:5,3.7);

\path [draw=darkorange25512714, semithick]
(axis cs:6,1.55)
--(axis cs:6,3.15);

\path [draw=darkorange25512714, semithick]
(axis cs:7,2.05)
--(axis cs:7,3.9);

\path [draw=darkorange25512714, semithick]
(axis cs:8,1.95)
--(axis cs:8,3.15);

\path [draw=darkorange25512714, semithick]
(axis cs:9,1.95)
--(axis cs:9,3.35);

\path [draw=darkorange25512714, semithick]
(axis cs:10,2.15)
--(axis cs:10,3.3);

\path [draw=darkorange25512714, semithick]
(axis cs:11,2.05)
--(axis cs:11,3.2);

\path [draw=darkorange25512714, semithick]
(axis cs:12,1.8)
--(axis cs:12,3.35);

\path [draw=darkorange25512714, semithick]
(axis cs:13,2.05)
--(axis cs:13,3.75);

\path [draw=darkorange25512714, semithick]
(axis cs:14,1.4)
--(axis cs:14,3.15);

\path [draw=darkorange25512714, semithick]
(axis cs:15,2.15)
--(axis cs:15,3.2);

\path [draw=darkorange25512714, semithick]
(axis cs:16,1.8)
--(axis cs:16,3.6);

\path [draw=darkorange25512714, semithick]
(axis cs:17,2.2)
--(axis cs:17,3.25);

\path [draw=darkorange25512714, semithick]
(axis cs:18,2)
--(axis cs:18,3.3);

\path [draw=darkorange25512714, semithick]
(axis cs:19,2.1)
--(axis cs:19,3.3);

\path [draw=darkorange25512714, semithick]
(axis cs:20,2)
--(axis cs:20,3.05);

\addplot [thick, darkorange25512714, mark=-, mark size=5, mark options={solid}, only marks]
table {%
2 2.05
3 3.2
4 3.4
5 3.7
6 3.15
7 3.9
8 3.15
9 3.35
10 3.3
11 3.2
12 3.35
13 3.75
14 3.15
15 3.2
16 3.6
17 3.25
18 3.3
19 3.3
20 3.05
};

\addplot [thick, steelblue31119180, mark=*, mark size=2, mark options={solid}]
table {%
2 1.55
3 3.7
4 4.65
5 5.25
6 6.55
7 7.275
8 8.425
9 9.25
10 10
11 11.425
12 11.525
13 12.45
14 13.6
15 14.25
16 14.5
17 15.65
18 16.775
19 17.125
20 18.1
};

\addplot [thick, darkorange25512714, mark=*, mark size=2, mark options={solid}]
table {%
2 1.55
3 2.6
4 2.5
5 2.6
6 2.55
7 2.8
8 2.775
9 2.725
10 2.75
11 2.625
12 2.6
13 2.75
14 2.575
15 2.6
16 2.7
17 2.775
18 2.525
19 2.75
20 2.45
};
\end{axis}

\end{tikzpicture}}
    \end{minipage}
    \begin{minipage}[c]{0.5\textwidth}
    \caption{Error at $x=1$ averaged over 20 repetitions using QCL circuits structured as given in Figure \ref{circ_simple_example_2} on a simulator with 2000 shots and a CNOT error of 1 \%. The errors for different qubit numbers with circular entanglement (blue line) and linear entanglement (orange line) are compared.}
    \label{fig:circ_ent}
    \end{minipage}
\end{figure*}
\subsection{Comparison of Classical Optimizers}
\label{sec_comp_opt}
As a last step before implementing QCL for different examples, we want to investigate which classical optimizer is best suited to optimize the parameters in the case of exact statevector simulations and for noise model 3. To do this, we use the circuit in Figure \ref{circ_simple_example_depth} with $R_Y(\arcsin(x))$ data encoding and the cost function introduced in Equation \eqref{cost_function_post} to learn the function $f(x) = x^3$.
The cost function is evaluated on 10 equidistant training points and is classically minimized using the Simultaneous Perturbation Stochastic Approximation (SPSA) \cite{Spall1998ANOO}, Sequential Least Squares Programming (SLSQP) \cite{Kraft.1988}, and Constrained Optimization BY Linear Approximation (COBYLA) \cite{Powell.1994}. The SPSA algorithm is run with the default Qiskit settings, while the SLSQP and COBYLA algorithms are executed with the default SciPy minimizer settings \cite{Virtanen.2020}. The final learned functions are plotted in Figure \ref{x^3_optimizers} (a) and (b).
The initial function resulting from the randomly selected start parameters is also shown (black dashed lines in Figure \ref{x^3_optimizers} (a) and (b)). The values of the cost function versus the number of cost function evaluations are shown in Figure \ref{x^3_optimizers} (c) and (d).
\begin{figure*}[!tp]
\centering
\subfloat[]{%
\scalebox{0.6}{\input{figures/optimizer_comparison/function_approximation}}}
\hfill
\subfloat[]{%
\scalebox{0.6}{\input{figures/optimizer_comparison/function_approximation_noise}}}
\\
\subfloat[]{%
\scalebox{0.6}{\input{figures/optimizer_comparison/convergence}}}
\hfill
\subfloat[]{%
\scalebox{0.6}{
\begin{tikzpicture}
\definecolor{green01270}{RGB}{0,127,0}
\definecolor{dimgray85}{RGB}{85,85,85}
\definecolor{gainsboro229}{RGB}{229,229,229}
\definecolor{lightgray204}{RGB}{204,204,204}

\begin{axis}[
width=10cm,  
height=8cm,  
axis line style={lightgray204},
legend cell align={left},
legend style={
  fill opacity=0.8,
  draw opacity=1,
  text opacity=1,
  anchor=north east,
  draw=none
},
log basis y={10},
tick align=outside,
tick pos=left,
x grid style={lightgray204},
xlabel=\textcolor{dimgray85}{Cost Function Evaluations},
xmajorgrids,
xmin=-2, xmax=130,
xtick style={color=dimgray85},
y grid style={lightgray204},
ylabel=\textcolor{dimgray85}{Cost Function Value},
ymajorgrids,
ymode=log,
ytick style={color=dimgray85},
ymin=0.003, ymax=20,
ytick={0.01,0.1,1,10},
yticklabels={
  \(\displaystyle {10^{-2}}\),
  \(\displaystyle {10^{-1}}\),
  \(\displaystyle {10^{0}}\),
  \(\displaystyle {10^{1}}\),
}
]
\addplot [semithick, red]
table {%
1 4.65608315643663
2 3.9894357007439
3 5.42439233701276
4 4.92721809679329
5 5.16466687155323
6 3.86466981366571
7 5.43873031056558
8 3.72445479210193
9 7.87244361514719
10 3.9413545177535
11 4.93383995555872
12 3.56728612642291
13 5.5723982638166
14 3.76287101728711
15 2.92655224922127
16 7.97512540741057
17 9.0510269843653
18 3.6620052976712
19 3.6464400218413
20 7.29930561020892
21 3.33302644356969
22 6.84778820510604
23 6.35721949267806
24 3.9031372873557
25 5.99983661926242
26 3.48620731029123
27 4.8440962540498
28 3.63961177317189
29 5.88515940905666
30 3.57430807912525
31 3.60692194140234
32 5.75463207100453
33 5.86630631714994
34 5.36326906238999
35 5.86526769811016
36 4.30453095714994
37 7.27756495308958
38 4.51668399177271
39 4.66613511934472
40 5.26079015276036
41 4.56081162041468
42 4.82229575495515
43 6.68644542387147
44 3.29257682206077
45 5.0883149110594
46 4.14455039769864
47 7.0070122189332
48 3.99819377986599
49 4.78844803259575
50 4.82223626392634
51 5.6205482757782
52 3.32360088724596
53 4.94649797929373
54 4.14540467405511
55 3.05120120785421
56 3.92654474197411
57 1.75537229933352
58 4.61918519831784
59 2.58212044259997
60 7.97000264799247
61 2.20281091658632
62 1.60113828831042
63 1.99941683047995
64 1.57530947689951
65 1.86743707535124
66 1.64303837290034
67 1.63091013681076
68 2.54507448245749
69 1.67741664478396
70 2.08325594102266
71 1.25861775430316
72 2.19412328998082
73 2.59122556032619
74 1.68309938777048
75 1.56332660806103
76 1.34403246434431
77 1.37262277521834
78 1.52487449453932
79 1.25624202701928
80 1.81474244621369
81 1.78110308813357
82 1.55326456749462
83 1.47695833980683
84 1.557148406799
85 1.44235419930466
86 1.8006493687682
87 1.79561741799916
88 1.38167355876138
89 1.98660577915425
90 1.18762199807579
91 1.28293971949781
92 1.43728449949718
93 1.21734547372149
94 1.29960819082395
95 1.49831012206145
96 1.54802365299005
97 1.47498734983841
98 1.22361926360973
99 1.52937971887284
100 1.16574300147107
101 1.2795947960557
102 1.85646329430912
103 1.42045519619645
104 2.12172361614165
105 1.92175539814344
106 1.77786841982481
107 1.47494463410437
108 1.23562704046769
109 1.38059267334621
110 1.75096387744135
111 1.95924862780499
112 1.13746473407325
113 1.23814676477314
114 1.27257723991547
115 1.33166119926837
116 1.25283106133039
117 1.75368559722362
118 1.74867553344159
119 1.20866879466815
120 0.950108293230737
 121 1.08919716468839
 122 1.37333778802421
 123 1.10818896523412
};
\addlegendentry{SPSA}
\addplot [thick, green01270]
table {%
1 5.31261006321304
2 5.25816619901551
3 5.24153532521578
4 5.26357692466708
5 5.09490923056557
6 5.15750145553128
7 5.31711333618972
8 5.30338490409095
9 5.13848887939959
10 5.17382937597024
11 5.23230826241179
};
\addlegendentry{SLSQP}
\addplot [semithick, orange]
table {%
1 5.33115936636804
2 5.72164003166297
3 1.59799110985227
4 3.78907688900179
5 1.81123388488656
6 2.36550013454363
7 3.84734999874116
8 2.31670350354226
9 2.42416117432415
10 1.47278153920755
11 2.00714310025008
12 1.49454683170305
13 2.53305022698971
14 1.34791596993457
15 1.69887838361285
16 1.4084893128701
17 1.07003008683588
18 1.0572924731273
19 0.833238266619272
20 0.960702413270554
21 0.762438140104275
22 1.62059460772611
23 0.792300761729863
24 0.908027148162983
25 0.827204658349246
26 0.872389874910613
27 0.815756739716235
28 0.696500430808693
29 0.663375945208272
30 0.758085745575921
31 0.806142696283698
32 0.649598379633308
33 0.409809307511124
34 0.231489614516032
35 0.290332400501044
36 0.242042759419381
37 0.214966781915937
38 0.282089261426598
39 0.205597028891283
40 0.182063222236736
41 0.15666189366469
42 0.135773384621321
43 0.204832165361301
44 0.0772819864071298
45 0.177714186676404
46 0.0697601716977507
47 0.15036847072012
48 0.0504103979827076
49 0.0960481430705634
50 0.08610036525199
51 0.111568624870841
52 0.0385316048145559
53 0.0578926705627873
54 0.0201034267013836
55 0.0446923538314815
56 0.00756385639820557
57 0.0219060675513088
58 0.0312673911006726
59 0.0742082929420485
60 0.0118183897390146
61 0.0251357722185154
62 0.0261157790772747
63 0.0115068811017937
64 0.0239350851201452
65 0.0117233513018207
66 0.0298271985079385
67 0.013395472441893
68 0.0142640847928418
69 0.0140837830072539
70 0.00790582051754976
71 0.015946019127422
72 0.0145709575828619
73 0.0121377524740777
74 0.0142047674454065
75 0.0116139133596558
76 0.00591283995239283
77 0.0180277972007703
78 0.0182365742799372
79 0.0166600348895428
80 0.0182961271470644
81 0.00484258621034575
82 0.015101856474291
83 0.0330077618210398
84 0.0163581365852004
85 0.0303814292413617
86 0.0151895964863137
87 0.0122902537750388
88 0.0210675488984206
89 0.0156895009090403
90 0.00655363049824793
91 0.0132947648786687
92 0.0122522036066974
93 0.0157726417959809
94 0.0141736702257863
95 0.0159594022609325
96 0.0148843731629415
97 0.0130286881056379
98 0.00904750421628901
99 0.00926706591130939
100 0.0183016731209623
101 0.0197101811269874
102 0.0139611715252022
103 0.0120401279968802
104 0.00666180467271732
105 0.0138155556285714
106 0.0197882168124454
107 0.0137598646117389
108 0.0154505838092984
109 0.012370234810402
110 0.0191536530762879
111 0.0139205723054572
112 0.0168788752347777
113 0.0102804093685272
114 0.020212000351885
115 0.0201459689843365
116 0.00537720362891258
117 0.00838910426221866
118 0.0136521388720999
119 0.0117765971951952
120 0.0242276509778675
121 0.0153253008799998
122 0.00938335626684724
};
\addlegendentry{COBYLA}
\end{axis}

\end{tikzpicture}}}
\caption{$f(x) = x^3$ learned using QCL circuits with a post-processing parameter $\theta_{\text{post}}$, $R_Y(\arcsin(x))$ data encoding and a qubit number of $N=3$ on a statevector simulator (a) and a simulator with noise model 3 (derived from calibration data) and 2000 shots (b). The cost function is evaluated on 10 equidistant training points and is classically minimized using different optimizers. The initial function (dashed black line) shows $f^{\text{post}}_{QC}(x)$ with the randomly chosen starting parameters before the optimization process and $\theta_{\text{post}} = 1$. In (c) and (d) the respective values of the cost function versus the number of cost function evaluations during the optimization process are plotted.}
\label{x^3_optimizers}
\end{figure*}
In the case of exact statevector simulations (Figure \ref{x^3_optimizers} (a) and (c)), the SLSQP algorithm demonstrates the best performance in terms of final accuracy. While its convergence speed may be comparable to that of COBYLA, SLSQP achieves a notably higher precision in approximating the target function $f(x) = x^3$. This superior accuracy can be attributed to SLSQP's ability to use gradient information, which can be precisely estimated in noiseless simulations.
However, the situation changes when considering shot noise and hardware noise (Figure \ref{x^3_optimizers} (b) and (d)). In this more realistic scenario, SLSQP no longer works effectively and the optimization breaks. This is primarily due to the optimizer's reliance on gradient calculations, which become very unreliable in the presence of noise. The stochastic nature of shot noise and the additional hardware errors introduce fluctuations that can mislead gradient-based optimizers.
In contrast, the COBYLA optimizer is more effective under noisy conditions. COBYLA does not rely on gradient information, making it more robust to noise. Hence, COBYLA achieves a better approximation of the target function and maintains a stable convergence trajectory in the presence of noise.
SPSA is a popular optimizer since it determines the gradients with fewer measurements and is therefore more stable in the presence of noise. In our example, however, it is outperformed by SLSQP in the noise-free case, while COBYLA proves to be the better choice in the noisy scenario.
\section{Function Learning}
In this section, we present the learning of different example functions with a simulator as well as on ibmq\_ehningen.
We use the cost function in Equation \eqref{cost_function_post} and the circuit in Figure \ref{circ_simple_example_depth} with $R_Y(\arcsin(x))$ data encoding and a depth of $D=3$ as this depth has proven to be suitable for representing complicated functions without creating a circuit that is too deep.
\subsection{Simulator}
\label{chap:QCL_sim}
We now learn the functions ${f_1(x) = x^3}$, ${f_2(x) = x^3-x^2+1}$ and ${f_3(x) = \sin(2x)}$ on the interval $[-1,1]$ with the help of an exact statevector simulator without shot noise.
\begin{figure*}[!tp]
    \centering
  \subfloat[]{%
       \scalebox{0.53}{
\begin{tikzpicture}

\definecolor{dimgray85}{RGB}{85,85,85}
\definecolor{gainsboro229}{RGB}{229,229,229}
\definecolor{green01270}{RGB}{0,127,0}
\definecolor{lightgray204}{RGB}{204,204,204}

\begin{axis}[
nodes={scale=0.9, transform shape},
axis line style={lightgray204},
legend cell align={left},
legend style={
  fill opacity=0.8,
  draw opacity=1,
  text opacity=1,
  at={(0.03,0.97)},
  anchor=north west,
  draw=none
},
tick align=outside,
tick pos=left,
x grid style={lightgray204},
xlabel=\textcolor{dimgray85}{$x$},
xmajorgrids,
xmin=-1, xmax=1,
xtick style={color=dimgray85},
y grid style={lightgray204},
ylabel=\textcolor{dimgray85}{$f^{\text{post}}_{QC}(x)$},
ymajorgrids,
ymin=-1.1, ymax=1.1,
ytick style={color=dimgray85}
]
\addplot [thick, red]
table {%
-1 -1.00070189919001
-0.97979797979798 -0.937439948791941
-0.95959595959596 -0.880291235386097
-0.939393939393939 -0.825907747333592
-0.919191919191919 -0.773965086486258
-0.898989898989899 -0.724318928735796
-0.878787878787879 -0.676877455549146
-0.858585858585859 -0.631570026291805
-0.838383838383838 -0.588336133690739
-0.818181818181818 -0.547120660659018
-0.797979797979798 -0.50787156625581
-0.777777777777778 -0.470538650746018
-0.757575757575758 -0.435072851221004
-0.737373737373737 -0.401425818821727
-0.717171717171717 -0.369549654637614
-0.696969696969697 -0.339396739409421
-0.676767676767677 -0.310919620905076
-0.656565656565657 -0.284070937927368
-0.636363636363636 -0.258803368232315
-0.616161616161616 -0.235069592418003
-0.595959595959596 -0.212822268690216
-0.575757575757576 -0.192014015158647
-0.555555555555556 -0.172597397419494
-0.535353535353535 -0.154524919891699
-0.515151515151515 -0.137749019843172
-0.494949494949495 -0.122222063358404
-0.474747474747475 -0.107896342714007
-0.454545454545454 -0.0947240747778576
-0.434343434343434 -0.082657400152279
-0.414141414141414 -0.0716483828562407
-0.393939393939394 -0.0616490103950298
-0.373737373737374 -0.0526111941047931
-0.353535353535353 -0.0444867696877657
-0.333333333333333 -0.0372274978750224
-0.313131313131313 -0.0307850651692234
-0.292929292929293 -0.0251110846314725
-0.272727272727273 -0.0201570966851902
-0.252525252525252 -0.0158745699165684
-0.232323232323232 -0.0122149018561303
-0.212121212121212 -0.00912941972980884
-0.191919191919192 -0.00656938117080303
-0.171717171717172 -0.00448597488572936
-0.151515151515151 -0.00283032127025025
-0.131313131313131 -0.00155347297064208
-0.111111111111111 -0.00060641538876263
-0.0909090909090908 5.99328714290604e-05
-0.0707070707070706 0.000494719617046077
-0.0505050505050504 0.000747158740021168
-0.0303030303030303 0.000866529951970403
-0.0101010101010101 0.000902178580509094
0.0101010101010102 0.00090351542717647
0.0303030303030305 0.000920016687009648
0.0505050505050506 0.00100122392969083
0.0707070707070707 0.00119674414210962
0.0909090909090911 0.00155624983208693
0.111111111111111 0.00212947919287713
0.131313131313131 0.0029662363279466
0.151515151515152 0.00411639153531927
0.171717171717172 0.00562988165053511
0.191919191919192 0.00755671044690914
0.212121212121212 0.00994694909134318
0.232323232323232 0.0128507366532985
0.252525252525253 0.0163182806637631
0.272727272727273 0.0203998577199618
0.292929292929293 0.0251458141301517
0.313131313131313 0.0306065665910776
0.333333333333333 0.0368326028881994
0.353535353535354 0.0438744826057778
0.373737373737374 0.0517828378297521
0.393939393939394 0.0606083738210758
0.414141414141414 0.0704018696301084
0.434343434343434 0.0812141786133921
0.454545454545455 0.0930962288018861
0.474747474747475 0.106099023053244
0.494949494949495 0.120273638898759
0.515151515151515 0.135671227965756
0.535353535353535 0.152343014815617
0.555555555555556 0.170340294981781
0.575757575757576 0.189714431914459
0.595959595959596 0.210516852429853
0.616161616161616 0.232799040106832
0.636363636363636 0.256612525850813
0.656565656565657 0.282008874518284
0.676767676767677 0.309039666009972
0.696969696969697 0.337756468505232
0.717171717171717 0.368210800372219
0.737373737373737 0.400454075485458
0.757575757575758 0.434537523748278
0.777777777777778 0.470512073693312
0.797979797979798 0.50842817547193
0.818181818181818 0.548335527025578
0.838383838383838 0.590282636697385
0.858585858585859 0.634316095924707
0.878787878787879 0.68047930631422
0.898989898989899 0.728810098155533
0.919191919191919 0.779335855929516
0.939393939393939 0.83206218279268
0.95959595959596 0.886940786682983
0.97979797979798 0.94373886544551
1 0.999132909749222
};
\addlegendentry{Final Function}
\addplot [thick, dashed, black]
table {%
-1 -0.188376708321095
-0.97979797979798 -0.354660111384183
-0.95959595959596 -0.413031523734479
-0.939393939393939 -0.45206741406779
-0.919191919191919 -0.480608843795883
-0.898989898989899 -0.502201052099808
-0.878787878787879 -0.518724717814281
-0.858585858585859 -0.531335117711849
-0.838383838383838 -0.540810446761827
-0.818181818181818 -0.547709355467066
-0.797979797979798 -0.552452046791891
-0.777777777777778 -0.555366055440026
-0.757575757575758 -0.556713948017231
-0.737373737373737 -0.556711019687414
-0.717171717171717 -0.555537105727158
-0.696969696969697 -0.553344754509808
-0.676767676767677 -0.55026505640772
-0.656565656565657 -0.546411909302546
-0.636363636363636 -0.541885210031847
-0.616161616161616 -0.536773288786576
-0.595959595959596 -0.531154797815427
-0.575757575757576 -0.525100198935006
-0.555555555555556 -0.518672950855246
-0.535353535353535 -0.51193046833912
-0.515151515151515 -0.504924905464173
-0.494949494949495 -0.497703801529684
-0.474747474747475 -0.490310618447503
-0.454545454545454 -0.48278519147904
-0.434343434343434 -0.475164110093488
-0.414141414141414 -0.467481041961737
-0.393939393939394 -0.459767010286037
-0.373737373737374 -0.452050632535013
-0.353535353535353 -0.444358327023747
-0.333333333333333 -0.436714492519356
-0.313131313131313 -0.42914166507068
-0.292929292929293 -0.421660655488442
-0.272727272727273 -0.414290670290087
-0.252525252525252 -0.407049418434339
-0.232323232323232 -0.399953205777027
-0.212121212121212 -0.393017018860758
-0.191919191919192 -0.386254599391007
-0.171717171717172 -0.379678510537607
-0.151515151515151 -0.373300196024357
-0.131313131313131 -0.367130032822808
-0.111111111111111 -0.361177378143775
-0.0909090909090908 -0.3554506113169
-0.0707070707070706 -0.349957171061203
-0.0505050505050504 -0.344703588574979
-0.0303030303030303 -0.339695516809234
-0.0101010101010101 -0.334937756233316
0.0101010101010102 -0.330434277352616
0.0303030303030305 -0.326188240195214
0.0505050505050506 -0.32220201094566
0.0707070707070707 -0.31847717586895
0.0909090909090911 -0.315014552635304
0.111111111111111 -0.311814199125695
0.131313131313131 -0.308875419768675
0.151515151515152 -0.306196769430039
0.171717171717172 -0.303776054847711
0.191919191919192 -0.301610333574148
0.212121212121212 -0.299695910356766
0.232323232323232 -0.298028330852559
0.252525252525253 -0.296602372535299
0.272727272727273 -0.295412032611084
0.292929292929293 -0.294450512709467
0.313131313131313 -0.293710200060925
0.333333333333333 -0.29318264480507
0.353535353535354 -0.292858532994974
0.373737373737374 -0.29272765476783
0.393939393939394 -0.292778867036474
0.414141414141414 -0.293000049914189
0.434343434343434 -0.293378055909101
0.454545454545455 -0.293898650704131
0.474747474747475 -0.294546444060294
0.494949494949495 -0.295304809026495
0.515151515151515 -0.296155787182459
0.535353535353535 -0.297079977047645
0.555555555555556 -0.298056402008101
0.575757575757576 -0.299062353074216
0.595959595959596 -0.300073200382394
0.616161616161616 -0.301062165441545
0.636363636363636 -0.302000043474625
0.656565656565657 -0.302854861471043
0.676767676767677 -0.303591452209996
0.696969696969697 -0.304170916680985
0.717171717171717 -0.304549935615781
0.737373737373737 -0.30467987290103
0.757575757575758 -0.304505585373674
0.777777777777778 -0.303963807550666
0.797979797979798 -0.302980902393767
0.818181818181818 -0.301469633057739
0.838383838383838 -0.299324359063802
0.858585858585859 -0.296413567339504
0.878787878787879 -0.292567609473583
0.898989898989899 -0.287557117656505
0.919191919191919 -0.281051337290349
0.939393939393939 -0.272526558338426
0.95959595959596 -0.261020694098913
0.97979797979798 -0.244190088872919
1 -0.199157333644476
};
\addlegendentry{Initial Function}
\addplot [semithick, blue, mark=*, mark size=2, mark options={solid}, only marks]
table {%
-1 -1
-0.894736842105263 -0.716285172765709
-0.789473684210526 -0.49205423531127
-0.684210526315789 -0.320309082956699
-0.578947368421053 -0.194051611022015
-0.473684210526316 -0.106283714827234
-0.368421052631579 -0.050007289692375
-0.263157894736842 -0.0182242309374544
-0.157894736842105 -0.00393643388249016
-0.0526315789473685 -0.000145793847499636
0.0526315789473684 0.000145793847499635
0.157894736842105 0.00393643388249016
0.263157894736842 0.0182242309374544
0.368421052631579 0.0500072896923749
0.473684210526316 0.106283714827234
0.578947368421053 0.194051611022015
0.684210526315789 0.320309082956699
0.789473684210526 0.492054235311269
0.894736842105263 0.716285172765709
1 1
};
\addlegendentry{Training Points}
\end{axis}

\end{tikzpicture}}}
    \hfill
  \subfloat[]{%
        \scalebox{0.53}{
\begin{tikzpicture}

\definecolor{dimgray85}{RGB}{85,85,85}
\definecolor{gainsboro229}{RGB}{229,229,229}
\definecolor{green01270}{RGB}{0,127,0}
\definecolor{lightgray204}{RGB}{204,204,204}

\begin{axis}[
nodes={scale=0.9, transform shape},
axis line style={lightgray204},
legend cell align={left},
legend style={
  fill opacity=0.8,
  draw opacity=1,
  text opacity=1,
  at={(0.97,0.03)},
  anchor=south east,
  draw=none
},
tick align=outside,
tick pos=left,
x grid style={lightgray204},
xlabel=\textcolor{dimgray85}{$x$},
xmajorgrids,
xmin=-1, xmax=1,
xtick style={color=dimgray85},
y grid style={lightgray204},
ylabel=\textcolor{dimgray85}{$f^{\text{post}}_{QC}(x)$},
ymajorgrids,
ymin=-1.1, ymax=1.1,
ytick style={color=dimgray85}
]
\addplot [thick, red]
table {%
-1 -1.00036764913141
-0.97979797979798 -0.860673057806911
-0.95959595959596 -0.763978008551108
-0.939393939393939 -0.674742019599352
-0.919191919191919 -0.589990224125955
-0.898989898989899 -0.5086749224262
-0.878787878787879 -0.430307619324061
-0.858585858585859 -0.354624153152445
-0.838383838383838 -0.281467025215997
-0.818181818181818 -0.210735044259115
-0.797979797979798 -0.14235885526082
-0.777777777777778 -0.0762879362266724
-0.757575757575758 -0.0124832173478751
-0.737373737373737 0.0490873248886889
-0.717171717171717 0.108451402885653
-0.696969696969697 0.165634169053193
-0.676767676767677 0.220659364314779
-0.656565656565657 0.27355008553798
-0.636363636363636 0.324329305437909
-0.616161616161616 0.37302022804807
-0.595959595959596 0.419646532884454
-0.575757575757576 0.464232542575031
-0.555555555555556 0.506803337183058
-0.535353535353535 0.547384831019952
-0.515151515151515 0.586003822857735
-0.494949494949495 0.622688027179912
-0.474747474747475 0.657466091883216
-0.454545454545454 0.690367606305058
-0.434343434343434 0.721423102375479
-0.414141414141414 0.750664050930554
-0.393939393939394 0.778122854679124
-0.373737373737374 0.803832838920948
-0.353535353535353 0.827828240827728
-0.333333333333333 0.850144197888006
-0.313131313131313 0.870816735961793
-0.292929292929293 0.889882757275346
-0.272727272727273 0.907380028600767
-0.252525252525252 0.923347169800693
-0.232323232323232 0.937823642870254
-0.212121212121212 0.95084974157224
-0.191919191919192 0.96246658173441
-0.171717171717172 0.972716092257418
-0.151515151515151 0.981641006866834
-0.131313131313131 0.989284856631446
-0.111111111111111 0.995691963261796
-0.0909090909090908 1.00090743319701
-0.0707070707070706 1.00497715248372
-0.0505050505050504 1.00794778244797
-0.0303030303030303 1.00986675615909
-0.0101010101010101 1.01078227568343
0.0101010101010102 1.01074331012488
0.0303030303030305 1.00979959444919
0.0505050505050506 1.00800162908856
0.0707070707070707 1.00540068032331
0.0909090909090911 1.00204878143688
0.111111111111111 0.997998734640414
0.131313131313131 0.993304113762536
0.151515151515152 0.988019267698444
0.171717171717172 0.982199324611028
0.191919191919192 0.975900196873779
0.212121212121212 0.969178586741352
0.232323232323232 0.962091992728218
0.252525252525253 0.954698716668295
0.272727272727273 0.947057871418351
0.292929292929293 0.9392293891544
0.313131313131313 0.931274030192353
0.333333333333333 0.923253392240343
0.353535353535354 0.915229919958756
0.373737373737374 0.907266914662507
0.393939393939394 0.899428543945314
0.414141414141414 0.891779850933204
0.434343434343434 0.884386762778222
0.454545454545455 0.877316097874888
0.474747474747475 0.870635571109997
0.494949494949495 0.864413796224451
0.515151515151515 0.858720284051234
0.535353535353535 0.853625434963402
0.555555555555556 0.849200523272464
0.575757575757576 0.845517670491104
0.595959595959596 0.842649803210378
0.616161616161616 0.840670589683685
0.636363636363636 0.839654346815659
0.656565656565657 0.839675905745519
0.676767676767677 0.840810418986529
0.696969696969697 0.843133084147732
0.717171717171717 0.846718746965022
0.737373737373737 0.851641326852389
0.757575757575758 0.857972976375772
0.777777777777778 0.865782832595715
0.797979797979798 0.87513512515083
0.818181818181818 0.886086237060591
0.838383838383838 0.898679992392444
0.858585858585859 0.912939794556043
0.878787878787879 0.928854826458247
0.898989898989899 0.946354164767764
0.919191919191919 0.965253670331038
0.939393939393939 0.985132215040456
0.95959595959596 1.00498032174154
0.97979797979798 1.0217683103223
1 0.998493752602622
};
\addlegendentry{Final Function}
\addplot [thick, dashed,black]
table {%
-1 -0.188376708321095
-0.97979797979798 -0.354660111384183
-0.95959595959596 -0.413031523734479
-0.939393939393939 -0.45206741406779
-0.919191919191919 -0.480608843795883
-0.898989898989899 -0.502201052099808
-0.878787878787879 -0.518724717814281
-0.858585858585859 -0.531335117711849
-0.838383838383838 -0.540810446761827
-0.818181818181818 -0.547709355467066
-0.797979797979798 -0.552452046791891
-0.777777777777778 -0.555366055440026
-0.757575757575758 -0.556713948017231
-0.737373737373737 -0.556711019687414
-0.717171717171717 -0.555537105727158
-0.696969696969697 -0.553344754509808
-0.676767676767677 -0.55026505640772
-0.656565656565657 -0.546411909302546
-0.636363636363636 -0.541885210031847
-0.616161616161616 -0.536773288786576
-0.595959595959596 -0.531154797815427
-0.575757575757576 -0.525100198935006
-0.555555555555556 -0.518672950855246
-0.535353535353535 -0.51193046833912
-0.515151515151515 -0.504924905464173
-0.494949494949495 -0.497703801529684
-0.474747474747475 -0.490310618447503
-0.454545454545454 -0.48278519147904
-0.434343434343434 -0.475164110093488
-0.414141414141414 -0.467481041961737
-0.393939393939394 -0.459767010286037
-0.373737373737374 -0.452050632535013
-0.353535353535353 -0.444358327023747
-0.333333333333333 -0.436714492519356
-0.313131313131313 -0.42914166507068
-0.292929292929293 -0.421660655488442
-0.272727272727273 -0.414290670290087
-0.252525252525252 -0.407049418434339
-0.232323232323232 -0.399953205777027
-0.212121212121212 -0.393017018860758
-0.191919191919192 -0.386254599391007
-0.171717171717172 -0.379678510537607
-0.151515151515151 -0.373300196024357
-0.131313131313131 -0.367130032822808
-0.111111111111111 -0.361177378143775
-0.0909090909090908 -0.3554506113169
-0.0707070707070706 -0.349957171061203
-0.0505050505050504 -0.344703588574979
-0.0303030303030303 -0.339695516809234
-0.0101010101010101 -0.334937756233316
0.0101010101010102 -0.330434277352616
0.0303030303030305 -0.326188240195214
0.0505050505050506 -0.32220201094566
0.0707070707070707 -0.31847717586895
0.0909090909090911 -0.315014552635304
0.111111111111111 -0.311814199125695
0.131313131313131 -0.308875419768675
0.151515151515152 -0.306196769430039
0.171717171717172 -0.303776054847711
0.191919191919192 -0.301610333574148
0.212121212121212 -0.299695910356766
0.232323232323232 -0.298028330852559
0.252525252525253 -0.296602372535299
0.272727272727273 -0.295412032611084
0.292929292929293 -0.294450512709467
0.313131313131313 -0.293710200060925
0.333333333333333 -0.29318264480507
0.353535353535354 -0.292858532994974
0.373737373737374 -0.29272765476783
0.393939393939394 -0.292778867036474
0.414141414141414 -0.293000049914189
0.434343434343434 -0.293378055909101
0.454545454545455 -0.293898650704131
0.474747474747475 -0.294546444060294
0.494949494949495 -0.295304809026495
0.515151515151515 -0.296155787182459
0.535353535353535 -0.297079977047645
0.555555555555556 -0.298056402008101
0.575757575757576 -0.299062353074216
0.595959595959596 -0.300073200382394
0.616161616161616 -0.301062165441545
0.636363636363636 -0.302000043474625
0.656565656565657 -0.302854861471043
0.676767676767677 -0.303591452209996
0.696969696969697 -0.304170916680985
0.717171717171717 -0.304549935615781
0.737373737373737 -0.30467987290103
0.757575757575758 -0.304505585373674
0.777777777777778 -0.303963807550666
0.797979797979798 -0.302980902393767
0.818181818181818 -0.301469633057739
0.838383838383838 -0.299324359063802
0.858585858585859 -0.296413567339504
0.878787878787879 -0.292567609473583
0.898989898989899 -0.287557117656505
0.919191919191919 -0.281051337290349
0.939393939393939 -0.272526558338426
0.95959595959596 -0.261020694098913
0.97979797979798 -0.244190088872919
1 -0.199157333644476
};
\addlegendentry{Initial Function}
\addplot [semithick, blue, mark=*, mark size=2, mark options={solid}, only marks]
table {%
-1 -1
-0.894736842105263 -0.516839189386208
-0.789473684210526 -0.115322933372212
-0.684210526315789 0.211546872721971
-0.578947368421053 0.470768333576323
-0.473684210526316 0.669339553870827
-0.368421052631579 0.814258638285464
-0.263157894736842 0.912523691500219
-0.157894736842105 0.971132818195072
-0.0526315789473685 0.997084123050007
0.0526315789473684 0.997375710745007
0.157894736842105 0.979005685960052
0.263157894736842 0.948972153375128
0.368421052631579 0.914273217670214
0.473684210526316 0.881906983525295
0.578947368421053 0.858871555620353
0.684210526315789 0.85216503863537
0.789473684210526 0.868785537250328
0.894736842105263 0.915731156145211
1 1
};
\addlegendentry{Training Points}

\end{axis}

\end{tikzpicture}}}
    \hfill
  \subfloat[]{%
        \scalebox{0.53}{
\begin{tikzpicture}

\definecolor{dimgray85}{RGB}{85,85,85}
\definecolor{gainsboro229}{RGB}{229,229,229}
\definecolor{green01270}{RGB}{0,127,0}
\definecolor{lightgray204}{RGB}{204,204,204}

\begin{axis}[
nodes={scale=0.9, transform shape},
axis line style={lightgray204},
legend cell align={left},
legend style={
  fill opacity=0.8,
  draw opacity=1,
  text opacity=1,
  at={(0.03,0.97)},
  anchor=north west,
  draw=none
},
tick align=outside,
tick pos=left,
x grid style={lightgray204},
xlabel=\textcolor{dimgray85}{$x$},
xmajorgrids,
xmin=-1, xmax=1,
xtick style={color=dimgray85},
y grid style={lightgray204},
ylabel=\textcolor{dimgray85}{$f^{\text{post}}_{QC}(x)$},
ymajorgrids,
ymin=-1.1, ymax=1.1,
ytick style={color=dimgray85}
]
\addplot [thick, red]
table {%
-1 -0.910656959628129
-0.97979797979798 -0.913542193668109
-0.95959595959596 -0.928125571809869
-0.939393939393939 -0.942843129675381
-0.919191919191919 -0.956532071266527
-0.898989898989899 -0.968744838813313
-0.878787878787879 -0.979263352454794
-0.858585858585859 -0.987971430625463
-0.838383838383838 -0.994806809575284
-0.818181818181818 -0.99973915535193
-0.797979797979798 -1.00275860076516
-0.777777777777778 -1.00386919479839
-0.757575757575758 -1.00308489153976
-0.737373737373737 -1.00042695777451
-0.717171717171717 -0.995922223000136
-0.696969696969697 -0.98960185505056
-0.676767676767677 -0.981500477363293
-0.656565656565657 -0.971655516099883
-0.636363636363636 -0.960106706528694
-0.616161616161616 -0.946895712602122
-0.595959595959596 -0.932065828793018
-0.575757575757576 -0.915661742890503
-0.555555555555556 -0.897729344761705
-0.535353535353535 -0.878315570315325
-0.515151515151515 -0.857468272801711
-0.494949494949495 -0.835236115610045
-0.474747474747475 -0.811668482164343
-0.454545454545454 -0.786815399561581
-0.434343434343434 -0.760727473359116
-0.414141414141414 -0.733455831486283
-0.393939393939394 -0.705052075682238
-0.373737373737374 -0.675568239187006
-0.353535353535353 -0.645056749662638
-0.333333333333333 -0.613570396515201
-0.313131313131313 -0.581162301940269
-0.292929292929293 -0.547885895134396
-0.272727272727273 -0.513794889210476
-0.252525252525252 -0.478943260431201
-0.232323232323232 -0.443385229436471
-0.212121212121212 -0.407175244190398
-0.191919191919192 -0.370367964414241
-0.171717171717172 -0.333018247304793
-0.151515151515151 -0.295181134364995
-0.131313131313131 -0.256911839195914
-0.111111111111111 -0.218265736117683
-0.0909090909090908 -0.179298349502028
-0.0707070707070706 -0.140065343711398
-0.0505050505050504 -0.100622513549723
-0.0303030303030303 -0.0610257751377754
-0.0101010101010101 -0.0213311571323556
0.0101010101010102 0.0184052077868024
0.0303030303030305 0.0581270912353469
0.0505050505050506 0.0977781773188496
0.0707070707070707 0.137302070774094
0.0909090909090911 0.176642305128301
0.111111111111111 0.215742350949649
0.131313131313131 0.254545624265853
0.151515151515152 0.292995495232716
0.171717171717172 0.331035297141328
0.191919191919192 0.36860833586134
0.212121212121212 0.405657899828862
0.232323232323232 0.442127270701341
0.252525252525253 0.477959734818863
0.272727272727273 0.513098595632292
0.292929292929293 0.547487187284668
0.313131313131313 0.581068889563972
0.333333333333333 0.613787144484718
0.353535353535354 0.64558547480442
0.373737373737374 0.676407504841266
0.393939393939394 0.706196984034705
0.414141414141414 0.734897813785205
0.434343434343434 0.762454078228963
0.454545454545455 0.78881007975526
0.474747474747475 0.813910380269032
0.494949494949495 0.837699849452935
0.515151515151515 0.860123721611392
0.535353535353535 0.881127663111182
0.555555555555556 0.90065785300779
0.575757575757576 0.91866108021981
0.595959595959596 0.935084861666424
0.616161616161616 0.949877587235845
0.636363636363636 0.962988699487255
0.656565656565657 0.97436891888389
0.676767676767677 0.983970529548306
0.696969696969697 0.991747746724027
0.717171717171717 0.997657196476617
0.737373737373737 1.00165855262793
0.757575757575758 1.0037153989113
0.777777777777778 1.00379642206446
0.797979797979798 1.00187710576043
0.818181818181818 0.997942209145946
0.838383838383838 0.991989526025037
0.858585858585859 0.984035840595807
0.878787878787879 0.974126888622541
0.898989898989899 0.962355213131788
0.919191919191919 0.948895259172908
0.939393939393939 0.934081880797375
0.95959595959596 0.91862452770836
0.97979797979798 0.904445749626445
1 0.912232341249451
};
\addlegendentry{Final Function}
\addplot [thick, dashed, black]
table {%
-1 -0.188376708321095
-0.97979797979798 -0.354660111384183
-0.95959595959596 -0.413031523734479
-0.939393939393939 -0.45206741406779
-0.919191919191919 -0.480608843795883
-0.898989898989899 -0.502201052099808
-0.878787878787879 -0.518724717814281
-0.858585858585859 -0.531335117711849
-0.838383838383838 -0.540810446761827
-0.818181818181818 -0.547709355467066
-0.797979797979798 -0.552452046791891
-0.777777777777778 -0.555366055440026
-0.757575757575758 -0.556713948017231
-0.737373737373737 -0.556711019687414
-0.717171717171717 -0.555537105727158
-0.696969696969697 -0.553344754509808
-0.676767676767677 -0.55026505640772
-0.656565656565657 -0.546411909302546
-0.636363636363636 -0.541885210031847
-0.616161616161616 -0.536773288786576
-0.595959595959596 -0.531154797815427
-0.575757575757576 -0.525100198935006
-0.555555555555556 -0.518672950855246
-0.535353535353535 -0.51193046833912
-0.515151515151515 -0.504924905464173
-0.494949494949495 -0.497703801529684
-0.474747474747475 -0.490310618447503
-0.454545454545454 -0.48278519147904
-0.434343434343434 -0.475164110093488
-0.414141414141414 -0.467481041961737
-0.393939393939394 -0.459767010286037
-0.373737373737374 -0.452050632535013
-0.353535353535353 -0.444358327023747
-0.333333333333333 -0.436714492519356
-0.313131313131313 -0.42914166507068
-0.292929292929293 -0.421660655488442
-0.272727272727273 -0.414290670290087
-0.252525252525252 -0.407049418434339
-0.232323232323232 -0.399953205777027
-0.212121212121212 -0.393017018860758
-0.191919191919192 -0.386254599391007
-0.171717171717172 -0.379678510537607
-0.151515151515151 -0.373300196024357
-0.131313131313131 -0.367130032822808
-0.111111111111111 -0.361177378143775
-0.0909090909090908 -0.3554506113169
-0.0707070707070706 -0.349957171061203
-0.0505050505050504 -0.344703588574979
-0.0303030303030303 -0.339695516809234
-0.0101010101010101 -0.334937756233316
0.0101010101010102 -0.330434277352616
0.0303030303030305 -0.326188240195214
0.0505050505050506 -0.32220201094566
0.0707070707070707 -0.31847717586895
0.0909090909090911 -0.315014552635304
0.111111111111111 -0.311814199125695
0.131313131313131 -0.308875419768675
0.151515151515152 -0.306196769430039
0.171717171717172 -0.303776054847711
0.191919191919192 -0.301610333574148
0.212121212121212 -0.299695910356766
0.232323232323232 -0.298028330852559
0.252525252525253 -0.296602372535299
0.272727272727273 -0.295412032611084
0.292929292929293 -0.294450512709467
0.313131313131313 -0.293710200060925
0.333333333333333 -0.29318264480507
0.353535353535354 -0.292858532994974
0.373737373737374 -0.29272765476783
0.393939393939394 -0.292778867036474
0.414141414141414 -0.293000049914189
0.434343434343434 -0.293378055909101
0.454545454545455 -0.293898650704131
0.474747474747475 -0.294546444060294
0.494949494949495 -0.295304809026495
0.515151515151515 -0.296155787182459
0.535353535353535 -0.297079977047645
0.555555555555556 -0.298056402008101
0.575757575757576 -0.299062353074216
0.595959595959596 -0.300073200382394
0.616161616161616 -0.301062165441545
0.636363636363636 -0.302000043474625
0.656565656565657 -0.302854861471043
0.676767676767677 -0.303591452209996
0.696969696969697 -0.304170916680985
0.717171717171717 -0.304549935615781
0.737373737373737 -0.30467987290103
0.757575757575758 -0.304505585373674
0.777777777777778 -0.303963807550666
0.797979797979798 -0.302980902393767
0.818181818181818 -0.301469633057739
0.838383838383838 -0.299324359063802
0.858585858585859 -0.296413567339504
0.878787878787879 -0.292567609473583
0.898989898989899 -0.287557117656505
0.919191919191919 -0.281051337290349
0.939393939393939 -0.272526558338426
0.95959595959596 -0.261020694098913
0.97979797979798 -0.244190088872919
1 -0.199157333644476
};
\addlegendentry{Initial Function}
\addplot [semithick, blue, mark=*, mark size=2, mark options={solid}, only marks]
table {%
-1 -0.909297426825682
-0.894736842105263 -0.976185235429829
-0.789473684210526 -0.999966780444129
-0.684210526315789 -0.979591919425577
-0.578947368421053 -0.915960362898154
-0.473684210526316 -0.811881945049832
-0.368421052631579 -0.671952547431521
-0.263157894736842 -0.502351154603513
-0.157894736842105 -0.310567003320375
-0.0526315789473685 -0.105068873765949
0.0526315789473684 0.105068873765949
0.157894736842105 0.310567003320375
0.263157894736842 0.502351154603513
0.368421052631579 0.671952547431521
0.473684210526316 0.811881945049832
0.578947368421053 0.915960362898154
0.684210526315789 0.979591919425577
0.789473684210526 0.999966780444129
0.894736842105263 0.976185235429829
1 0.909297426825682
};
\addlegendentry{Training Points}
\end{axis}

\end{tikzpicture}}}
    \\
  \subfloat[]{%
        \scalebox{0.53}{
\begin{tikzpicture}

\definecolor{dimgray85}{RGB}{85,85,85}
\definecolor{gainsboro229}{RGB}{229,229,229}
\definecolor{lightgray204}{RGB}{204,204,204}

\begin{axis}[
scaled ticks=false,
axis line style={lightgray204},
legend cell align={left},
legend style={
  fill opacity=0.8,
  draw opacity=1,
  text opacity=1,,
  draw=none
},
tick align=outside,
tick pos=left,
x grid style={lightgray204},
xlabel=\textcolor{dimgray85}{$x$},
xmajorgrids,
xmin=-1, xmax=1,
xtick style={color=dimgray85},
y grid style={lightgray204},
ylabel=\textcolor{dimgray85}{Error},
ymajorgrids,
ymin=-0.005, ymax=0.045,
ytick={0,0.01,0.02,0.03,0.04},
yticklabels={0,0.01,0.02,0.03,0.04},
ytick style={color=dimgray85}
]
\addplot [thick, red]
table {%
-1 0.000701899190011845
-0.97979797979798 0.00317011058150996
-0.95959595959596 0.00332814379495938
-0.939393939393939 0.00307074280192321
-0.919191919191919 0.00267283646326577
-0.898989898989899 0.00222927959998542
-0.878787878787879 0.00178242145783836
-0.858585858585859 0.00135343338402694
-0.838383838383838 0.000953353364282483
-0.818181818181818 0.000587829198231971
-0.797979797979798 0.000259432539406745
-0.777777777777778 3.11061643989974e-05
-0.757575757575758 0.000284193291850199
-0.737373737373737 0.000500949271207252
-0.717171717171717 0.000682944479198588
-0.696969696969697 0.000832028943884233
-0.676767676767677 0.000950219720492973
-0.656565656565657 0.00103962489911569
-0.636363636363636 0.00110239152307345
-0.616161616161616 0.00114066947775446
-0.595959595959596 0.0011565862562444
-0.575757575757576 0.00115222925553857
-0.555555555555556 0.00112963335913768
-0.535353535353535 0.00109077227328497
-0.515151515151515 0.00103755255319188
-0.494949494949495 0.000971809570653792
-0.474747474747475 0.000895304889583601
-0.454545454545454 0.000809724665160438
-0.434343434343434 0.000716678787009165
-0.414141414141414 0.000617700561401729
-0.393939393939394 0.000514246780927326
-0.373737373737374 0.000407698069035081
-0.353535353535353 0.000299359415262067
-0.333333333333333 0.0001904608379854
-0.313131313131313 8.21581271673068e-05
-0.292929292929293 2.44663687861586e-05
-0.272727272727273 0.000128402939152415
-0.252525252525252 0.000228713710437339
-0.232323232323232 0.000324531864815472
-0.212121212121212 0.000415060889051938
-0.191919191919192 0.000499573862645408
-0.171717171717172 0.000577412791677282
-0.151515151515151 0.000647987993182976
-0.131313131313131 0.00071077753358394
-0.111111111111111 0.000765326723720221
-0.0909090909090908 0.000811247672330636
-0.0707070707070706 0.000848218899226105
-0.0505050505050504 0.000875985009037213
-0.0303030303030303 0.000894356426077869
-0.0101010101010101 0.000903209190661222
0.0101010101010102 0.000902484817024341
0.0303030303030305 0.000892190212902182
0.0505050505050506 0.000872397660674787
0.0707070707070707 0.000843244859929592
0.0909090909090911 0.000804935031185349
0.111111111111111 0.000757737080394272
0.131313131313131 0.000701985823720574
0.151515151515152 0.000638082271886037
0.171717171717172 0.000566493973128453
0.191919191919192 0.000487755413460671
0.212121212121212 0.000402468472482388
0.232323232323232 0.000311302932352674
0.252525252525253 0.000214997036757428
0.272727272727273 0.000114358095619192
0.292929292929293 1.02631298929748e-05
0.313131313131313 9.63404509785049e-05
0.333333333333333 0.000204434148837647
0.353535353535354 0.000312927666725846
0.373737373737374 0.00042065820600598
0.393939393939394 0.000526389793026681
0.414141414141414 0.000628812664730694
0.434343434343434 0.000726542751877821
0.454545454545455 0.00081812131081116
0.474747474747475 0.000902014771179627
0.494949494949495 0.000976614888990784
0.515151515151515 0.00104023932422387
0.535353535353535 0.00109113280279757
0.555555555555556 0.00112746907857528
0.575757575757576 0.0011473539886489
0.595959595959596 0.00114883000411803
0.616161616161616 0.00112988283341628
0.636363636363636 0.00108845085842829
0.656565656565657 0.00102243850996847
0.676767676767677 0.000929735174611868
0.696969696969697 0.00080824196030449
0.717171717171717 0.000655909786196507
0.737373737373737 0.00047079406506223
0.757575757575758 0.000251134180875801
0.777777777777778 4.52911169307457e-06
0.797979797979798 0.000297176676712474
0.818181818181818 0.000627037168327127
0.838383838383838 0.000993149642363744
0.858585858585859 0.00139263624887498
0.878787878787879 0.00181942930723533
0.898989898989899 0.00226188981975139
0.919191919191919 0.00269793297999199
0.939393939393939 0.00308369265716457
0.95959595959596 0.00332140750192644
0.97979797979798 0.00312880607205901
1 0.000867090250777802
};
\end{axis}

\end{tikzpicture}}}
    \hfill
  \subfloat[]{%
        \scalebox{0.53}{
\begin{tikzpicture}

\definecolor{dimgray85}{RGB}{85,85,85}
\definecolor{gainsboro229}{RGB}{229,229,229}
\definecolor{lightgray204}{RGB}{204,204,204}

\begin{axis}[
scaled ticks=false,
axis line style={lightgray204},
legend cell align={left},
legend style={
  fill opacity=0.8,
  draw opacity=1,
  text opacity=1,,
  draw=none
},
tick align=outside,
tick pos=left,
x grid style={lightgray204},
xlabel=\textcolor{dimgray85}{$x$},
xmajorgrids,
xmin=-1, xmax=1,
xtick style={color=dimgray85},
y grid style={lightgray204},
ylabel=\textcolor{dimgray85}{Error},
ymajorgrids,
ymin=-0.005, ymax=0.045,
ytick={0,0.01,0.02,0.03,0.04},
yticklabels={0,0.01,0.02,0.03,0.04},
ytick style={color=dimgray85}
]
\addplot [thick, red]
table {%
-1 0.000367649131407299
-0.97979797979798 0.0399410827827426
-0.95959595959596 0.0404657763028385
-0.939393939393939 0.0366974439062273
-0.919191919191919 0.0315614831312921
-0.898989898989899 0.0260561243954499
-0.878787878787879 0.0206203935874226
-0.858585858585859 0.015468983087003
-0.838383838383838 0.0107099223022424
-0.818181818181818 0.00639493320144027
-0.797979797979798 0.00254390151827602
-0.777777777777778 0.000842120040115454
-0.757575757575758 0.00377353095223821
-0.737373737373737 0.00626777699227766
-0.717171717171717 0.00834661504490286
-0.696969696969697 0.0100343619872388
-0.676767676767677 0.0113567461831193
-0.656565656565657 0.0123401400522765
-0.636363636363636 0.0130110401668993
-0.616161616161616 0.0133957117807867
-0.595959595959596 0.0135199446652499
-0.575757575757576 0.0134088854796204
-0.555555555555556 0.0130869234479433
-0.535353535353535 0.0125776135461046
-0.515151515151515 0.0119036262893837
-0.494949494949495 0.0110867164815778
-0.474747474747475 0.0101477055132572
-0.454545454545454 0.00910647333431136
-0.434343434343434 0.0079819573020018
-0.414141414141414 0.0067921558675571
-0.393939393939394 0.00555413560943652
-0.373737373737374 0.00428404051518494
-0.353535353535353 0.00299710270040121
-0.333333333333333 0.00170765396384576
-0.313131313131313 0.000429137732810059
-0.292929292929293 0.000825878931660196
-0.272727272727273 0.00204569351436612
-0.252525252525252 0.00321945659064093
-0.232323232323232 0.00433716086831482
-0.212121212121212 0.00538963082287347
-0.191919191919192 0.00636851299477392
-0.171717171717172 0.00726626699736888
-0.151515151515151 0.0080761572689263
-0.131313131313131 0.00879224559093161
-0.111111111111111 0.00940938438662409
-0.0909090909090908 0.00992321080783032
-0.0707070707070706 0.0103301416138752
-0.0505050505050504 0.0106273688434988
-0.0303030303030303 0.0108128562787455
-0.0101010101010101 0.0108853366986428
0.0101010101010102 0.0108443099197921
0.0303030303030305 0.0106900416206247
0.0505050505050506 0.0104235629460619
0.0707070707070707 0.0100466708891093
0.0909090909090911 0.00956192944589196
0.111111111111111 0.00897267154027659
0.131313131313131 0.00828300171356977
0.151515151515152 0.00749779957367003
0.171717171717172 0.00662272399616615
0.191919191919192 0.00566421806724582
0.212121212121212 0.00462951475426399
0.232323232323232 0.00352664328438701
0.252525252525253 0.00236443620423166
0.272727272727273 0.00115253708326513
0.292929292929293 9.85911898031544e-05
0.313131313131313 0.00137765758636266
0.333333333333333 0.00267253368558285
0.353535353535354 0.00397024411437985
0.373737373737374 0.00525695684514194
0.393939393939394 0.00651797357145123
0.414141414141414 0.00773772045458421
0.434343434343434 0.00889973962979895
0.454545454545455 0.00998668198987551
0.474747474747475 0.0109803019353216
0.494949494949495 0.0118614550125383
0.515151515151515 0.0126100996758439
0.535353535353535 0.0132053048394838
0.555555555555556 0.0136252654792506
0.575757575757576 0.0138473293697637
0.595959595959596 0.0138480392072686
0.616161616161616 0.0136031960256688
0.636363636363636 0.0130879522076316
0.656565656565657 0.0122769459012415
0.676767676767677 0.0111444938805361
0.696969696969697 0.00966486782377363
0.717171717171717 0.00781269128236384
0.737373737373737 0.00556351412961753
0.757575757575758 0.00289465308689851
0.777777777777778 0.000213559619034909
0.797979797979798 0.00377588433949272
0.818181818181818 0.0077992348066469
0.838383838383838 0.0122779658006409
0.858585858585859 0.0171860114438267
0.878787878787879 0.0224630853557617
0.898989898989899 0.0279887949178514
0.919191919191919 0.0335295316892374
0.939393939393939 0.0386146982750055
0.95959595959596 0.0421853482333706
0.97979797979798 0.0411623321650476
1 0.00150624739737848
};
\end{axis}

\end{tikzpicture}}}
    \hfill
  \subfloat[]{%
        \scalebox{0.53}{
\begin{tikzpicture}

\definecolor{dimgray85}{RGB}{85,85,85}
\definecolor{gainsboro229}{RGB}{229,229,229}
\definecolor{lightgray204}{RGB}{204,204,204}

\begin{axis}[
scaled ticks=false,
axis line style={lightgray204},
legend cell align={left},
legend style={
  fill opacity=0.8,
  draw opacity=1,
  text opacity=1,,
  draw=none
},
tick align=outside,
tick pos=left,
x grid style={lightgray204},
xlabel=\textcolor{dimgray85}{$x$},
xmajorgrids,
xmin=-1, xmax=1,
xtick style={color=dimgray85},
y grid style={lightgray204},
ylabel=\textcolor{dimgray85}{Error},
ymajorgrids,
ymin=-0.005, ymax=0.045,
ytick={0,0.01,0.02,0.03,0.04},
yticklabels={0,0.01,0.02,0.03,0.04},
ytick style={color=dimgray85}
]
\addplot [thick, red]
table {%
-1 0.00135953280244727
-0.97979797979798 0.0118225654404824
-0.95959595959596 0.0117960796202617
-0.939393939393939 0.0101012134176807
-0.919191919191919 0.00787950635475065
-0.898989898989899 0.00555979864267753
-0.878787878787879 0.00334402205294559
-0.858585858585859 0.00133480588887747
-0.838383838383838 0.000416520413524135
-0.818181818181818 0.00188792142011818
-0.797979797979798 0.00307517911763744
-0.777777777777778 0.00398533310429183
-0.757575757575758 0.0046326646393684
-0.737373737373737 0.00503610370160024
-0.717171717171717 0.00521748281857426
-0.696969696969697 0.00520032084728561
-0.676767676767677 0.00500895272646884
-0.656565656565657 0.00466789339058338
-0.636363636363636 0.00420136522713432
-0.616161616161616 0.00363294297500782
-0.595959595959596 0.00298528508727613
-0.575757575757576 0.00227993021045136
-0.555555555555556 0.00153714373174851
-0.535353535353535 0.000775803568571098
-0.515151515151515 1.33172666517689e-05
-0.494949494949495 0.000734435507326103
-0.474747474747475 0.00145313955801751
-0.454545454545454 0.00213006328267651
-0.434343434343434 0.00275406300762171
-0.414141414141414 0.00331557466441335
-0.393939393939394 0.00380659450880017
-0.373737373737374 0.00422065026621321
-0.353535353535353 0.00455276384306824
-0.333333333333333 0.00479940655453637
-0.313131313131313 0.00495844767252496
-0.292929292929293 0.00502909698265219
-0.272727272727273 0.00501184194815119
-0.252525252525252 0.00490838000673388
-0.232323232323232 0.00472154646898676
-0.212121212121212 0.00445523844096868
-0.191919191919192 0.00411433515669396
-0.171717171717172 0.00370461507642716
-0.151515151515151 0.00323267008264638
-0.131313131313131 0.00270581708607892
-0.111111111111111 0.00213200733843913
-0.0909090909090908 0.00151973373581904
-0.0707070707070706 0.000877936386453226
-0.0505050505050504 0.000215906708381605
-0.0303030303030303 0.000456809683835935
-0.0101010101010101 0.00113051104916429
0.0101010101010102 0.0017954382963891
0.0303030303030305 0.00244187421859304
0.0505050505050506 0.00306024293925519
0.0707070707070707 0.00364120932375797
0.0909090909090911 0.00417577810954661
0.111111111111111 0.00465539250647293
0.131313131313131 0.00507203201614032
0.151515151515152 0.00541830921492553
0.171717171717172 0.00568756523989206
0.191919191919192 0.00587396370959503
0.212121212121212 0.00597258280250473
0.232323232323232 0.00597950520411633
0.252525252525253 0.00589190561907177
0.272727272727273 0.00570813552633576
0.292929292929293 0.00542780483238048
0.313131313131313 0.00505186004882185
0.333333333333333 0.00458265858501916
0.353535353535354 0.00402403870128665
0.373737373737374 0.00338138461195314
0.393939393939394 0.00266168615633366
0.414141414141414 0.00187359236549167
0.434343434343434 0.00102745813777505
0.454545454545455 0.000135383088997409
0.474747474747475 0.000788758546671708
0.494949494949495 0.00172929833556412
0.515151515151515 0.00266876607633237
0.535353535353535 0.00358789636442802
0.555555555555556 0.00446565197783388
0.575757575757576 0.00527926753975905
0.595959595959596 0.00600431796068179
0.616161616161616 0.00661481760873017
0.636363636363636 0.00708335818569494
0.656565656565657 0.00738129617459038
0.676767676767677 0.00747900491148112
0.696969696969697 0.0073462125207524
0.717171717171717 0.00695245629505548
0.737373737373737 0.00626769855502674
0.757575757575758 0.00526317201091231
0.777777777777778 0.0039125603703567
0.797979797979798 0.00219368411290299
0.818181818181818 9.09752141344811e-05
0.838383838383838 0.00240076313672277
0.858585858585859 0.00527039591853351
0.878787878787879 0.00848048588519823
0.898989898989899 0.0119494243242018
0.919191919191919 0.0155163184483691
0.939393939393939 0.018862462295687
0.95959595959596 0.0212971237217713
0.97979797979798 0.0209190094821469
1 0.00293491442376881
};
\end{axis}

\end{tikzpicture}}}
    \\
  \subfloat[]{%
        \scalebox{0.53}{\input{figures/QCL-sim-post/convergence-3-qubits-function-approx-d3-arcsin-tp20-xhoch3.tex}}}
    \hfill
  \subfloat[]{%
        \scalebox{0.53}{
\begin{tikzpicture}

\definecolor{dimgray85}{RGB}{85,85,85}
\definecolor{gainsboro229}{RGB}{229,229,229}
\definecolor{lightgray204}{RGB}{204,204,204}

\begin{axis}[
axis line style={lightgray204},
log basis y={10},
axis line style={lightgray204},
legend cell align={left},
legend style={
  fill opacity=0.8,
  draw opacity=1,
  text opacity=1,,
  draw=none
},
tick align=outside,
tick pos=left,
x grid style={lightgray204},
xlabel=\textcolor{dimgray85}{Cost Function Evaluations},
xmajorgrids,
xmin=-10, xmax=800,
xtick style={color=dimgray85},
y grid style={lightgray204},
ylabel=\textcolor{dimgray85}{Cost Function Value},
ymajorgrids,
ymin=10e-06, ymax=1000,
ymode=log,
ytick style={color=dimgray85},
ytick={0.0001,0.01,1,100},
yticklabels={
  \(\displaystyle {10^{-4}}\),
  \(\displaystyle {10^{-2}}\),
  \(\displaystyle {10^{0}}\),
  \(\displaystyle {10^{2}}\)
}
]
\addplot [semithick, red]
table {%
1 25.4563514037246
2 25.4563514749905
3 25.4563510791121
4 25.4563518150829
5 25.4563514755418
6 25.4563513594672
7 25.4563515174245
8 25.4563513278842
9 25.456351327311
10 25.4563514746263
11 25.4563516184261
12 13.9524162806065
13 23.6619120923164
14 15.6787034767895
15 7.09495765451862
16 11.1737671456173
17 11.1737673332849
18 11.1737672430212
19 11.1737670621767
20 11.1737671867688
21 11.173767333494
22 11.1737671518272
23 11.1737671582194
24 11.1737671417636
25 11.1737671276638
26 11.173767113154
27 11.8363239031237
28 8.11196653070471
29 11.0649780070011
30 6.7491253429041
31 6.74912521286994
32 6.74912530057274
33 6.74912547608895
34 6.74912534541546
35 6.74912532013652
36 6.7491254463827
37 6.74912535311721
38 6.749125361166
39 6.74912535092579
40 6.74912524297685
41 7.51553565727904
42 19.7455799262241
43 40.5988795824775
44 3.19056013441131
45 3.19056009142459
46 3.19056011880967
47 3.1905601696666
48 3.19056007540134
49 3.19056012086371
50 3.19056015474476
51 3.19056012219684
52 3.19056022096235
53 3.19056015332804
54 3.19056005866766
55 28.957210586032
56 71.0518494265979
57 3.50603651717006
58 1.95744723323036
59 1.95744722529426
60 1.95744721607541
61 1.95744723911166
62 1.95744723375024
63 1.95744717989773
64 1.95744720143385
65 1.95744722901528
66 1.95744723842412
67 1.95744724482729
68 1.95744725372469
69 8.55864064337378
70 1.62055913437967
71 1.62055914208974
72 1.62055912759241
73 1.62055912402912
74 1.62055912296322
75 1.62055908973922
76 1.62055915091554
77 1.62055915276948
78 1.6205591121384
79 1.62055911429261
80 1.62055914855191
81 27.8304906916077
82 3.22253194274986
83 1.09425122739274
84 1.09425123693616
85 1.09425121315212
86 1.09425120802945
87 1.09425122123949
88 1.09425120714096
89 1.0942512580372
90 1.09425124320079
91 1.0942512005852
92 1.09425120859678
93 1.09425122329119
94 7.60431921445982
95 11.6713182301578
96 0.869711890413369
97 0.869711874587672
98 0.869711888998879
99 0.869711893159732
100 0.869711882970657
101 0.869711894032269
102 0.869711928892932
103 0.869711901139628
104 0.869711871172445
105 0.869711878773153
106 0.869711864766398
107 1.31612936242166
108 0.706633406960918
109 0.706633378613295
110 0.706633401738207
111 0.706633429712802
112 0.706633406426693
113 0.706633401987687
114 0.706633423362173
115 0.706633412125104
116 0.70663340684282
117 0.706633403089913
118 0.706633387627218
119 1.09594561221473
120 0.577578117006844
121 0.577578102365787
122 0.577578112402242
123 0.577578124500697
124 0.577578125583466
125 0.577578109057295
126 0.577578140596413
127 0.577578119231662
128 0.577578109600933
129 0.577578111976111
130 0.577578105033972
131 2.02562342050088
132 0.414315139337544
133 0.414315122004758
134 0.414315130340295
135 0.414315155998772
136 0.414315149222055
137 0.414315132279686
138 0.414315148845311
139 0.414315138946836
140 0.41431514081096
141 0.414315138515192
142 0.414315132083339
143 0.418025636132706
144 0.223578065407503
145 0.22357806549972
146 0.223578061062386
147 0.223578065977412
148 0.223578080482995
149 0.223578066344336
150 0.22357807581008
151 0.22357806368235
152 0.223578065542222
153 0.223578063962797
154 0.223578065702334
155 0.134226297002804
156 0.134226240272175
157 0.134226319828349
158 0.134226364612889
159 0.134226254644153
160 0.13422628247816
161 0.134226251582262
162 0.134226302218004
163 0.134226343313912
164 0.134226291166696
165 0.134226298415119
166 0.856088160279421
167 0.0352232503169609
168 0.0352232437247941
169 0.0352232488472851
170 0.0352232565063728
171 0.0352232409590498
172 0.0352232392274391
173 0.0352232432491036
174 0.0352232554302867
175 0.0352232500569771
176 0.0352232452134259
177 0.0352232460694124
178 2.22609412443973
179 0.0127446170185072
180 0.0127446142812351
181 0.0127446170883288
182 0.0127446200663352
183 0.0127446126659401
184 0.0127446141102491
185 0.0127446129878931
186 0.0127446186213392
187 0.0127446184179738
188 0.0127446154053135
189 0.0127446165512684
190 0.00858956066451272
191 0.00858956224576586
192 0.00858955987407884
193 0.00858955890405894
194 0.00858956058791158
195 0.00858956031462517
196 0.00858956072017393
197 0.00858956103173368
198 0.00858955957217871
199 0.00858956033867571
200 0.00858956080246354
201 0.00770411513840932
202 0.00770411483176835
203 0.00770411510620732
204 0.00770411573671123
205 0.00770411489620408
206 0.00770411565123466
207 0.00770411469458144
208 0.00770411528360303
209 0.00770411601584347
210 0.00770411495505569
211 0.00770411556504731
212 0.00722611259357959
213 0.00722611407641635
214 0.00722611198305641
215 0.00722611103694873
216 0.00722611320167022
217 0.00722611256002473
218 0.00722611327015602
219 0.00722611288650758
220 0.00722611173639141
221 0.00722611230744774
222 0.00722611274059444
223 0.00653031380020279
224 0.00653031695885952
225 0.00653031263664304
226 0.00653031026661578
227 0.00653031522807774
228 0.00653031311383633
229 0.00653031558922449
230 0.00653031422561796
231 0.00653031121636724
232 0.00653031342397742
233 0.00653031361374697
234 0.00553306692200627
235 0.00553307158457335
236 0.00553306490552522
237 0.00553306167681563
238 0.00553306933419285
239 0.00553306509285826
240 0.005533070102853
241 0.00553306744225553
242 0.00553306200541391
243 0.00553306648600643
244 0.00553306607021511
245 0.00418493557448011
246 0.00418493869471064
247 0.0041849340176598
248 0.00418493210687265
249 0.00418493716314775
250 0.00418493408654986
251 0.00418493772081848
252 0.00418493590950253
253 0.00418493195998228
254 0.00418493528995603
255 0.00418493490150652
256 0.00324924796268961
257 0.00324924502465568
258 0.00324924885284562
259 0.00324925106348825
260 0.0032492462019178
261 0.00324924743336684
262 0.00324924605256504
263 0.00324924836638114
264 0.00324924979078774
265 0.00324924755871049
266 0.00324924799044281
267 0.0029497099410075
268 0.00294970924606861
269 0.0029497098971994
270 0.00294971064065006
271 0.00294970951936245
272 0.0029497095717054
273 0.00294970955272555
274 0.00294971007953164
275 0.00294970989655318
276 0.0029497098098799
277 0.00294970979084423
278 0.00274034138149426
279 0.00274034154401354
280 0.00274034094627213
281 0.0027403411752396
282 0.00274034182624707
283 0.00274034082423662
284 0.00274034196643354
285 0.0027403414972772
286 0.00274034046069201
287 0.00274034127625902
288 0.00274034103428821
289 0.00279531621153216
290 0.00267218872013294
291 0.00267218947574686
292 0.0026721882350495
293 0.00267218791925259
294 0.00267218926213959
295 0.00267218886321232
296 0.00267218932102231
297 0.00267218864369147
298 0.00267218804967967
299 0.00267218879854915
300 0.00267218863437877
301 0.00259249857530276
302 0.00259249869261893
303 0.00259249831631455
304 0.00259249843993854
305 0.00259249883608666
306 0.00259249863479524
307 0.00259249886466174
308 0.00259249850300245
309 0.00259249821402025
310 0.00259249864652886
311 0.00259249844781994
312 0.00257246837442311
313 0.00257246819727111
314 0.00257246831763632
315 0.0025724685510597
316 0.00257246840576346
317 0.00257246846094723
318 0.00257246839022968
319 0.0025724683411816
320 0.00257246841749377
321 0.00257246840560166
322 0.00257246833423197
323 0.0025597390817941
324 0.00255973897376248
325 0.00255973908762214
326 0.00255973919279114
327 0.00255973905213714
328 0.00255973915025942
329 0.00255973903333634
330 0.00255973906372145
331 0.00255973919099652
332 0.00255973909784771
333 0.00255973909510424
334 0.00255729411879633
335 0.0025572941075063
336 0.00255729411508323
337 0.0025572941312033
338 0.00255729411283568
339 0.00255729412044031
340 0.00255729410917743
341 0.00255729411306904
342 0.0025572941416615
343 0.00255729412352968
344 0.00255729412803146
};
\end{axis}

\end{tikzpicture}}}
    \hfill
  \subfloat[]{%
        \scalebox{0.53}{\input{figures/QCL-sim-post/convergence-3-qubits-function-approx-d3-arcsin-tp20-sin2x.tex}}}
  \caption{Three learned functions using QCL circuits on a statevector simulator with a post-processing parameter $\theta_{\text{post}}$, $R_Y(\arcsin(x))$ data encoding, a qubit number of $N=3$ and a depth of $D = 3$. The learned functions are $f_1(x) = x^3$ (a), $f_2(x) = x^3-x^2+1$ (b) and $f_3(x) = \sin(2x)$ (c). The cost function is evaluated on 20 equidistant training points and is classically minimized using SLSQP. The initial function (dashed black line) shows $f_{QC}(x)$ with the randomly chosen starting parameters for the optimization process and $\theta_\text{post} = 1$. Additionally, in (d)-(f) the absolute values of the respective errors $|f(x) - f^{\text{post}}_{QC}(x)|$ are shown on a fine grid. In (g)-(i) the respective values of the cost function versus the number of cost function evaluations in the optimization process are plotted.}
  \label{function_approx_example} 
\end{figure*}
The cost function is evaluated on 20 equidistant training points and is classically minimized using SLSQP with the default settings of the SciPy minimizer. The final learned functions are plotted in Figure \ref{function_approx_example} (a)-(c) (red lines).
The initial function resulting from the randomly selected start parameters at the beginning is also shown (black dashed lines in Figure \ref{function_approx_example} (a)-(c)).
We see that the target function can be approximated on a simulator.
The next step is to reproduce these results on actual quantum hardware, in particular on ibmq\_ehningen, before investigating the possibility of solving differential equations.
\subsection{IBM Quantum Computer}
\label{chap:ehningen}
In the literature, QCL has been executed entirely on a simulator or only the circuit with the final optimized parameters has been tested on a quantum computer \cite{HatakeyamaSato.2022}.
In this work, the full algorithm is executed on a quantum computer.
This means that also every circuit evaluation during the optimization process is performed on the quantum computer.
The QCL circuits that have been simulated in Section \ref{chap:QCL_sim} are now executed on ibmq\_ehningen which was introduced in Section \ref{hardware_introduction}.
As we analyzed in Section \ref{sec_comp_opt}, due to the harware and shot noise, a gradient-based optimization runs into problems and the gradient-free COBYLA optimizer proves to be suitable.\\
\begin{figure*}[!tp]
    \centering
  \subfloat[]{%
       \scalebox{0.53}{
\begin{tikzpicture}

\definecolor{dimgray85}{RGB}{85,85,85}
\definecolor{gainsboro229}{RGB}{229,229,229}
\definecolor{green01270}{RGB}{0,127,0}
\definecolor{lightgray204}{RGB}{204,204,204}
\definecolor{lightgray204}{RGB}{204,204,204}

\begin{axis}[
axis line style={lightgray204},
legend cell align={left},
legend style={
  fill opacity=0.8,
  draw opacity=1,
  text opacity=1,
  at={(0.03,0.97)},
  anchor=north west,
  draw=none
},
tick align=outside,
tick pos=left,
x grid style={lightgray204},
xlabel=\textcolor{dimgray85}{$x$},
xmajorgrids,
xmin=-1, xmax=1,
xtick style={color=dimgray85},
y grid style={lightgray204},
ylabel=\textcolor{dimgray85}{$f^{\text{post}}_{QC}(x)$},
ymajorgrids,
ymin=-1.1, ymax=1.1,
ytick style={color=dimgray85}
]
\addplot [thick, dashed, black, mark=x, mark size=3, mark options={solid}]
table {%
-1 0.378452473505463
-0.777777777777778 0.421056933829239
-0.555555555555556 0.410229132569558
-0.333333333333333 0.332679658454876
-0.111111111111111 0.130064663992797
0.111111111111111 -0.0429103713712485
0.333333333333333 -0.206219782020815
0.555555555555555 -0.385297270553311
0.777777777777778 -0.266427718040621
1 -0.142545663978937
};
\addlegendentry{Initial Function}
\addplot [thick, blue]
table {%
-1 -1
-0.97979797979798 -0.940610059373451
-0.95959595959596 -0.883619379181056
-0.939393939393939 -0.828978490135515
-0.919191919191919 -0.776637922949524
-0.898989898989899 -0.726548208335781
-0.878787878787879 -0.678659877006984
-0.858585858585859 -0.632923459675832
-0.838383838383838 -0.589289487055021
-0.818181818181818 -0.54770848985725
-0.797979797979798 -0.508130998795217
-0.777777777777778 -0.470507544581619
-0.757575757575758 -0.434788657929154
-0.737373737373737 -0.40092486955052
-0.717171717171717 -0.368866710158415
-0.696969696969697 -0.338564710465537
-0.676767676767677 -0.309969401184583
-0.656565656565657 -0.283031313028252
-0.636363636363636 -0.257700976709241
-0.616161616161616 -0.233928922940248
-0.595959595959596 -0.211665682433971
-0.575757575757576 -0.190861785903108
-0.555555555555556 -0.171467764060357
-0.535353535353535 -0.153434147618414
-0.515151515151515 -0.13671146728998
-0.494949494949495 -0.12125025378775
-0.474747474747475 -0.107001037824423
-0.454545454545454 -0.0939143501126972
-0.434343434343434 -0.0819407213652698
-0.414141414141414 -0.071030682294839
-0.393939393939394 -0.0611347636141025
-0.373737373737374 -0.052203496035758
-0.353535353535353 -0.0441874102725036
-0.333333333333333 -0.037037037037037
-0.313131313131313 -0.0307029070420561
-0.292929292929293 -0.0251355510002587
-0.272727272727273 -0.0202854996243426
-0.252525252525252 -0.0161032836270057
-0.232323232323232 -0.0125394337209458
-0.212121212121212 -0.00954448061886077
-0.191919191919192 -0.00706895503344844
-0.171717171717172 -0.00506338767740664
-0.151515151515151 -0.00347830926343323
-0.131313131313131 -0.00226425050422601
-0.111111111111111 -0.00137174211248285
-0.0909090909090908 -0.000751314800901576
-0.0707070707070706 -0.000353499282180028
-0.0505050505050504 -0.000128826269016045
-0.0303030303030303 -2.78264741074658e-05
-0.0101010101010101 -1.03061015212835e-06
0.0101010101010102 1.03061015212838e-06
0.0303030303030305 2.78264741074664e-05
0.0505050505050506 0.000128826269016046
0.0707070707070707 0.000353499282180029
0.0909090909090911 0.000751314800901581
0.111111111111111 0.00137174211248286
0.131313131313131 0.00226425050422603
0.151515151515152 0.00347830926343324
0.171717171717172 0.00506338767740665
0.191919191919192 0.00706895503344847
0.212121212121212 0.00954448061886079
0.232323232323232 0.0125394337209458
0.252525252525253 0.0161032836270057
0.272727272727273 0.0202854996243426
0.292929292929293 0.0251355510002587
0.313131313131313 0.0307029070420561
0.333333333333333 0.0370370370370371
0.353535353535354 0.0441874102725036
0.373737373737374 0.0522034960357581
0.393939393939394 0.0611347636141025
0.414141414141414 0.0710306822948391
0.434343434343434 0.08194072136527
0.454545454545455 0.0939143501126972
0.474747474747475 0.107001037824423
0.494949494949495 0.12125025378775
0.515151515151515 0.13671146728998
0.535353535353535 0.153434147618415
0.555555555555556 0.171467764060357
0.575757575757576 0.190861785903108
0.595959595959596 0.211665682433971
0.616161616161616 0.233928922940249
0.636363636363636 0.257700976709241
0.656565656565657 0.283031313028252
0.676767676767677 0.309969401184583
0.696969696969697 0.338564710465537
0.717171717171717 0.368866710158415
0.737373737373737 0.40092486955052
0.757575757575758 0.434788657929154
0.777777777777778 0.470507544581619
0.797979797979798 0.508130998795217
0.818181818181818 0.547708489857251
0.838383838383838 0.589289487055021
0.858585858585859 0.632923459675832
0.878787878787879 0.678659877006985
0.898989898989899 0.726548208335782
0.919191919191919 0.776637922949524
0.939393939393939 0.828978490135515
0.95959595959596 0.883619379181057
0.97979797979798 0.940610059373451
1 1
};
\addlegendentry{$f_1(x)$}
\addplot [semithick, red, mark=x, mark size=3, mark options={solid}]
table {%
-1 -0.99555364725667
-0.777777777777778 -0.430045181087775
-0.555555555555556 -0.11429850585146
-0.333333333333333 -0.0679104934508952
-0.111111111111111 0.00338437198318291
0.111111111111111 -0.0174168978030645
0.333333333333333 0.0789626255236452
0.555555555555555 0.141930841538714
0.777777777777778 0.521617598946525
1 0.983667379577951
};
\addlegendentry{Final Function}
\end{axis}

\end{tikzpicture}}}
    \hfill
  \subfloat[]{%
        \scalebox{0.53}{
\begin{tikzpicture}

\definecolor{dimgray85}{RGB}{85,85,85}
\definecolor{gainsboro229}{RGB}{229,229,229}
\definecolor{green01270}{RGB}{0,127,0}
\definecolor{lightgray204}{RGB}{204,204,204}

\begin{axis}[
axis line style={lightgray204},
legend cell align={left},
legend style={
  fill opacity=0.8,
  draw opacity=1,
  text opacity=1,
  at={(0.97,0.03)},
  anchor=south east,
  draw=none
},
tick align=outside,
tick pos=left,
x grid style={lightgray204},
xlabel=\textcolor{dimgray85}{$x$},
xmajorgrids,
xmin=-1, xmax=1,
xtick style={color=dimgray85},
y grid style={lightgray204},
ylabel=\textcolor{dimgray85}{$f^{\text{post}}_{QC}(x)$},
ymajorgrids,
ymin=-1.1, ymax=1.1,
ytick style={color=dimgray85}
]
\addplot [thick, dashed, black, mark=x, mark size=3, mark options={solid}]
table {%
-1 -0.0625357172014042
-0.777777777777778 -0.506445910404001
-0.555555555555556 -0.335344476151651
-0.333333333333333 -0.401549439347604
-0.111111111111111 -0.330122842100967
0.111111111111111 -0.321997874601488
0.333333333333333 -0.249683453825103
0.555555555555555 -0.296131377915764
0.777777777777778 -0.340466291676857
1 -0.10507216831118
};
\addlegendentry{Initial Function}
\addplot [thick, blue]
table {%
-1 -1
-0.97979797979798 -0.900614140589653
-0.95959595959596 -0.804443784853947
-0.939393939393939 -0.711439463505579
-0.919191919191919 -0.621551707257248
-0.898989898989899 -0.53473104682165
-0.878787878787879 -0.450928012911484
-0.858585858585859 -0.370093136239448
-0.838383838383838 -0.292176947518239
-0.818181818181818 -0.217129977460556
-0.797979797979798 -0.144902756779096
-0.777777777777778 -0.075445816186557
-0.757575757575758 -0.00870968639563685
-0.737373737373737 0.0553551018809666
-0.717171717171717 0.116798017930556
-0.696969696969697 0.175668531040432
-0.676767676767677 0.232016110497898
-0.656565656565657 0.285890225590256
-0.636363636363636 0.337340345604809
-0.616161616161616 0.386415939828857
-0.595959595959596 0.433166477549704
-0.575757575757576 0.477641428054651
-0.555555555555556 0.519890260631001
-0.535353535353535 0.559962444566057
-0.515151515151515 0.597907449147119
-0.494949494949495 0.63377474366149
-0.474747474747475 0.667613797396473
-0.454545454545454 0.699474079639369
-0.434343434343434 0.729405059677481
-0.414141414141414 0.757456206798111
-0.393939393939394 0.783676990288561
-0.373737373737374 0.808116879436133
-0.353535353535353 0.830825343528129
-0.333333333333333 0.851851851851852
-0.313131313131313 0.871245873694603
-0.292929292929293 0.889056878343686
-0.272727272727273 0.905334335086401
-0.252525252525252 0.920127713210052
-0.232323232323232 0.93348648200194
-0.212121212121212 0.945460110749367
-0.191919191919192 0.956098068739636
-0.171717171717172 0.965449825260049
-0.151515151515151 0.973564849597907
-0.131313131313131 0.980492611040514
-0.111111111111111 0.986282578875171
-0.0909090909090908 0.990984222389181
-0.0707070707070706 0.994647010869845
-0.0505050505050504 0.997320413604466
-0.0303030303030303 0.999053899880346
-0.0101010101010101 0.999896938984787
0.0101010101010102 0.999899000205091
0.0303030303030305 0.999109552828561
0.0505050505050506 0.997578066142498
0.0707070707070707 0.995354009434205
0.0909090909090911 0.992486851990984
0.111111111111111 0.989026063100137
0.131313131313131 0.985021112048966
0.151515151515152 0.980521468124774
0.171717171717172 0.975576600614862
0.191919191919192 0.970235978806533
0.212121212121212 0.964549071987089
0.232323232323232 0.958565349443831
0.252525252525253 0.952334280464063
0.272727272727273 0.945905334335086
0.292929292929293 0.939327980344203
0.313131313131313 0.932651687778716
0.333333333333333 0.925925925925926
0.353535353535354 0.919200164073136
0.373737373737374 0.912523871507649
0.393939393939394 0.905946517516765
0.414141414141414 0.899517571387789
0.434343434343434 0.893286502408021
0.454545454545455 0.887302779864763
0.474747474747475 0.881615873045319
0.494949494949495 0.87627525123699
0.515151515151515 0.871330383727078
0.535353535353535 0.866830739802885
0.555555555555556 0.862825788751715
0.575757575757576 0.859364999860868
0.595959595959596 0.856497842417646
0.616161616161616 0.854273785709353
0.636363636363636 0.852742299023291
0.656565656565657 0.85195285164676
0.676767676767677 0.851954912867065
0.696969696969697 0.852797951971506
0.717171717171717 0.854531438247386
0.737373737373737 0.857204840982007
0.757575757575758 0.860867629462671
0.777777777777778 0.86556927297668
0.797979797979798 0.871359240811338
0.818181818181818 0.878287002253945
0.838383838383838 0.886402026591803
0.858585858585859 0.895753783112216
0.878787878787879 0.906391741102485
0.898989898989899 0.918365369849912
0.919191919191919 0.9317241386418
0.939393939393939 0.946517516765451
0.95959595959596 0.962794973508166
0.97979797979798 0.980605978157248
1 1
};
\addlegendentry{$f_2(x)$}
\addplot [semithick, red, mark=x, mark size=3, mark options={solid}]
table {%
-1 -0.98411117622488
-0.777777777777778 -0.0341198110787897
-0.555555555555556 0.533066535118702
-0.333333333333333 0.897266238710822
-0.111111111111111 0.985651275556391
0.111111111111111 1.02519034839855
0.333333333333333 0.957429976976924
0.555555555555555 0.919976593766161
0.777777777777778 0.875466795125362
1 1.00330410299214
};
\addlegendentry{Final Function}
\end{axis}

\end{tikzpicture}}}
    \hfill
  \subfloat[]{%
        \scalebox{0.53}{
\begin{tikzpicture}

\definecolor{dimgray85}{RGB}{85,85,85}
\definecolor{gainsboro229}{RGB}{229,229,229}
\definecolor{green01270}{RGB}{0,127,0}
\definecolor{lightgray204}{RGB}{204,204,204}

\begin{axis}[
axis line style={lightgray204},
legend cell align={left},
legend style={
  fill opacity=0.8,
  draw opacity=1,
  text opacity=1,
  at={(0.03,0.97)},
  anchor=north west,
  draw=none
},
tick align=outside,
tick pos=left,
x grid style={lightgray204},
xlabel=\textcolor{dimgray85}{$x$},
xmajorgrids,
xmin=-1, xmax=1,
xtick style={color=dimgray85},
y grid style={lightgray204},
ylabel=\textcolor{dimgray85}{$f^{\text{post}}_{QC}(x)$},
ymajorgrids,
ymin=-1.2, ymax=1.2,
ytick style={color=dimgray85}
]
\addplot [thick, dashed, black, mark=x, mark size=3, mark options={solid}]
table {%
-1 -0.0625357172014042
-0.777777777777778 -0.506445910404001
-0.555555555555556 -0.335344476151651
-0.333333333333333 -0.401549439347604
-0.111111111111111 -0.330122842100967
0.111111111111111 -0.321997874601488
0.333333333333333 -0.249683453825103
0.555555555555555 -0.296131377915764
0.777777777777778 -0.340466291676857
1 -0.10507216831118
};
\addlegendentry{Initial Function}
\addplot [thick, blue]
table {%
-1 -0.909297426825682
-0.97979797979798 -0.925364759108592
-0.95959595959596 -0.939921651430131
-0.939393939393939 -0.952944343093062
-0.919191919191919 -0.964411577621277
-0.898989898989899 -0.97430463745599
-0.878787878787879 -0.982607374507739
-0.858585858585859 -0.98930623651434
-0.838383838383838 -0.99439028916176
-0.818181818181818 -0.997851233931811
-0.797979797979798 -0.999683421647525
-0.777777777777778 -0.999883861694102
-0.757575757575758 -0.998452226900389
-0.737373737373737 -0.995390854072907
-0.717171717171717 -0.990704740181562
-0.696969696969697 -0.984401534203274
-0.676767676767677 -0.976491524636824
-0.656565656565657 -0.9669876227093
-0.636363636363636 -0.95590534130156
-0.616161616161616 -0.943262769627114
-0.595959595959596 -0.929080543705742
-0.575757575757576 -0.913381812680051
-0.555555555555556 -0.896192201029956
-0.535353535353535 -0.877539766746754
-0.515151515151515 -0.85745495553506
-0.494949494949495 -0.835970551117371
-0.474747474747475 -0.81312162172236
-0.454545454545454 -0.788945462844257
-0.434343434343434 -0.763481536366738
-0.414141414141414 -0.736771406150697
-0.393939393939394 -0.708858670191038
-0.373737373737374 -0.679788889453219
-0.353535353535353 -0.649609513505707
-0.333333333333333 -0.618369803069737
-0.313131313131313 -0.586120749612793
-0.292929292929293 -0.552914992117049
-0.272727272727273 -0.518806731158627
-0.252525252525252 -0.483851640437935
-0.232323232323232 -0.448106775905457
-0.212121212121212 -0.411630482631366
-0.191919191919192 -0.374482299570935
-0.171717171717172 -0.33672286238122
-0.151515151515151 -0.298413804447641
-0.131313131313131 -0.259617656281993
-0.111111111111111 -0.220397743456122
-0.0909090909090908 -0.180818083237847
-0.0707070707070706 -0.140943280097851
-0.0505050505050504 -0.100838420258104
-0.0303030303030303 -0.0605689654539395
-0.0101010101010101 -0.0202006460831913
0.0101010101010102 0.0202006460831915
0.0303030303030305 0.0605689654539399
0.0505050505050506 0.100838420258105
0.0707070707070707 0.140943280097852
0.0909090909090911 0.180818083237848
0.111111111111111 0.220397743456122
0.131313131313131 0.259617656281993
0.151515151515152 0.298413804447641
0.171717171717172 0.336722862381221
0.191919191919192 0.374482299570935
0.212121212121212 0.411630482631367
0.232323232323232 0.448106775905458
0.252525252525253 0.483851640437935
0.272727272727273 0.518806731158628
0.292929292929293 0.552914992117049
0.313131313131313 0.586120749612794
0.333333333333333 0.618369803069737
0.353535353535354 0.649609513505707
0.373737373737374 0.679788889453219
0.393939393939394 0.708858670191038
0.414141414141414 0.736771406150697
0.434343434343434 0.763481536366738
0.454545454545455 0.788945462844257
0.474747474747475 0.813121621722361
0.494949494949495 0.835970551117371
0.515151515151515 0.85745495553506
0.535353535353535 0.877539766746754
0.555555555555556 0.896192201029956
0.575757575757576 0.913381812680051
0.595959595959596 0.929080543705742
0.616161616161616 0.943262769627115
0.636363636363636 0.95590534130156
0.656565656565657 0.9669876227093
0.676767676767677 0.976491524636825
0.696969696969697 0.984401534203274
0.717171717171717 0.990704740181562
0.737373737373737 0.995390854072907
0.757575757575758 0.99845222690039
0.777777777777778 0.999883861694102
0.797979797979798 0.999683421647525
0.818181818181818 0.997851233931811
0.838383838383838 0.99439028916176
0.858585858585859 0.98930623651434
0.878787878787879 0.982607374507739
0.898989898989899 0.97430463745599
0.919191919191919 0.964411577621277
0.939393939393939 0.952944343093062
0.95959595959596 0.939921651430131
0.97979797979798 0.925364759108592
1 0.909297426825682
};
\addlegendentry{$f_3(x)$}
\addplot [semithick, red, mark=x, mark size=3, mark options={solid}]
table {%
-1 -1.08963845626539
-0.777777777777778 -1.06905696291104
-0.555555555555556 -0.87630953284329
-0.333333333333333 -0.51411782078073
-0.111111111111111 -0.2287209985276
0.111111111111111 0.118866974592606
0.333333333333333 0.544090360904338
0.555555555555555 0.771157031683338
0.777777777777778 1.04547267629185
1 1.0472633107518
};
\addlegendentry{Final Function}
\end{axis}

\end{tikzpicture}}}
    \\
  \subfloat[]{%
        \scalebox{0.53}{
\begin{tikzpicture}

\definecolor{dimgray85}{RGB}{85,85,85}
\definecolor{gainsboro229}{RGB}{229,229,229}
\definecolor{lightgray204}{RGB}{204,204,204}

\begin{axis}[
axis line style={lightgray204},
tick align=outside,
tick pos=left,
x grid style={lightgray204},
xlabel=\textcolor{dimgray85}{$x$},
xmajorgrids,
xmin=-1, xmax=1,
xtick style={color=dimgray85},
y grid style={lightgray204},
ylabel=\textcolor{dimgray85}{Error},
ymajorgrids,
ymin=-0.01, ymax=0.21,
ytick={0,0.1,0.2},
yticklabels={0,0.1,0.2},
ytick style={color=dimgray85}
]
\addplot [semithick, red, mark=x, mark size=3, mark options={solid}]
table {%
-1 0.00444635274333027
-0.777777777777778 0.0404623634938434
-0.555555555555556 0.0571692582088962
-0.333333333333333 0.0308734564138581
-0.111111111111111 0.00475611409566577
0.111111111111111 0.0187886399155474
0.333333333333333 0.0419255884866082
0.555555555555555 0.0295369225216423
0.777777777777778 0.051110054364907
1 0.0163326204220485
};
\end{axis}

\end{tikzpicture}}}
    \hfill
  \subfloat[]{%
        \scalebox{0.53}{
\begin{tikzpicture}

\definecolor{dimgray85}{RGB}{85,85,85}
\definecolor{gainsboro229}{RGB}{229,229,229}
\definecolor{lightgray204}{RGB}{204,204,204}

\begin{axis}[
axis line style={lightgray204},
tick align=outside,
tick pos=left,
x grid style={lightgray204},
xlabel=\textcolor{dimgray85}{$x$},
xmajorgrids,
xmin=-1, xmax=1,
xtick style={color=dimgray85},
y grid style={lightgray204},
ylabel=\textcolor{dimgray85}{Error},
ymajorgrids,
ymin=-0.01, ymax=0.21,
ytick={0,0.1,0.2},
yticklabels={0,0.1,0.2},
ytick style={color=dimgray85}
]
\addplot [semithick, red, mark=x, mark size=3, mark options={solid}]
table {%
-1 0.0158888237751202
-0.777777777777778 0.0413260051077673
-0.555555555555556 0.0131762744877011
-0.333333333333333 0.0454143868589706
-0.111111111111111 0.00063130331878003
0.111111111111111 0.0361642852984158
0.333333333333333 0.031504051050998
0.555555555555555 0.0571508050144462
0.777777777777778 0.00989752214868123
1 0.00330410299213901
};
\end{axis}

\end{tikzpicture}}}
    \hfill
  \subfloat[]{%
        \scalebox{0.53}{
\begin{tikzpicture}

\definecolor{dimgray85}{RGB}{85,85,85}
\definecolor{gainsboro229}{RGB}{229,229,229}
\definecolor{lightgray204}{RGB}{204,204,204}

\begin{axis}[
scaled ticks=false,
ytick={0, 0.05, 0.1 , 0.15},
yticklabels={0, 0.05, 0.1 , 0.15},
axis line style={lightgray204},
tick align=outside,
tick pos=left,
x grid style={lightgray204},
xlabel=\textcolor{dimgray85}{$x$},
xmajorgrids,
xmin=-1, xmax=1,
xtick style={color=dimgray85},
y grid style={lightgray204},
ylabel=\textcolor{dimgray85}{Error},
ymajorgrids,
ymin=-0.01, ymax=0.21,
ytick={0,0.1,0.2},
yticklabels={0,0.1,0.2},
ytick style={color=dimgray85}
]
\addplot [semithick, red, mark=x, mark size=3, mark options={solid}]
table {%
-1 0.180341029439713
-0.777777777777778 0.0691731012169337
-0.555555555555556 0.0198826681866663
-0.333333333333333 0.104251982289008
-0.111111111111111 0.00832325507147783
0.111111111111111 0.101530768863516
0.333333333333333 0.0742794421653986
0.555555555555555 0.125035169346618
0.777777777777778 0.0455888145977479
1 0.137965883926116
};
\end{axis}

\end{tikzpicture}}}
    \\
  \subfloat[]{%
        \scalebox{0.53}{
\begin{tikzpicture}

\definecolor{dimgray85}{RGB}{85,85,85}
\definecolor{gainsboro229}{RGB}{229,229,229}
\definecolor{lightgray204}{RGB}{204,204,204}

\begin{axis}[
axis line style={lightgray204},
log basis y={10},
tick align=outside,
tick pos=left,
x grid style={lightgray204},
xlabel=\textcolor{dimgray85}{Cost Function Evaluations},
xmajorgrids,
xmin=-2, xmax=75,
xtick style={color=dimgray85},
y grid style={lightgray204},
ylabel=\textcolor{dimgray85}{Cost Function Value},
ymajorgrids,
ymode=log,
ytick style={color=dimgray85},
ymin=0.005, ymax=20,
ytick={0.01,0.1,1,10},
yticklabels={
  \(\displaystyle {10^{-2}}\),
  \(\displaystyle {10^{-1}}\),
  \(\displaystyle {10^{0}}\),
  \(\displaystyle {10^{1}}\)
}
]
\addplot [semithick, red]
table {%
1 5.40696194913583
2 0.737536785826964
3 0.408622328698449
4 0.459756048365609
5 0.454256962862124
6 0.414685064199861
7 0.503676224217821
8 0.369022123333716
9 0.697578745468171
10 0.330226449790945
11 0.370181336408092
12 0.342163389740349
13 0.879910638837099
14 0.372303021282159
15 0.280630850301831
16 0.221835664771188
17 0.249342543308743
18 0.683573007175272
19 0.155351283983273
20 0.13429562270515
21 0.255415773093831
22 0.417486742493946
23 0.167888795540626
24 0.136773730530438
25 0.515926688484895
26 0.295090641494416
27 0.159508173501251
28 0.113439758235804
29 0.128554399388233
30 0.136711480728846
31 0.109910931940847
32 0.153635011366437
33 0.206149393679078
34 0.169300255312662
35 0.0876237625578264
36 0.0603491053294626
37 0.0571265899331743
38 0.0868862816854606
39 0.0362586250162867
40 0.0300644178409954
41 0.0568156815169799
42 0.0511722978233059
43 0.123679556162936
44 0.0720613330703924
45 0.0323867053312996
46 0.0771919654241586
47 0.0548770092324072
48 0.0237085203273416
49 0.0495888324508889
50 0.0593543524951007
51 0.0347661492243777
52 0.0572640531735564
53 0.0412158046116314
54 0.0652672778988109
55 0.0501930110270817
56 0.0333725555038981
57 0.0553874246439632
58 0.0327945954999563
59 0.0586754366209914
60 0.0591077217170663
61 0.0359315619217508
62 0.0206870236893715
63 0.030476488930351
64 0.0562863495900583
65 0.081576895435914
66 0.0241585249103437
67 0.027188128642961
68 0.0381194242469867
69 0.0618915959728116
70 0.0495465692992056
71 0.0592570298290872
72 0.0347721694042299
73 0.0117632777573113
};
\end{axis}

\end{tikzpicture}}}
    \hfill
  \subfloat[]{%
        \scalebox{0.53}{
\begin{tikzpicture}

\definecolor{dimgray85}{RGB}{85,85,85}
\definecolor{gainsboro229}{RGB}{229,229,229}
\definecolor{lightgray204}{RGB}{204,204,204}

\begin{axis}[
axis line style={lightgray204},
log basis y={10},
tick align=outside,
tick pos=left,
x grid style={lightgray204},
xlabel=\textcolor{dimgray85}{Cost Function Evaluations},
xmajorgrids,
xmin=-2, xmax=75,
xtick style={color=dimgray85},
y grid style={lightgray204},
ylabel=\textcolor{dimgray85}{Cost Function Value},
ymajorgrids,
ymode=log,
ytick style={color=dimgray85},
ymin=0.005, ymax=20,
ytick={0.01,0.1,1,10},
yticklabels={
  \(\displaystyle {10^{-2}}\),
  \(\displaystyle {10^{-1}}\),
  \(\displaystyle {10^{0}}\),
  \(\displaystyle {10^{1}}\)
}
]
\addplot [semithick, red]
table {%
1 12.2196940204344
2 8.50642990911791
3 7.15774449690972
4 7.9238567280843
5 8.97338570179804
6 13.063983381938
7 9.45834718498536
8 8.17764070198974
9 5.58913087892023
10 6.30928396948973
11 4.60306851677106
12 2.9866152679826
13 6.66071569348729
14 5.88365378676922
15 2.93066919528721
16 5.85570608521965
17 2.90381352936292
18 2.71511676006414
19 2.30020054580999
20 3.98222491960295
21 3.81131624819383
22 2.70068959994936
23 1.90201786736769
24 2.07227283795384
25 12.227934871409
26 0.75038338817748
27 0.625296112941326
28 0.509910683571213
29 0.373706728093237
30 0.530397719665686
31 0.498395738921441
32 0.482525027228367
33 0.191196378884678
34 0.575188634925582
35 0.563778874243258
36 0.0627459490307499
37 0.389084452270396
38 0.137689117278054
39 0.0967281396608875
40 0.0962578659241066
41 0.135372392886962
42 0.150246713974823
43 0.0366470620921148
44 0.129159415351736
45 0.231509949292863
46 0.250493633462368
47 0.0581049436664939
48 0.0893871448796905
49 0.0802065744798299
50 0.0990694456458618
51 0.269574995008237
52 0.0751548988454262
53 0.0736509178697775
54 0.066825045279853
55 0.0694379218357981
56 0.0615550626402274
57 0.0469576841552412
58 0.0297773426024587
59 0.0975917544238757
60 0.0480405920817883
61 0.0532608755420464
62 0.00987222602499047
};
\end{axis}

\end{tikzpicture}}}
    \hfill
  \subfloat[]{%
        \scalebox{0.53}{
\begin{tikzpicture}

\definecolor{dimgray85}{RGB}{85,85,85}
\definecolor{gainsboro229}{RGB}{229,229,229}
\definecolor{lightgray204}{RGB}{204,204,204}

\begin{axis}[
axis line style={lightgray204},
log basis y={10},
tick align=outside,
tick pos=left,
x grid style={lightgray204},
xlabel=\textcolor{dimgray85}{Cost Function Evaluations},
xmajorgrids,
xmin=-2, xmax=75,
xtick style={color=dimgray85},
y grid style={lightgray204},
ylabel=\textcolor{dimgray85}{Cost Function Value},
ymajorgrids,
ymode=log,
ytick style={color=dimgray85},
ymin=0.005, ymax=20,
ytick={0.01,0.1,1,10},
yticklabels={
  \(\displaystyle {10^{-2}}\),
  \(\displaystyle {10^{-1}}\),
  \(\displaystyle {10^{0}}\),
  \(\displaystyle {10^{1}}\)
}
]
\addplot [semithick, red]
table {%
1 6.14837742400344
2 5.17482828610874
3 6.34699379837576
4 3.68489689825144
5 3.3599664363026
6 3.97425155419246
7 5.94548845764557
8 3.15882173572293
9 6.01022754873385
10 3.08659346312234
11 3.31416489444035
12 4.26797984742432
13 6.27643021841851
14 3.13869447270189
15 3.60196784672089
16 3.69576664980256
17 3.30497007105304
18 3.12716898455269
19 1.90951611157309
20 1.1310105926558
21 0.935678437751293
22 1.73502122069951
23 0.818267056060597
24 0.729251575883113
25 0.496572091315838
26 0.285924888699338
27 0.685114366883796
28 0.245871998948051
29 0.322839587485163
30 0.258389997652233
31 0.443804403815985
32 0.178859332850413
33 0.151262721279291
34 0.179037121950484
35 0.256330374893246
36 0.207492774217945
37 0.325356058333067
38 0.242375502394296
39 0.127005225563187
40 0.159884715651393
41 0.150214314894484
42 0.101213528983636
};
\end{axis}

\end{tikzpicture}}}
  \caption{Three learned functions using QCL circuits on the ibmq\_ehningen with a post-processing parameter $\theta_{\text{post}}$, $R_Y(\arcsin(x))$ data encoding and a qubit number of $N=3$. The learned functions are $f_1(x) = x^3$ (a), $f_2(x) = x^3-x^2+1$ (b) and $f_3(x) = \sin(2x)$ (c). The cost function is evaluated on 10 equidistant training points with 2000 shots for every circuit and is classically minimized using COBYLA. The initial function (dashed black line) shows $f^{\text{post}}_{QC}(x)$ with the randomly chosen starting parameters for the optimization process and $\theta_\text{post} = 1$. Additionally, in (d)-(f) the absolute values of the respective errors $|f(x) - f^{\text{post}}_{QC}(x)|$ are shown. In (g)-(i) the respective values of the cost function versus the number of cost function evaluations during the optimization process are plotted.}
  \label{x^3_real_backend} 
\end{figure*}
The same functions that have already been learned in Section \ref{chap:QCL_sim} on a simulator are now learned on ibmq\_ehningen.
The cost function is evaluated on 10 equidistant training points with 2000 shots for every circuit and is classically minimized using COBYLA (red markers in Figure \ref{x^3_real_backend} (a)-(c)).
Additionally, the initial function resulting from the randomly selected start parameters at the beginning of the optimization and $\theta_\text{post} = 1$ is plotted (black dashed lines in Figure \ref{x^3_real_backend} (a)-(c)).
The functions can be learned with good accuracy on the quantum computer which becomes clear in the error diagrams (Figure \ref{x^3_real_backend} (d)-(f)), where the absolute values of the respective errors $|f(x) - f^{\text{post}}_{QC}(x)|$ are plotted.
It can be observed that the two polynomials (${f_1(x) = x^3}$, ${f_2(x) = x^3-x^2+1}$) can be learned with higher accuracy than the trigonometric function (${f_3(x) = \sin(2x)}$). This can be explained by the fact that the selected data encoding layer generates polynomial-like functions (see Equation \eqref{arcsin_function}) which are more suited to learn polynomial functions.\\
The values of the cost function versus the number of cost function evaluations (Figure \ref{x^3_real_backend} (g)-(i)) shows that the cost function is evaluated less than one hundred times during the optimization process.\\
If we compare the results presented here with those in Section \ref{chap:hardware_noise}, we can see that the error is smaller than expected from the investigation with fixed parameters. We assume that this is due to the fact that systematic hardware errors are learned by the variational algorithm and thus the results are closer to the target function compared to the experiments without variational optimization.
\section{Differential Equations}
This section investigates the possibility of solving differential equations with QCL circuits in combination with the parameter shift rule (PSR).
\subsection{Parameter Shift Rule}
\label{chap:parameter_shift_rule}
The PSR is a method to determine the exact derivative of a parameterized quantum circuit \cite{Mitarai.2018, Schuld.2019}.
To obtain the first derivative w.r.t. $x$ of our QCL circuits, the PSR boils down to evaluating $2N$ expectation values.
As the name suggests, in each of these expectation values the variable $x$ in one of the data encoding gates is shifted by $\pm \tfrac{\pi}{2}$.
Also note that, due to the chain rule, the derivative of the inner function $\varphi^\prime(x)$ also enters in the calculation.
Higher derivatives are obtained similarly but require additional evaluations of the expectation value.
For example, we need to execute $4N^2-2N$ QCL circuits for the second derivative.\\
To investigate the applicability of the PSR for QCL circuits on current quantum hardware, it is tested on ibmq\_ehningen. 
For this purpose, the circuit and the optimized parameters from the learning of $f(x) = x^3$ in Figure \ref{function_approx_example} (a) are used and the derivatives w.r.t $x$ are calculated with the PSR.
Since the derivative of the inner function $\varphi(x) = \arcsin(x)$ is required, which diverges for $x=-1$ and $x=1$, a value range of $x \in  [-0.9,0.9]$ is selected in this example.
The results can be seen in Figure \ref{parameter_shift_derivative_real_qc}.
\begin{figure*}[!tp]
    \centering
  \subfloat[]{%
       \scalebox{0.52}{
\begin{tikzpicture}

\definecolor{dimgray85}{RGB}{85,85,85}
\definecolor{gainsboro229}{RGB}{229,229,229}
\definecolor{lightgray204}{RGB}{204,204,204}

\begin{axis}[
axis line style={lightgray204},
legend cell align={left},
legend style={
  fill opacity=0.8,
  draw opacity=1,
  text opacity=1,
  at={(0.03,0.97)},
  anchor=north west,
  draw=none
},
tick align=outside,
tick pos=left,
x grid style={lightgray204},
xlabel=\textcolor{dimgray85}{$x$},
xmajorgrids,
xmin=-1, xmax=1,
xtick style={color=dimgray85},
y grid style={lightgray204},
ylabel=\textcolor{dimgray85}{$f^{\text{post}}_{QC}(x)$},
ymajorgrids,
ymin=-1.1, ymax=1.1,
ytick style={color=dimgray85}
]
\addplot [thick, blue]
table {%
-1 -1
-0.97979797979798 -0.940610059373451
-0.95959595959596 -0.883619379181056
-0.939393939393939 -0.828978490135515
-0.919191919191919 -0.776637922949524
-0.898989898989899 -0.726548208335781
-0.878787878787879 -0.678659877006984
-0.858585858585859 -0.632923459675832
-0.838383838383838 -0.589289487055021
-0.818181818181818 -0.54770848985725
-0.797979797979798 -0.508130998795217
-0.777777777777778 -0.470507544581619
-0.757575757575758 -0.434788657929154
-0.737373737373737 -0.40092486955052
-0.717171717171717 -0.368866710158415
-0.696969696969697 -0.338564710465537
-0.676767676767677 -0.309969401184583
-0.656565656565657 -0.283031313028252
-0.636363636363636 -0.257700976709241
-0.616161616161616 -0.233928922940248
-0.595959595959596 -0.211665682433971
-0.575757575757576 -0.190861785903108
-0.555555555555556 -0.171467764060357
-0.535353535353535 -0.153434147618414
-0.515151515151515 -0.13671146728998
-0.494949494949495 -0.12125025378775
-0.474747474747475 -0.107001037824423
-0.454545454545454 -0.0939143501126972
-0.434343434343434 -0.0819407213652698
-0.414141414141414 -0.071030682294839
-0.393939393939394 -0.0611347636141025
-0.373737373737374 -0.052203496035758
-0.353535353535353 -0.0441874102725036
-0.333333333333333 -0.037037037037037
-0.313131313131313 -0.0307029070420561
-0.292929292929293 -0.0251355510002587
-0.272727272727273 -0.0202854996243426
-0.252525252525252 -0.0161032836270057
-0.232323232323232 -0.0125394337209458
-0.212121212121212 -0.00954448061886077
-0.191919191919192 -0.00706895503344844
-0.171717171717172 -0.00506338767740664
-0.151515151515151 -0.00347830926343323
-0.131313131313131 -0.00226425050422601
-0.111111111111111 -0.00137174211248285
-0.0909090909090908 -0.000751314800901576
-0.0707070707070706 -0.000353499282180028
-0.0505050505050504 -0.000128826269016045
-0.0303030303030303 -2.78264741074658e-05
-0.0101010101010101 -1.03061015212835e-06
0.0101010101010102 1.03061015212838e-06
0.0303030303030305 2.78264741074664e-05
0.0505050505050506 0.000128826269016046
0.0707070707070707 0.000353499282180029
0.0909090909090911 0.000751314800901581
0.111111111111111 0.00137174211248286
0.131313131313131 0.00226425050422603
0.151515151515152 0.00347830926343324
0.171717171717172 0.00506338767740665
0.191919191919192 0.00706895503344847
0.212121212121212 0.00954448061886079
0.232323232323232 0.0125394337209458
0.252525252525253 0.0161032836270057
0.272727272727273 0.0202854996243426
0.292929292929293 0.0251355510002587
0.313131313131313 0.0307029070420561
0.333333333333333 0.0370370370370371
0.353535353535354 0.0441874102725036
0.373737373737374 0.0522034960357581
0.393939393939394 0.0611347636141025
0.414141414141414 0.0710306822948391
0.434343434343434 0.08194072136527
0.454545454545455 0.0939143501126972
0.474747474747475 0.107001037824423
0.494949494949495 0.12125025378775
0.515151515151515 0.13671146728998
0.535353535353535 0.153434147618415
0.555555555555556 0.171467764060357
0.575757575757576 0.190861785903108
0.595959595959596 0.211665682433971
0.616161616161616 0.233928922940249
0.636363636363636 0.257700976709241
0.656565656565657 0.283031313028252
0.676767676767677 0.309969401184583
0.696969696969697 0.338564710465537
0.717171717171717 0.368866710158415
0.737373737373737 0.40092486955052
0.757575757575758 0.434788657929154
0.777777777777778 0.470507544581619
0.797979797979798 0.508130998795217
0.818181818181818 0.547708489857251
0.838383838383838 0.589289487055021
0.858585858585859 0.632923459675832
0.878787878787879 0.678659877006985
0.898989898989899 0.726548208335782
0.919191919191919 0.776637922949524
0.939393939393939 0.828978490135515
0.95959595959596 0.883619379181057
0.97979797979798 0.940610059373451
1 1
};
\addlegendentry{$f(x) = x^3$}
\addplot [semithick, red, mark=x, mark size=3, mark options={solid}]
table {%
-0.9 -0.417445414452927
-0.7 -0.517243204700341
-0.5 -0.167888377504228
-0.3 0.107163389163609
-0.1 -0.0884652290796645
0.1 0.073491018433947
0.3 0.0370283500817575
0.5 0.0865722465061352
0.7 0.332701898662561
0.9 0.759545632750206
};
\addlegendentry{Function}
\end{axis}

\end{tikzpicture}}}
    \hfill
  \subfloat[]{%
        \scalebox{0.52}{
\begin{tikzpicture}

\definecolor{dimgray85}{RGB}{85,85,85}
\definecolor{gainsboro229}{RGB}{229,229,229}
\definecolor{lightgray204}{RGB}{204,204,204}

\begin{axis}[
axis line style={lightgray204},
legend cell align={left},
legend style={
  fill opacity=0.8,
  draw opacity=1,
  text opacity=1,
  at={(0.5,0.91)},
  anchor=north,
  draw=none
},
tick align=outside,
tick pos=left,
x grid style={lightgray204},
xlabel=\textcolor{dimgray85}{$x$},
xmajorgrids,
xmin=-1, xmax=1,
xtick style={color=dimgray85},
y grid style={lightgray204},
ylabel=\textcolor{dimgray85}{$f^{\text{post}}_{QC}(x)$},
ymajorgrids,
ymin=-0.4, ymax=3.1,
ytick style={color=dimgray85}
]
\addplot [thick, blue]
table {%
-1 3
-0.97979797979798 2.88001224364861
-0.95959595959596 2.76247321701867
-0.939393939393939 2.64738292011019
-0.919191919191919 2.53474135292317
-0.898989898989899 2.42454851545761
-0.878787878787879 2.3168044077135
-0.858585858585859 2.21150902969085
-0.838383838383838 2.10866238138965
-0.818181818181818 2.00826446280992
-0.797979797979798 1.91031527395164
-0.777777777777778 1.81481481481481
-0.757575757575758 1.72176308539945
-0.737373737373737 1.63116008570554
-0.717171717171717 1.54300581573309
-0.696969696969697 1.45730027548209
-0.676767676767677 1.37404346495256
-0.656565656565657 1.29323538414448
-0.636363636363636 1.21487603305785
-0.616161616161616 1.13896541169268
-0.595959595959596 1.06550352004897
-0.575757575757576 0.994490358126722
-0.555555555555556 0.925925925925926
-0.535353535353535 0.859810223446587
-0.515151515151515 0.796143250688705
-0.494949494949495 0.73492500765228
-0.474747474747475 0.676155494337312
-0.454545454545454 0.619834710743801
-0.434343434343434 0.565962656871748
-0.414141414141414 0.514539332721151
-0.393939393939394 0.465564738292011
-0.373737373737374 0.419038873584328
-0.353535353535353 0.374961738598102
-0.333333333333333 0.333333333333333
-0.313131313131313 0.294153657790021
-0.292929292929293 0.257422711968166
-0.272727272727273 0.223140495867769
-0.252525252525252 0.191307009488828
-0.232323232323232 0.161922252831344
-0.212121212121212 0.134986225895317
-0.191919191919192 0.110498928680747
-0.171717171717172 0.0884603611876338
-0.151515151515151 0.0688705234159779
-0.131313131313131 0.051729415365779
-0.111111111111111 0.037037037037037
-0.0909090909090908 0.024793388429752
-0.0707070707070706 0.014998469543924
-0.0505050505050504 0.00765228037955307
-0.0303030303030303 0.00275482093663911
-0.0101010101010101 0.000306091215182122
0.0101010101010102 0.000306091215182128
0.0303030303030305 0.00275482093663915
0.0505050505050506 0.00765228037955314
0.0707070707070707 0.0149984695439241
0.0909090909090911 0.0247933884297521
0.111111111111111 0.0370370370370371
0.131313131313131 0.0517294153657791
0.151515151515152 0.068870523415978
0.171717171717172 0.0884603611876339
0.191919191919192 0.110498928680747
0.212121212121212 0.134986225895317
0.232323232323232 0.161922252831344
0.252525252525253 0.191307009488828
0.272727272727273 0.223140495867769
0.292929292929293 0.257422711968167
0.313131313131313 0.294153657790021
0.333333333333333 0.333333333333334
0.353535353535354 0.374961738598102
0.373737373737374 0.419038873584329
0.393939393939394 0.465564738292011
0.414141414141414 0.514539332721151
0.434343434343434 0.565962656871748
0.454545454545455 0.619834710743802
0.474747474747475 0.676155494337313
0.494949494949495 0.734925007652281
0.515151515151515 0.796143250688706
0.535353535353535 0.859810223446588
0.555555555555556 0.925925925925926
0.575757575757576 0.994490358126722
0.595959595959596 1.06550352004897
0.616161616161616 1.13896541169269
0.636363636363636 1.21487603305785
0.656565656565657 1.29323538414448
0.676767676767677 1.37404346495256
0.696969696969697 1.45730027548209
0.717171717171717 1.54300581573309
0.737373737373737 1.63116008570554
0.757575757575758 1.72176308539945
0.777777777777778 1.81481481481482
0.797979797979798 1.91031527395164
0.818181818181818 2.00826446280992
0.838383838383838 2.10866238138965
0.858585858585859 2.21150902969085
0.878787878787879 2.3168044077135
0.898989898989899 2.42454851545761
0.919191919191919 2.53474135292317
0.939393939393939 2.64738292011019
0.95959595959596 2.76247321701867
0.97979797979798 2.88001224364861
1 3
};
\addlegendentry{$f'(x) = 3x^2$}
\addplot [semithick, red, mark=x, mark size=3, mark options={solid}]
table {%
-0.9 1.42669903189264
-0.7 1.23653717753064
-0.5 0.545354692230121
-0.3 0.0536413163146281
-0.1 0.00938539059644089
0.1 -0.117765093928388
0.3 0.496116617195415
0.5 0.770587221461891
0.7 1.23129523008381
0.9 1.2129381100184
};
\addlegendentry{First Derivative with PSR}
\end{axis}

\end{tikzpicture}}}
    \hfill
  \subfloat[]{%
        \scalebox{0.52}{
\begin{tikzpicture}

\definecolor{dimgray85}{RGB}{85,85,85}
\definecolor{gainsboro229}{RGB}{229,229,229}
\definecolor{lightgray204}{RGB}{204,204,204}

\begin{axis}[
axis line style={lightgray204},
legend cell align={left},
legend style={
  fill opacity=0.8,
  draw opacity=1,
  text opacity=1,
  at={(0.03,0.97)},
  anchor=north west,
  draw=none
},
tick align=outside,
tick pos=left,
x grid style={lightgray204},
xlabel=\textcolor{dimgray85}{$x$},
xmajorgrids,
xmin=-1, xmax=1,
xtick style={color=dimgray85},
y grid style={lightgray204},
ylabel=\textcolor{dimgray85}{$f^{\text{post}}_{QC}(x)$},
ymajorgrids,
ymin=-8.1, ymax=7.1,
ytick style={color=dimgray85}
]
\addplot [thick, blue]
table {%
-1 -6
-0.97979797979798 -5.87878787878788
-0.95959595959596 -5.75757575757576
-0.939393939393939 -5.63636363636364
-0.919191919191919 -5.51515151515152
-0.898989898989899 -5.39393939393939
-0.878787878787879 -5.27272727272727
-0.858585858585859 -5.15151515151515
-0.838383838383838 -5.03030303030303
-0.818181818181818 -4.90909090909091
-0.797979797979798 -4.78787878787879
-0.777777777777778 -4.66666666666667
-0.757575757575758 -4.54545454545454
-0.737373737373737 -4.42424242424242
-0.717171717171717 -4.3030303030303
-0.696969696969697 -4.18181818181818
-0.676767676767677 -4.06060606060606
-0.656565656565657 -3.93939393939394
-0.636363636363636 -3.81818181818182
-0.616161616161616 -3.6969696969697
-0.595959595959596 -3.57575757575758
-0.575757575757576 -3.45454545454545
-0.555555555555556 -3.33333333333333
-0.535353535353535 -3.21212121212121
-0.515151515151515 -3.09090909090909
-0.494949494949495 -2.96969696969697
-0.474747474747475 -2.84848484848485
-0.454545454545454 -2.72727272727273
-0.434343434343434 -2.60606060606061
-0.414141414141414 -2.48484848484848
-0.393939393939394 -2.36363636363636
-0.373737373737374 -2.24242424242424
-0.353535353535353 -2.12121212121212
-0.333333333333333 -2
-0.313131313131313 -1.87878787878788
-0.292929292929293 -1.75757575757576
-0.272727272727273 -1.63636363636364
-0.252525252525252 -1.51515151515151
-0.232323232323232 -1.39393939393939
-0.212121212121212 -1.27272727272727
-0.191919191919192 -1.15151515151515
-0.171717171717172 -1.03030303030303
-0.151515151515151 -0.909090909090909
-0.131313131313131 -0.787878787878788
-0.111111111111111 -0.666666666666666
-0.0909090909090908 -0.545454545454545
-0.0707070707070706 -0.424242424242424
-0.0505050505050504 -0.303030303030302
-0.0303030303030303 -0.181818181818182
-0.0101010101010101 -0.0606060606060603
0.0101010101010102 0.060606060606061
0.0303030303030305 0.181818181818183
0.0505050505050506 0.303030303030304
0.0707070707070707 0.424242424242424
0.0909090909090911 0.545454545454546
0.111111111111111 0.666666666666667
0.131313131313131 0.787878787878789
0.151515151515152 0.90909090909091
0.171717171717172 1.03030303030303
0.191919191919192 1.15151515151515
0.212121212121212 1.27272727272727
0.232323232323232 1.39393939393939
0.252525252525253 1.51515151515152
0.272727272727273 1.63636363636364
0.292929292929293 1.75757575757576
0.313131313131313 1.87878787878788
0.333333333333333 2
0.353535353535354 2.12121212121212
0.373737373737374 2.24242424242424
0.393939393939394 2.36363636363636
0.414141414141414 2.48484848484849
0.434343434343434 2.60606060606061
0.454545454545455 2.72727272727273
0.474747474747475 2.84848484848485
0.494949494949495 2.96969696969697
0.515151515151515 3.09090909090909
0.535353535353535 3.21212121212121
0.555555555555556 3.33333333333333
0.575757575757576 3.45454545454546
0.595959595959596 3.57575757575758
0.616161616161616 3.6969696969697
0.636363636363636 3.81818181818182
0.656565656565657 3.93939393939394
0.676767676767677 4.06060606060606
0.696969696969697 4.18181818181818
0.717171717171717 4.3030303030303
0.737373737373737 4.42424242424242
0.757575757575758 4.54545454545455
0.777777777777778 4.66666666666667
0.797979797979798 4.78787878787879
0.818181818181818 4.90909090909091
0.838383838383838 5.03030303030303
0.858585858585859 5.15151515151515
0.878787878787879 5.27272727272727
0.898989898989899 5.3939393939394
0.919191919191919 5.51515151515152
0.939393939393939 5.63636363636364
0.95959595959596 5.75757575757576
0.97979797979798 5.87878787878788
1 6
};
\addlegendentry{$f''(x) = 6x$}
\addplot [semithick, red, mark=x, mark size=3, mark options={solid}]
table {%
-0.9 -7.44647092458857
-0.7 -2.83139377084935
-0.5 -2.45191458160923
-0.3 -1.36870456002826
-0.1 -0.170383606486508
0.1 0.0787950591302489
0.3 1.83213926920688
0.5 2.22584229226111
0.7 3.21225955168179
0.9 2.77390350822483
};
\addlegendentry{Second Derivative with PSR}
\end{axis}

\end{tikzpicture}}}
    \\
  \subfloat[]{%
        \scalebox{0.53}{
\begin{tikzpicture}

\definecolor{dimgray85}{RGB}{85,85,85}
\definecolor{gainsboro229}{RGB}{229,229,229}
\definecolor{lightgray204}{RGB}{204,204,204}

\begin{axis}[
axis line style={lightgray204},
tick align=outside,
tick pos=left,
x grid style={lightgray204},
xlabel=\textcolor{dimgray85}{$x$},
xmajorgrids,
xmin=-1, xmax=1,
xtick={-1,-0.5,0,0.5,1},
xticklabels={-1,-0.5,0,0.5,1},
xtick style={color=dimgray85},
y grid style={lightgray204},
ylabel=\textcolor{dimgray85}{Error},
ymajorgrids,
ymin=-0.1, ymax=2.8,
ytick={0,1,2},
yticklabels={0,1,2},
ytick style={color=dimgray85}
]
\addplot [semithick, red, mark=x, mark size=3, mark options={solid}]
table {%
-0.9 0.311554585547073
-0.7 0.174243204700341
-0.5 0.0428883775042282
-0.3 0.134163389163609
-0.1 0.0874652290796645
0.1 0.072491018433947
0.3 0.0100283500817575
0.5 0.0384277534938648
0.7 0.0102981013374393
0.9 0.0305456327502064
};
\end{axis}

\end{tikzpicture}}}
    \hfill
  \subfloat[]{%
        \scalebox{0.53}{
\begin{tikzpicture}

\definecolor{dimgray85}{RGB}{85,85,85}
\definecolor{gainsboro229}{RGB}{229,229,229}
\definecolor{lightgray204}{RGB}{204,204,204}

\begin{axis}[
axis line style={lightgray204},
tick align=outside,
tick pos=left,
x grid style={lightgray204},
xlabel=\textcolor{dimgray85}{$x$},
xmajorgrids,
xmin=-1, xmax=1,
xtick={-1,-0.5,0,0.5,1},
xticklabels={-1,-0.5,0,0.5,1},
xtick style={color=dimgray85},
y grid style={lightgray204},
ylabel=\textcolor{dimgray85}{Error},
ymajorgrids,
ymin=-0.1, ymax=2.8,
ytick={0,1,2},
yticklabels={0,1,2},
ytick style={color=dimgray85}
]
\addplot [semithick, red, mark=x, mark size=3, mark options={solid}]
table {%
-0.9 1.00330096810736
-0.7 0.233462822469356
-0.5 0.204645307769879
-0.3 0.216358683685372
-0.1 0.0206146094035591
0.1 0.147765093928388
0.3 0.226116617195415
0.5 0.0205872214618906
0.7 0.238704769916189
0.9 1.2170618899816
};
\end{axis}

\end{tikzpicture}}}
    \hfill
  \subfloat[]{%
        \scalebox{0.53}{
\begin{tikzpicture}

\definecolor{dimgray85}{RGB}{85,85,85}
\definecolor{gainsboro229}{RGB}{229,229,229}
\definecolor{lightgray204}{RGB}{204,204,204}

\begin{axis}[
axis line style={lightgray204},
tick align=outside,
tick pos=left,
x grid style={lightgray204},
xlabel=\textcolor{dimgray85}{$x$},
xmajorgrids,
xmin=-1, xmax=1,
xtick={-1,-0.5,0,0.5,1},
xticklabels={-1,-0.5,0,0.5,1},
xtick style={color=dimgray85},
y grid style={lightgray204},
ylabel=\textcolor{dimgray85}{Error},
ymajorgrids,
ymin=-0.1, ymax=2.8,
ytick={0,1,2},
yticklabels={0,1,2},
ytick style={color=dimgray85}
]
\addplot [semithick, red, mark=x, mark size=3, mark options={solid}]
table {%
-0.9 2.04647092458857
-0.7 1.36860622915065
-0.5 0.548085418390774
-0.3 0.431295439971736
-0.1 0.429616393513491
0.1 0.521204940869751
0.3 0.032139269206878
0.5 0.774157707738892
0.7 0.987740448318215
0.9 2.62609649177517
};
\end{axis}

\end{tikzpicture}}}
  \caption{Resulting function values from executing the QCL circuit with trained parameters for $f(x) = x ^3$, see Figure \ref{function_approx_example}, on ibmq\_ehningen with 10 equidistant training points (a). Values of the first (b) and second (c) derivative obtained with the parameter shift rule applied to aforementioned circuit and executed on ibmq\_ehningen. The shot number is chosen as 1024 in all cases. In (d)-(f) we provide the absolute value of the errors between the QCL model and the exact functions on the grid points.}
  \label{parameter_shift_derivative_real_qc} 
\end{figure*}
The qualitative behavior of the derivatives can be determined well. However, there are large errors, especially near $x = -1$ and $x = 1$, which go far beyond shot noise.
The errors are mainly hardware errors that accumulate due to the high number of circuits that need to be evaluated.\\
In the next section, we focus on differential equations where these derivatives become crucial. Solving these differential equations on a real quantum computer is therefore associated with large errors.
\subsection{Differential Equation on IBM Quantum Computer}
\label{chap_DGL_real_qc}
In this section, we aim to solve a differential equation with QCL circuits and the PSR on ibmq\_ehningen. For this purpose, a simple example of a differential equation is considered to limit the required quantum resources, namely
\begin{equation}
\left\{
\begin{aligned}
f'(x) &= 3x^2\,, \\[0.5ex]
f(0)&= 0\,.
\end{aligned}
\right.
\label{dgl_for_qc}
\end{equation}
The solution of this differential equation is
\begin{equation}
    f(x) = x^3\,.
    \label{dgl_for_qc_solution}
\end{equation}
To be able to solve it with QCL circuits, we define the cost function
\begin{equation}
    \begin{split}
        L(\boldsymbol{\theta}) =& \sum_i\Bigl(\left|{f'}^{\text{post}}_{QC}(x_i, \boldsymbol{\theta}) - 3x_i^2\right|^2\\
        &+ \mu  \left|{f}^{\text{post}}_{QC}(0, \boldsymbol{\theta}) - 0\right|^2 \Bigr),
    \end{split}
\end{equation}
where $\mu$ is a weight factor and the derivatives are determined with the PSR.\\
The result for a circuit with $N=3$, $D=3$, a weight factor $\mu = 10$, 10 equidistant training points and 2000 shots is shown in Figure \ref{dgl_real_qc}. The value range is chosen to be $x \in [-0.9, 0.9]$ because the $R_Y(\arcsin(x))$-encoding is used.
It can be observed that the differential equation can be solved to a certain extent in the sense that the qualitative behavior can be reproduced approximately. The errors (see Figure \ref{dgl_real_qc} (d)-(f)) are higher than in the case of learned functions on ibmq\_ehningen (compare Figure \ref{x^3_real_backend} (d)-(f)). This is because the derivatives are included in the cost function which results in significantly more circuit evaluations and the errors accumulate.
\begin{figure*}[!tp]
    \centering
  \subfloat[]{%
       \scalebox{0.53}{
\begin{tikzpicture}

\definecolor{dimgray85}{RGB}{85,85,85}
\definecolor{gainsboro229}{RGB}{229,229,229}
\definecolor{lightgray204}{RGB}{204,204,204}

\begin{axis}[
axis line style={lightgray204},
legend cell align={left},
legend style={
  fill opacity=0.8,
  draw opacity=1,
  text opacity=1,
  at={(0.03,0.97)},
  anchor=north west,
  draw=none
},
tick align=outside,
tick pos=left,
x grid style={lightgray204},
xlabel=\textcolor{dimgray85}{$x$},
xmajorgrids,
xmin=-1, xmax=1,
xtick style={color=dimgray85},
y grid style={lightgray204},
ylabel=\textcolor{dimgray85}{$f^{\text{post}}_{QC}(x)$},
ymajorgrids,
ymin=-1, ymax=1,
ytick style={color=dimgray85}
]
\addplot [thick, blue]
table {%
-1 -1
-0.97979797979798 -0.940610059373451
-0.95959595959596 -0.883619379181056
-0.939393939393939 -0.828978490135515
-0.919191919191919 -0.776637922949524
-0.898989898989899 -0.726548208335781
-0.878787878787879 -0.678659877006984
-0.858585858585859 -0.632923459675832
-0.838383838383838 -0.589289487055021
-0.818181818181818 -0.54770848985725
-0.797979797979798 -0.508130998795217
-0.777777777777778 -0.470507544581619
-0.757575757575758 -0.434788657929154
-0.737373737373737 -0.40092486955052
-0.717171717171717 -0.368866710158415
-0.696969696969697 -0.338564710465537
-0.676767676767677 -0.309969401184583
-0.656565656565657 -0.283031313028252
-0.636363636363636 -0.257700976709241
-0.616161616161616 -0.233928922940248
-0.595959595959596 -0.211665682433971
-0.575757575757576 -0.190861785903108
-0.555555555555556 -0.171467764060357
-0.535353535353535 -0.153434147618414
-0.515151515151515 -0.13671146728998
-0.494949494949495 -0.12125025378775
-0.474747474747475 -0.107001037824423
-0.454545454545454 -0.0939143501126972
-0.434343434343434 -0.0819407213652698
-0.414141414141414 -0.071030682294839
-0.393939393939394 -0.0611347636141025
-0.373737373737374 -0.052203496035758
-0.353535353535353 -0.0441874102725036
-0.333333333333333 -0.037037037037037
-0.313131313131313 -0.0307029070420561
-0.292929292929293 -0.0251355510002587
-0.272727272727273 -0.0202854996243426
-0.252525252525252 -0.0161032836270057
-0.232323232323232 -0.0125394337209458
-0.212121212121212 -0.00954448061886077
-0.191919191919192 -0.00706895503344844
-0.171717171717172 -0.00506338767740664
-0.151515151515151 -0.00347830926343323
-0.131313131313131 -0.00226425050422601
-0.111111111111111 -0.00137174211248285
-0.0909090909090908 -0.000751314800901576
-0.0707070707070706 -0.000353499282180028
-0.0505050505050504 -0.000128826269016045
-0.0303030303030303 -2.78264741074658e-05
-0.0101010101010101 -1.03061015212835e-06
0.0101010101010102 1.03061015212838e-06
0.0303030303030305 2.78264741074664e-05
0.0505050505050506 0.000128826269016046
0.0707070707070707 0.000353499282180029
0.0909090909090911 0.000751314800901581
0.111111111111111 0.00137174211248286
0.131313131313131 0.00226425050422603
0.151515151515152 0.00347830926343324
0.171717171717172 0.00506338767740665
0.191919191919192 0.00706895503344847
0.212121212121212 0.00954448061886079
0.232323232323232 0.0125394337209458
0.252525252525253 0.0161032836270057
0.272727272727273 0.0202854996243426
0.292929292929293 0.0251355510002587
0.313131313131313 0.0307029070420561
0.333333333333333 0.0370370370370371
0.353535353535354 0.0441874102725036
0.373737373737374 0.0522034960357581
0.393939393939394 0.0611347636141025
0.414141414141414 0.0710306822948391
0.434343434343434 0.08194072136527
0.454545454545455 0.0939143501126972
0.474747474747475 0.107001037824423
0.494949494949495 0.12125025378775
0.515151515151515 0.13671146728998
0.535353535353535 0.153434147618415
0.555555555555556 0.171467764060357
0.575757575757576 0.190861785903108
0.595959595959596 0.211665682433971
0.616161616161616 0.233928922940249
0.636363636363636 0.257700976709241
0.656565656565657 0.283031313028252
0.676767676767677 0.309969401184583
0.696969696969697 0.338564710465537
0.717171717171717 0.368866710158415
0.737373737373737 0.40092486955052
0.757575757575758 0.434788657929154
0.777777777777778 0.470507544581619
0.797979797979798 0.508130998795217
0.818181818181818 0.547708489857251
0.838383838383838 0.589289487055021
0.858585858585859 0.632923459675832
0.878787878787879 0.678659877006985
0.898989898989899 0.726548208335782
0.919191919191919 0.776637922949524
0.939393939393939 0.828978490135515
0.95959595959596 0.883619379181057
0.97979797979798 0.940610059373451
1 1
};
\addlegendentry{Exact Solution}
\addplot [semithick, red, mark=x, mark size=3, mark options={solid}]
table {%
-0.9 -0.710598715154526
-0.7 -0.443886426138428
-0.5 -0.385148648494709
-0.3 -0.0854486503798252
-0.1 -0.148524713034089
0.1 -0.107908128227997
0.3 0.0941331809347994
0.5 0.262152057420894
0.7 0.413451086908256
0.9 0.800220952818011
};
\addlegendentry{Final Function}
\end{axis}

\end{tikzpicture}}}
    \hfill
  \subfloat[]{%
        \scalebox{0.53}{
\begin{tikzpicture}

\definecolor{dimgray85}{RGB}{85,85,85}
\definecolor{gainsboro229}{RGB}{229,229,229}
\definecolor{lightgray204}{RGB}{204,204,204}

\begin{axis}[
axis line style={lightgray204},
tick align=outside,
tick pos=left,
x grid style={lightgray204},
xlabel=\textcolor{dimgray85}{$x$},
xmajorgrids,
xmin=-0.99, xmax=0.99,
xtick style={color=dimgray85},
y grid style={lightgray204},
ylabel=\textcolor{dimgray85}{Error},
ymajorgrids,
ymin=0, ymax=0.3,
ytick={0,0.05,0.1,0.15,0.2,0.25,0.3},
yticklabels={0,0.05,0.1,0.15,0.2,0.25,0.3},
ytick style={color=dimgray85}
]
\addplot [semithick, red, mark=x, mark size=3, mark options={solid}]
table {%
-0.9 0.0184012848454737
-0.7 0.100886426138429
-0.5 0.260148648494709
-0.3 0.0584486503798252
-0.1 0.147524713034089
0.1 0.108908128227997
0.3 0.0671331809347994
0.5 0.137152057420894
0.7 0.0704510869082555
0.9 0.071220952818011
};
\end{axis}

\end{tikzpicture}}}
    \hfill
  \subfloat[]{%
        \scalebox{0.53}{
\begin{tikzpicture}

\definecolor{dimgray85}{RGB}{85,85,85}
\definecolor{gainsboro229}{RGB}{229,229,229}

\begin{axis}[
axis line style={gainsboro229},
log basis y={10},
tick align=outside,
tick pos=left,
x grid style={gainsboro229},
xlabel=\textcolor{dimgray85}{Cost Function Evaluations},
xmajorgrids,
xmin=-2.5, xmax=74.5,
xtick style={color=dimgray85},
y grid style={gainsboro229},
ylabel=\textcolor{dimgray85}{Cost Function Value},
ymajorgrids,
ymin=1, ymax=70,
ymode=log,
ytick style={color=dimgray85},
ytick={0.01,0.1,1,10,100,1000},
yticklabels={
  \(\displaystyle {10^{-2}}\),
  \(\displaystyle {10^{-1}}\),
  \(\displaystyle {10^{0}}\),
  \(\displaystyle {10^{1}}\),
  \(\displaystyle {10^{2}}\),
  \(\displaystyle {10^{3}}\)
}
]
\addplot [semithick, red]
table {%
1 26.1385873760395
2 25.9468436592913
3 26.465237861589
4 40.1505865426007
5 23.7715970505812
6 24.9819534633648
7 26.7623828692076
8 20.0550226094446
9 11.0572281379294
10 28.6664318263998
11 10.3800753016001
12 3.32297419845723
13 28.8634133926928
14 19.9952134652185
15 51.6011942207193
16 9.19923195344372
17 20.7748182826115
18 65.9428569943307
19 15.1429071407308
20 22.6648941348889
21 21.2307273470363
22 20.0945678519491
23 62.0204008516644
24 38.6969738597025
25 18.0921123775829
26 19.1981058135469
27 27.7084980010591
28 18.9199551660609
29 33.8356063832828
30 3.78007056599567
31 57.2836841219141
32 16.6781205011889
33 4.68033373916297
34 31.024140787762
35 19.7651557862311
36 35.7532281229376
37 20.5459577811055
38 30.823612945772
39 13.8890734808104
40 5.82269682809665
41 3.45205692752208
42 14.1221044505181
43 8.89417342428458
44 4.5358087708723
45 5.66402895082016
46 18.4675344867668
47 5.23598178933597
48 10.6066148306508
49 12.5937165305943
50 39.7483354549768
51 15.0860652580642
52 20.9092866626893
53 8.62876900051647
54 5.34897889189168
55 4.09335668727324
56 8.0987643664651
57 25.1064864801013
58 1.85389325216779
59 6.1198430870855
60 10.8007630511101
61 9.88908313870172
62 31.1694012580901
63 5.15469921877014
64 21.5239403387692
65 6.02431335518293
66 1.08170988202115
67 4.87325761335887
68 27.9968292308711
69 3.48083118308576
70 12.1911615219162
71 1.77012637244235
};
\end{axis}

\end{tikzpicture}}}
  \caption{(a) Solution of the differential equation \eqref{dgl_for_qc} using QCL circuits on ibmq\_ehningen with $\mu = 10$, $N = 3$, $D = 3$, 10 equidistant training points and 2000 shots with $R_Y(\arcsin(x))$ data encoding. The parameters are optimized using COBYLA. (b) The absolute values of the respective errors $|f(x) - f^{\text{post}}_{QC}(x)|$. (c) Values of the cost function versus the number of cost function evaluations during the optimization process.}
  \label{dgl_real_qc} 
\end{figure*}
\section{Multi-Qubit Measurements}
\label{chap:Multi-Qubit}
So far, we have only measured a single qubit in all our QCL circuits.
More specifically, we have used the $Z$ expectation value of the first qubit to learn different functions.
However, we found that it is possible to use the expectation values of different qubits to learn different functions simultaneously with the same circuit.
In this section, this idea will be investigated in more detail.
For this purpose, simulations are carried out where the expectation value of the first qubit
and the expectation value of the second qubit are used to learn two different functions simultaneously.
The quantum model function of the first qubit is $f^{\text{post}}_{QC}(x, \boldsymbol{\theta})$, as before. The quantum model function of the second qubit is
\begin{equation}
    g^{\text{post}}_{QC}(x, \boldsymbol{\theta}) = \braket{Z_1}(x, \boldsymbol{\theta})\cdot \tilde{\theta}_{\text{post}}\,,
\end{equation}
where $\braket{Z_1}(x, \boldsymbol{\theta})$ is the $Z$ expectation value of the second qubit and $\tilde{\theta}_{\text{post}}$ is a separate post-processing parameter.
We define the cost function
\begin{equation}
    \begin{split}
        L(\boldsymbol{\theta}) =& \sum_i \bigl(|f(x_i) - f^{\text{post}}_{QC}(x_i, \boldsymbol{\theta})|^2\\
       &+ |g(x_i) - g^{\text{post}}_{QC}(x_i, \boldsymbol{\theta})|^2\bigr)\,,
    \end{split}
\end{equation}
where $f(x)$ and $g(x)$ can be two different functions.\\
Three pairs of exemplary functions ($f_1(x) = x$ and $g_1(x) = x^2$; $f_2(x) = x^2$ and $g_2(x) = x^3$; $f_3(x) = x^3-x^2+1$ and $g_3(x) = x^3$) are now learned with a three qubit QCL circuit with a depth of $D = 3$ and $R_Y(\arcsin(x))$ data encoding (see Figure \ref{circ_simple_example_multi}) on a statevector simulator.
\begin{figure}[H]
    \centering
    \begin{tikzpicture}
    \node[scale=0.5] {
    \begin{quantikz}
    \lstick{$\ket{0}$}&  \gate{R_Y(\arcsin(x))}  & \ctrl{1} \gategroup[3,steps=6,style={dashed,rounded corners, inner xsep=2pt},background,label style={label position=below,anchor=north,yshift=-0.2cm}]{{$3\times$}}&  \qw & \targ{} & \gate{R_X(\theta_0 )}& \gate{R_Y(\theta_1 )}  & \gate{R_Z(\theta_2)}  &   \meter[]{ \braket{Z}} \\[0.2cm]
    \lstick{$\ket{0}$} &  \gate{R_Y(\arcsin(x))}  & \targ{} &  \ctrl{1} & \qw  & \gate{R_X(\theta_3 )}& \gate{R_Y(\theta_4)}  & \gate{R_Z(\theta_5)}  &   \meter[]{ \braket{Z}}\\
    \lstick{$\ket{0}$} &  \gate{R_Y(\arcsin(x))}  & \qw  &  \targ{} & \ctrl{-2}  & \gate{R_X(\theta_6 )}& \gate{R_Y(\theta_7)}  & \gate{R_Z(\theta_8)}  &     \qw
    \end{quantikz}
    };
    \end{tikzpicture}
    \caption{Three qubit QCL circuit with $R_Y(\arcsin(x))$ data encoding. The $Z$ expectation values of the first and second qubit are measured.}
    \label{circ_simple_example_multi}
\end{figure}
The cost function is evaluated on 20 equidistant training points and is classically minimized using SLSQP (see Figure \ref{multiple_functions} (a)-(c)).
In Figure \ref{multiple_functions} (d)-(f), the absolute values of the respective errors $|f_i(x) - f^{\text{post}}_{QC,i}(x, \boldsymbol{\theta})|$ are plotted.
These errors are higher than in the previous example in Figure \ref{function_approx_example}, in which only one function was learned.\\
To investigate whether it can be advantageous to use multiple qubits for different functions, a six qubit QCL circuit with a depth of $D=4$ is used to learn polynomials of the form
\begin{equation}
    f(x) = \sum_{i=0}^{6} c_i x^i
\end{equation}
where the coefficients $c_i$ are randomly chosen with
\begin{equation}
    \sum_{i=0}^{6} c_i^2 \le 1\,.
\end{equation}
These polynomials are first learned individually by measuring only the first qubit.
This is done for 100 random polynomials and the average of the resulting convergence curves of the cost function is shown in Figure \ref{convergence} (red line).
In addition, four different random polynomials are simultaneously learned by measuring four different qubits.
This is also repeated 100 times with four different polynomials (400 polynomials in total) and the average of the resulting convergence curves is plotted.
In order to adequately compare the convergence curves for the two cases, the cost function values and the number of function evaluations are divided by four for the multi-function case (blue line in Figure \ref{convergence}).
In order to have the necessary parameters to learn multiple functions, unlike in the rest of the paper, no parameters are repeated here with increasing depth, but new parameters are introduced.\\
In the beginning of the optimization the functions can be learned faster if multiple qubits of the same circuit are used.
However, the maximum accuracy that can be achieved with just one function is higher.
Therefore, the approach presented here could be valuable when aiming to quickly learn the qualitative behavior of several functions.
We only considered one example and there could be further advantages when approximating a higher number of functions.
\begin{figure*}[!tp]
    \centering
  \subfloat[]{%
       \scalebox{0.53}{\input{figures/multi_qubit_measurements/3_multuple_functions_qubits_d3_arcsin_tp30_second_qubit1_function_approx_x_xhoch2}}}
    \hfill
  \subfloat[]{%
        \scalebox{0.53}{\input{figures/multi_qubit_measurements/3_multuple_functions_qubits_d3_arcsin_tp30_second_qubit1_function_approx_3para}}}
    \hfill
  \subfloat[]{%
        \scalebox{0.53}{\input{figures/multi_qubit_measurements/3_multuple_functions_qubits_d3_arcsin_tp30_second_qubit1_function_approx_xhoch3_complex}}}
    \\
    \hspace{-0.3cm}
  \subfloat[]{%
        \scalebox{0.53}{
\pgfplotsset{scaled y ticks=false}
\begin{tikzpicture}

\definecolor{dimgray85}{RGB}{85,85,85}
\definecolor{gainsboro229}{RGB}{229,229,229}
\definecolor{lightgray204}{RGB}{204,204,204}

\begin{axis}[
axis line style={lightgray204},
legend cell align={left},
legend style={
    fill opacity=0.8,
    draw opacity=1,
    text opacity=1,
    draw = none
},
tick align=outside,
tick pos=left,
x grid style={lightgray204},
xlabel=\textcolor{dimgray85}{$x$},
xmajorgrids,
xmin=-1, xmax=1,
xtick style={color=dimgray85},
y grid style={lightgray204},
ylabel=\textcolor{dimgray85}{Error},
ymajorgrids,
ymin=-0.01, ymax=0.07,
ytick={0,0.02,0.04,0.06},
yticklabels={0,0.02,0.04,0.06},
ytick style={color=dimgray85}
]
\addplot [thick, olive]
table {%
-1 0.0101180597602069
-0.97979797979798 0.00676216833498999
-0.95959595959596 0.00955410753803654
-0.939393939393939 0.0103378509592315
-0.919191919191919 0.0102284966338754
-0.898989898989899 0.00964592783932239
-0.878787878787879 0.0087981036512933
-0.858585858585859 0.00780339340170288
-0.838383838383838 0.00673502760126798
-0.818181818181818 0.00564079642162552
-0.797979797979798 0.0045529838737981
-0.777777777777778 0.00349386005618879
-0.757575757575758 0.00247893474857253
-0.737373737373737 0.00151899165092373
-0.717171717171717 0.000621416598271196
-0.696969696969697 0.000208903750605116
-0.676767676767677 0.00096895544442599
-0.656565656565657 0.0016571493326239
-0.636363636363636 0.00227299377353357
-0.616161616161616 0.0028168509052412
-0.595959595959596 0.00328975210123728
-0.575757575757576 0.00369325616602789
-0.555555555555556 0.00402933892984153
-0.535353535353535 0.00430030626630828
-0.515151515151515 0.00450872482389436
-0.494949494949495 0.0046573663192645
-0.474747474747475 0.0047491623295384
-0.454545454545454 0.00478716729401626
-0.434343434343434 0.00477452799345962
-0.414141414141414 0.00471445818233612
-0.393939393939394 0.00461021735052886
-0.373737373737374 0.00446509281620777
-0.353535353535353 0.00428238452161737
-0.333333333333333 0.0040653920333114
-0.313131313131313 0.00381740334813598
-0.292929292929293 0.0035416851837342
-0.272727272727273 0.00324147449281224
-0.252525252525252 0.00291997098803334
-0.232323232323232 0.00258033050210173
-0.212121212121212 0.00222565903758325
-0.191919191919192 0.00185900738502872
-0.171717171717172 0.00148336620723655
-0.151515151515151 0.00110166150302547
-0.131313131313131 0.000716750376411369
-0.111111111111111 0.000331417047240871
-0.0909090909090908 5.16309525511495e-05
-0.0707070707070706 0.000429766446273939
-0.0505050505050504 0.000800446187442679
-0.0303030303030303 0.00116121438224739
-0.0101010101010101 0.00150970619629288
0.0101010101010102 0.00184365127999242
0.0303030303030305 0.00216087734540432
0.0505050505050506 0.00245931382635584
0.0707070707070707 0.00273699565331109
0.0909090909090911 0.00299206717472328
0.111111111111111 0.00322278625737374
0.131313131313131 0.00342752859962878
0.151515151515152 0.00360479229350717
0.171717171717172 0.00375320267413589
0.191919191919192 0.00387151749853246
0.212121212121212 0.00395863249984185
0.232323232323232 0.00401358736828367
0.252525252525253 0.00403557221628303
0.272727272727273 0.00402393459271314
0.292929292929293 0.00397818712025085
0.313131313131313 0.00389801584064632
0.333333333333333 0.00378328936584971
0.353535353535354 0.0036340689487549
0.373737373737374 0.00345061960654344
0.393939393939394 0.00323342245308905
0.414141414141414 0.0029831884256063
0.434343434343434 0.00270087362618024
0.454545454545455 0.00238769654284607
0.474747474747475 0.0020451574698773
0.494949494949495 0.00167506051628555
0.515151515151515 0.00127953867961655
0.535353535353535 0.00086108257511297
0.555555555555556 0.000422573556610573
0.575757575757576 3.26778428262342e-05
0.595959595959596 0.000500886968751002
0.616161616161616 0.000977742079221078
0.636363636363636 0.00145834136875089
0.656565656565657 0.0019371173556248
0.676767676767677 0.00240774510514291
0.696969696969697 0.00286302947079586
0.717171717171717 0.00329476464694178
0.737373737373737 0.0036935564969699
0.757575757575758 0.00404859376603606
0.777777777777778 0.0043473473770409
0.797979797979798 0.00457516565032612
0.818181818181818 0.00471471383802635
0.838383838383838 0.00474517140880071
0.858585858585859 0.00464103394310322
0.878787878787879 0.00437023036201989
0.898989898989899 0.00389096182085513
0.919191919191919 0.00314590402546333
0.939393939393939 0.0020501622146778
0.95959595959596 0.000460954861443708
0.97979797979798 0.00193058190611151
1 0.00749512174314604
};
\addlegendentry{Error Function Qubit 0}
\addplot [thick, red]
table {%
-1 0.000961933686857264
-0.97979797979798 0.00150041573156967
-0.95959595959596 0.00187543122196543
-0.939393939393939 0.00196120212749185
-0.919191919191919 0.00192059039873105
-0.898989898989899 0.0018141319540661
-0.878787878787879 0.00167143758311494
-0.858585858585859 0.00150915537179674
-0.838383838383838 0.0013374589649372
-0.818181818181818 0.00116290588484746
-0.797979797979798 0.00098987037362841
-0.777777777777778 0.000821330695580547
-0.757575757575758 0.000659332585001615
-0.737373737373737 0.000505277114163594
-0.717171717171717 0.00036010737815706
-0.696969696969697 0.000224433919195244
-0.676767676767677 9.86215170452054e-05
-0.656565656565657 1.71492347688096e-05
-0.636363636363636 0.000122837296770317
-0.616161616161616 0.000218508228365222
-0.595959595959596 0.000304309458557639
-0.575757575757576 0.000380451464182197
-0.555555555555556 0.000447193263663381
-0.535353535353535 0.00050483111155708
-0.515151515151515 0.000553689599532281
-0.494949494949495 0.000594114588802208
-0.474747474747475 0.000626467551879561
-0.454545454545454 0.000651121009639544
-0.434343434343434 0.000668454827443654
-0.414141414141414 0.000678853190564432
-0.393939393939394 0.000682702120849088
-0.373737373737374 0.000680387427554496
-0.353535353535353 0.00067229300868038
-0.333333333333333 0.000658799436856261
-0.313131313131313 0.000640282777466361
-0.292929292929293 0.000617113597225938
-0.272727272727273 0.000589656129605265
-0.252525252525252 0.000558267569980897
-0.232323232323232 0.000523297478442403
-0.212121212121212 0.000485087272280776
-0.191919191919192 0.000443969793392646
-0.171717171717172 0.000400268938491766
-0.151515151515151 0.000354299342120196
-0.131313131313131 0.000306366104199485
-0.111111111111111 0.000256764555285963
-0.0909090909090908 0.000205780053879598
-0.0707070707070706 0.000153687811131009
-0.0505050505050504 0.000100752739109783
-0.0303030303030303 4.72293195288367e-05
-0.0101010101010101 6.63850957836713e-06
0.0101010101010102 6.06174511795874e-05
0.0303030303030305 0.000114484932035003
0.0505050505050506 0.000168029171975413
0.0707070707070707 0.000221049240263199
0.0909090909090911 0.000273355099221925
0.111111111111111 0.000324767634898721
0.131313131313131 0.000375118674103475
0.151515151515152 0.000424250986691882
0.171717171717172 0.000472018271479052
0.191919191919192 0.000518285123566553
0.212121212121212 0.000562926980239992
0.232323232323232 0.000605830041805117
0.252525252525253 0.000646891162833285
0.272727272727273 0.00068601770817911
0.292929292929293 0.000723127366814735
0.313131313131313 0.000758147914897811
0.333333333333333 0.000791016917470966
0.353535353535354 0.000821681355715198
0.373737373737374 0.000850097163537622
0.393939393939394 0.000876228653349187
0.414141414141414 0.000900047805886039
0.434343434343434 0.000921533392537421
0.454545454545455 0.000940669890419243
0.474747474747475 0.000957446139742024
0.494949494949495 0.000971853679036794
0.515151515151515 0.000983884675335611
0.535353535353535 0.00099352934176844
0.555555555555556 0.00100077270183607
0.575757575757576 0.00100559051444293
0.595959595959596 0.00100794411140348
0.616161616161616 0.00100777381206296
0.636363636363636 0.00100499045622904
0.656565656565657 0.000999464418858675
0.676767676767677 0.000991011209482529
0.696969696969697 0.000979372370237375
0.717171717171717 0.000964189792342363
0.737373737373737 0.000944970641789111
0.757575757575758 0.000921038591517043
0.777777777777778 0.000891464580385692
0.797979797979798 0.000854966060765516
0.818181818181818 0.000809756062387645
0.838383838383838 0.000753309021869319
0.858585858585859 0.000681981596088233
0.878787878787879 0.000590364953170108
0.898989898989899 0.000470099796044732
0.919191919191919 0.000307500631955726
0.939393939393939 7.81368576753261e-05
0.95959595959596 0.00026824031192807
0.97979797979798 0.000867903174753493
1 0.00324578175484747
};
\addlegendentry{Error Function Qubit 1}
\end{axis}

\end{tikzpicture}}}
    \hfill
  \subfloat[]{%
        \scalebox{0.53}{
\pgfplotsset{scaled y ticks=false}
\begin{tikzpicture}

\definecolor{dimgray85}{RGB}{85,85,85}
\definecolor{gainsboro229}{RGB}{229,229,229}
\definecolor{lightgray204}{RGB}{204,204,204}

\begin{axis}[
axis line style={lightgray204},
legend cell align={left},
legend style={
    fill opacity=0.8,
    draw opacity=1,
    text opacity=1,
    draw = none
},
tick align=outside,
tick pos=left,
x grid style={lightgray204},
xlabel=\textcolor{dimgray85}{$x$},
xmajorgrids,
xmin=-1, xmax=1,
xtick style={color=dimgray85},
y grid style={lightgray204},
ylabel=\textcolor{dimgray85}{Error},
ymajorgrids,
ymin=-0.01, ymax=0.07,
ytick={0,0.02,0.04,0.06},
yticklabels={0,0.02,0.04,0.06},
ytick style={color=dimgray85}
]
\addplot [thick, olive]
table {%
-1 0.0013704847282211
-0.97979797979798 0.000592998417467183
-0.95959595959596 0.000351485997306167
-0.939393939393939 0.000204406530753931
-0.919191919191919 0.000107179782129152
-0.898989898989899 4.18401957300585e-05
-0.878787878787879 1.16519373949409e-06
-0.858585858585859 2.77042418034279e-05
-0.838383838383838 4.17197798597835e-05
-0.818181818181818 4.60285693747453e-05
-0.797979797979798 4.27340378787999e-05
-0.777777777777778 3.34600685998332e-05
-0.757575757575758 1.94929364989993e-05
-0.737373737373737 1.87228411774321e-06
-0.717171717171717 1.85479612000439e-05
-0.696969696969697 4.1057302014702e-05
-0.676767676767677 6.50584190269865e-05
-0.656565656565657 9.00449648915669e-05
-0.636363636363636 0.000115584898800147
-0.616161616161616 0.000141307735677637
-0.595959595959596 0.000166894623696623
-0.575757575757576 0.000192070505791508
-0.555555555555556 0.000216597843978678
-0.535353535353535 0.000240271534053454
-0.515151515151515 0.000262914739834724
-0.494949494949495 0.000284375446927623
-0.474747474747475 0.000304523585980737
-0.454545454545454 0.000323248611457008
-0.434343434343434 0.000340457448409923
-0.414141414141414 0.000356072739096172
-0.393939393939394 0.000370031335963257
-0.373737373737374 0.000382282998651368
-0.353535353535353 0.000392789260981963
-0.333333333333333 0.000401522440778082
-0.313131313131313 0.000408464770070456
-0.292929292929293 0.000413607627679141
-0.272727272727273 0.000416950859091384
-0.252525252525252 0.000418502171248458
-0.232323232323232 0.000418276591850265
-0.212121212121212 0.000416295984488019
-0.191919191919192 0.00041258861223966
-0.171717171717172 0.000407188743571521
-0.151515151515151 0.000400136295109445
-0.131313131313131 0.000391476506814771
-0.111111111111111 0.000381259645632937
-0.0909090909090908 0.000369540734147695
-0.0707070707070706 0.000356379301316348
-0.0505050505050504 0.000341839152687496
-0.0303030303030303 0.00032598815777293
-0.0101010101010101 0.000308898052539964
0.0101010101010102 0.000290644255215718
0.0303030303030305 0.000271305693704414
0.0505050505050506 0.000250964643165239
0.0707070707070707 0.000229706572263838
0.0909090909090911 0.00020761999691745
0.111111111111111 0.000184796340149242
0.131313131313131 0.000161329796953933
0.151515151515152 0.000137317202956132
0.171717171717172 0.000112857905668427
0.191919191919192 8.80536371845363e-05
0.212121212121212 6.3008386961768e-05
0.232323232323232 3.78282734293713e-05
0.252525252525253 1.26214129416886e-05
0.272727272727273 1.25022155798085e-05
0.292929292929293 3.74309120894506e-05
0.313131313131313 6.2051404978801e-05
0.333333333333333 8.62490112543968e-05
0.353535353535354 0.000109907810018781
0.373737373737374 0.000132910832654057
0.393939393939394 0.000155140273654132
0.414141414141414 0.000176477726900728
0.434343434343434 0.000196804453259217
0.454545454545455 0.000216001686741338
0.474747474747475 0.000233950988128839
0.494949494949495 0.00025053465736237
0.515151515151515 0.000265636218848037
0.535353535353535 0.000279140997861416
0.555555555555556 0.000290936811357362
0.575757575757576 0.000300914803733177
0.595959595959596 0.000308970467544134
0.616161616161616 0.000315004902830068
0.636363636363636 0.00031892638724168
0.656565656565657 0.000320652356222639
0.676767676767677 0.000320111931360278
0.696969696969697 0.000317249192808733
0.717171717171717 0.000312027479027477
0.737373737373737 0.000304435132707281
0.757575757575758 0.000294493327667456
0.777777777777778 0.000282266966907641
0.797979797979798 0.00026788024799318
0.818181818181818 0.000251539569333215
0.838383838383838 0.000233568464257594
0.858585858585859 0.00021446324054486
0.878787878787879 0.00019498650935823
0.898989898989899 0.000176335645026171
0.919191919191919 0.000160475403375715
0.939393939393939 0.000150885242512278
0.95959595959596 0.000154606797983181
0.97979797979798 0.000190293169249678
1 0.000457793252315875
};
\addlegendentry{Error Function Qubit 0}
\addplot [thick, red]
table {%
-1 0.000964219093251062
-0.97979797979798 0.00192787299987374
-0.95959595959596 0.00245533893741012
-0.939393939393939 0.00261347026785264
-0.919191919191919 0.00259597966507297
-0.898989898989899 0.00247857644678429
-0.878787878787879 0.00230022180847655
-0.858585858585859 0.00208400985207935
-0.838383838383838 0.00184486491609503
-0.818181818181818 0.0015930018693725
-0.797979797979798 0.00133569657899191
-0.777777777777778 0.00107827920879866
-0.757575757575758 0.000824731514136201
-0.737373737373737 0.000578066035424651
-0.717171717171717 0.000340577582224166
-0.696969696969697 0.00011401613223444
-0.676767676767677 0.000100290654077961
-0.656565656565657 0.000301348359485254
-0.636363636363636 0.000488429596656681
-0.616161616161616 0.000661023827204121
-0.595959595959596 0.000818798678061783
-0.575757575757576 0.000961569666645012
-0.555555555555556 0.00108927618411042
-0.535353535353535 0.00120196220928404
-0.515151515151515 0.00129976064895354
-0.494949494949495 0.00138288049341481
-0.474747474747475 0.0014515961826751
-0.454545454545454 0.00150623872674488
-0.434343434343434 0.00154718823097008
-0.414141414141414 0.00157486755664091
-0.393939393939394 0.00158973690610464
-0.373737373737374 0.00159228916627124
-0.353535353535353 0.00158304587829826
-0.333333333333333 0.00156255372735283
-0.313131313131313 0.0015313814667105
-0.292929292929293 0.00149011720626688
-0.272727272727273 0.00143936600807537
-0.252525252525252 0.00137974774149188
-0.232323232323232 0.0013118951584624
-0.212121212121212 0.00123645215583074
-0.191919191919192 0.00115407219679226
-0.171717171717172 0.00106541686773779
-0.151515151515151 0.000971154550184251
-0.131313131313131 0.000871959190267682
-0.111111111111111 0.000768509150539616
-0.0909090909090908 0.000661486130667108
-0.0707070707070706 0.000551574145186267
-0.0505050505050504 0.000439458547590684
-0.0303030303030303 0.000325825091142612
-0.0101010101010101 0.000211359017436183
0.0101010101010102 9.67441643859028e-05
0.0303030303030305 1.73379142498592e-05
0.0505050505050506 0.000130208825514618
0.0707070707070707 0.000241194217587555
0.0909090909090911 0.000349624671336186
0.111111111111111 0.000454836610883224
0.131313131313131 0.000556173241135976
0.151515151515152 0.000652985520797011
0.171717171717172 0.000744633179976591
0.191919191919192 0.000830485792311014
0.212121212121212 0.000909923912551555
0.232323232323232 0.000982340291883071
0.252525252525253 0.00104714118481974
0.272727272727273 0.00110374776339274
0.292929292929293 0.00115159765685395
0.313131313131313 0.00119014663791422
0.333333333333333 0.00121887048014216
0.353535353535354 0.00123726701554171
0.373737373737374 0.00124485842668196
0.393939393939394 0.00124119381448029
0.414141414141414 0.00122585209107307
0.434343434343434 0.00119844525763568
0.454545454545455 0.00115862214027465
0.474747474747475 0.00110607267385851
0.494949494949495 0.00104053284522063
0.515151515151515 0.000961790435060783
0.535353535353535 0.00086969173428858
0.555555555555556 0.000764149458613639
0.575757575757576 0.000645152149462869
0.595959595959596 0.000512775436007418
0.616161616161616 0.000367195652019203
0.636363636363636 0.000208706466456376
0.656565656565657 3.77394200846703e-05
0.676767676767677 0.000145110404126902
0.696969696969697 0.000339051857234818
0.717171717171717 0.000543055890811606
0.737373737373737 0.000755800825942399
0.757575757575758 0.000975599351559442
0.777777777777778 0.00120029897102303
0.797979797979798 0.00142714270597333
0.818181818181818 0.00165256822068283
0.838383838383838 0.00187190753398736
0.858585858585859 0.00207891808329719
0.878787878787879 0.00226500962507048
0.898989898989899 0.00241787823820983
0.919191919191919 0.0025188600062106
0.939393939393939 0.00253709694936233
0.95959595959596 0.00241385652130255
0.97979797979798 0.002002119419794
1 0.000163617563517349
};
\addlegendentry{Error Function Qubit 1}
\end{axis}

\end{tikzpicture}}}
    \hfill
  \subfloat[]{%
        \scalebox{0.53}{
\begin{tikzpicture}

\definecolor{dimgray85}{RGB}{85,85,85}
\definecolor{gainsboro229}{RGB}{229,229,229}
\definecolor{lightgray204}{RGB}{204,204,204}

\begin{axis}[
scaled ticks=false,
ytick={0, 0.05, 0.1},
yticklabels={0,0.05,0.1},
axis line style={lightgray204},
legend cell align={left},
legend style={
    fill opacity=0.8,
    draw opacity=1,
    text opacity=1,
    draw = none
},
tick align=outside,
tick pos=left,
x grid style={lightgray204},
xlabel=\textcolor{dimgray85}{$x$},
xmajorgrids,
xmin=-1, xmax=1,
xtick style={color=dimgray85},
y grid style={lightgray204},
ylabel=\textcolor{dimgray85}{Error},
ymajorgrids,
ymin=-0.01, ymax=0.07,
ytick={0,0.02,0.04,0.06},
yticklabels={0,0.02,0.04,0.06},
ytick style={color=dimgray85}
]
\addplot [thick, olive]
table {%
-1 0.0149565499703486
-0.97979797979798 0.0180691411428026
-0.95959595959596 0.0221135146959649
-0.939393939393939 0.0224262660266322
-0.919191919191919 0.0212295292283482
-0.898989898989899 0.0193283329334623
-0.878787878787879 0.0171045361178337
-0.858585858585859 0.0147648663928277
-0.838383838383838 0.0124297967511998
-0.818181818181818 0.0101723732240794
-0.797979797979798 0.00803750817451532
-0.777777777777778 0.00605248288942545
-0.757575757575758 0.00423306885571784
-0.737373737373737 0.00258728987431711
-0.717171717171717 0.00111783434335649
-0.696969696969697 0.000176343578052407
-0.676767676767677 0.00129893030464462
-0.656565656565657 0.00225548762480277
-0.636363636363636 0.0030529062928078
-0.616161616161616 0.00369900900329795
-0.595959595959596 0.00420225756223266
-0.575757575757576 0.00457153410509825
-0.555555555555556 0.00481597580464765
-0.535353535353535 0.00494484877848991
-0.515151515151515 0.00496745109156427
-0.494949494949495 0.00489303759703369
-0.474747474747475 0.00473076133168539
-0.454545454545454 0.00448962756954366
-0.434343434343434 0.00417845762758569
-0.414141414141414 0.00380586023317286
-0.393939393939394 0.00338020878632651
-0.373737373737374 0.00290962323716101
-0.353535353535353 0.00240195558783484
-0.333333333333333 0.00186477824639685
-0.313131313131313 0.00130537462547978
-0.292929292929293 0.000730731505641662
-0.272727272727273 0.000147532781084869
-0.252525252525252 0.000437845718671628
-0.232323232323232 0.00101934057818662
-0.212121212121212 0.00159120451712375
-0.191919191919192 0.00214800964679239
-0.171717171717172 0.00268465039706835
-0.151515151515151 0.00319634622561423
-0.131313131313131 0.00367864420303783
-0.111111111111111 0.00412742155315885
-0.0909090909090908 0.00453888821603019
-0.0707070707070706 0.00490958949237173
-0.0505050505050504 0.00523640882089884
-0.0303030303030303 0.00551657073458175
-0.0101010101010101 0.00574764403773033
0.0101010101010102 0.00592754524275173
0.0303030303030305 0.00605454230333222
0.0505050505050506 0.00612725867957668
0.0707070707070707 0.00614467777004402
0.0909090909090911 0.00610614774578096
0.111111111111111 0.00601138682207588
0.131313131313131 0.00586048900502048
0.151515151515152 0.00565393035159726
0.171717171717172 0.00539257578451544
0.191919191919192 0.00507768650565468
0.212121212121212 0.00471092805548599
0.232323232323232 0.00429437906967584
0.252525252525253 0.00383054078860735
0.272727272727273 0.0033223473805704
0.292929292929293 0.00277317714510928
0.313131313131313 0.00218686466934639
0.333333333333333 0.00156771401700811
0.353535353535354 0.000920513037518229
0.373737373737374 0.000250548890668639
0.393939393939394 0.000436375109053741
0.414141414141414 0.00113392121545937
0.434343434343434 0.00183519741937627
0.454545454545455 0.00253273593330608
0.474747474747475 0.0032184698560636
0.494949494949495 0.00388370787814608
0.515151515151515 0.00451910689105753
0.535353535353535 0.00511464237531964
0.555555555555556 0.00565957646849496
0.575757575757576 0.0061424236651936
0.595959595959596 0.00655091418987774
0.616161616161616 0.00687195523363782
0.636363636363636 0.00709159049431968
0.656565656565657 0.007194958866649
0.676767676767677 0.00716625379555963
0.696969696969697 0.00698868590343016
0.717171717171717 0.0066444533274459
0.737373737373737 0.00611472728603651
0.757575757575758 0.00537966570425163
0.777777777777778 0.00441847710821724
0.797979797979798 0.00320957408223099
0.818181818181818 0.00173088787617626
0.838383838383838 3.95203627577834e-05
0.858585858585859 0.00212228262324343
0.878787878787879 0.00453441255286535
0.898989898989899 0.00728455653140414
0.919191919191919 0.0103624835030085
0.939393939393939 0.013712955828867
0.95959595959596 0.0171572335918408
0.97979797979798 0.0200560768732087
1 0.0132109935396528
};
\addlegendentry{Error Function Qubit 0}
\addplot [thick, red]
table {%
-1 0.118133086348966
-0.97979797979798 0.00827854864642563
-0.95959595959596 0.0364552320775712
-0.939393939393939 0.0497090969505802
-0.919191919191919 0.0558854548013584
-0.898989898989899 0.0579893093230216
-0.878787878787879 0.0575385284442215
-0.858585858585859 0.0554172442493584
-0.838383838383838 0.0521875780598332
-0.818181818181818 0.0482285191015432
-0.797979797979798 0.0438063457543851
-0.777777777777778 0.0391137784740174
-0.757575757575758 0.034293309144565
-0.737373737373737 0.0294518753403738
-0.717171717171717 0.0246705015510011
-0.696969696969697 0.0200108648792218
-0.676767676767677 0.0155199026907329
-0.656565656565657 0.0112331298642818
-0.636363636363636 0.00717708018946372
-0.616161616161616 0.0033711379421999
-0.595959595959596 0.000171064684641342
-0.575757575757576 0.00344056533895748
-0.555555555555556 0.00643222210258798
-0.535353535353535 0.00914403877233827
-0.515151515151515 0.0115766230403627
-0.494949494949495 0.0137327470598619
-0.474747474747475 0.0156169877908301
-0.454545454545454 0.0172354301577654
-0.434343434343434 0.0185954201286205
-0.414141414141414 0.0197053578128712
-0.393939393939394 0.0205745228943639
-0.373737373737374 0.0212129263789553
-0.353535353535353 0.0216311838991794
-0.333333333333333 0.0218404067846691
-0.313131313131313 0.0218521078537954
-0.292929292929293 0.0216781194638093
-0.272727272727273 0.0213305218134862
-0.252525252525252 0.0208215798535111
-0.232323232323232 0.0201636874474265
-0.212121212121212 0.0193693176562255
-0.191919191919192 0.0184509782052518
-0.171717171717172 0.0174211713422625
-0.151515151515151 0.0162923574176789
-0.131313131313131 0.0150769216178336
-0.111111111111111 0.0137871433637533
-0.0909090909090908 0.0124351679551641
-0.0707070707070706 0.0110329800946094
-0.0505050505050504 0.00959237897204313
-0.0303030303030303 0.00812495462755292
-0.0101010101010101 0.00664206534049172
0.0101010101010102 0.00515481581808097
0.0303030303030305 0.00367403597662197
0.0505050505050506 0.00221026012413761
0.0707070707070707 0.000773706365432503
0.0909090909090911 0.000625743940753987
0.111111111111111 0.00197856683702757
0.131313131313131 0.00327561669669585
0.151515151515152 0.00450814758866236
0.171717171717172 0.00566783524181391
0.191919191919192 0.00674679977814496
0.212121212121212 0.0077376293919702
0.232323232323232 0.00863340516380418
0.252525252525253 0.00942772721211116
0.272727272727273 0.0101147424045899
0.292929292929293 0.0106891738734547
0.313131313131313 0.0111463526071053
0.333333333333333 0.0114822514245095
0.353535353535354 0.0116935216796525
0.373737373737374 0.0117775330931632
0.393939393939394 0.0117324171683707
0.414141414141414 0.0115571147221363
0.434343434343434 0.0112514281498027
0.454545454545455 0.0108160791524942
0.474747474747475 0.0102527727888899
0.494949494949495 0.0095642688792914
0.515151515151515 0.00875446199600149
0.535353535353535 0.00782847153294752
0.555555555555556 0.00679274367454796
0.575757575757576 0.00565516750141792
0.595959595959596 0.00442520800806401
0.616161616161616 0.00311405950740395
0.636363636363636 0.00173482381779611
0.656565656565657 0.000302718855470507
0.676767676767677 0.00116467508738205
0.696969696969697 0.00264712182621885
0.717171717171717 0.00412137581050598
0.737373737373737 0.00556075148101909
0.757575757575758 0.00693459174245359
0.777777777777778 0.00820760220830252
0.797979797979798 0.00933900326468173
0.818181818181818 0.0102814292014716
0.838383838383838 0.0109794669619468
0.858585858585859 0.0113676659322521
0.878787878787879 0.0113677445690018
0.898989898989899 0.0108845306664873
0.919191919191919 0.00979982756431663
0.939393939393939 0.00796281188369519
0.95959595959596 0.00517521420588474
0.97979797979798 0.00118034376588372
1 0.0029031628210725
};
\addlegendentry{Error Function Qubit 1}
\end{axis}

\end{tikzpicture}}}
    \\
  \subfloat[]{%
        \scalebox{0.53}{\input{figures/multi_qubit_measurements/convergence_3_multuple_functions_qubits_d3_arcsin_tp30_second_qubit1_function_approx_x_xhoch2.tex}}}
    \hfill
  \subfloat[]{%
        \scalebox{0.53}{\input{figures/multi_qubit_measurements/convergence_3_multuple_functions_qubits_d3_arcsin_tp30_second_qubit1_function_approx_3para.tex}}}
    \hfill
  \subfloat[]{%
        \scalebox{0.53}{\input{figures/multi_qubit_measurements/convergence_3_multuple_functions_qubits_d3_arcsin_tp30_second_qubit1_function_approx_xhoch3_complex.tex}}}
  \caption{Learning two different functions per QCL circuit on a statevector simulator with $R_Y(\arcsin(x))$ data encoding, a qubit number of $N=3$ and a depth of $D = 3$. The learned functions are $f_1(x) = x$ and $g_1(x) = x^2$ (a), $f_2(x) = x^2$ and $g_2(x) = x^3$ (b) and $f_3(x) = x^3-x^2+1$ and $g_3(x) = x^3$ (c). The cost function is evaluated on 30 equidistant training points and is classically minimized using SLSQP. Additionally, in (d)-(f) the absolute values of the respective errors are shown. In (g)-(i) the respective values of the cost function versus the number of cost function evaluations are plotted.}
  \label{multiple_functions} 
\end{figure*}

\begin{figure*}[!tp]
    \begin{minipage}[c]{0.6\textwidth}
        \scalebox{1}{\input{figures/multi_qubit_measurements/aver/q6_d4_tp10_func100_eval1000_more_para_4_func_2}}
    \end{minipage}
    \begin{minipage}[c]{0.4\textwidth}
      \caption{Average of the convergence curves of the training of 100 QCL circuits with $N=6$ qubits and a depth of $D=4$: For the red line 
      100 random polynomials are learned, i.e. one polynomial per circuit. For the blue line four polynomials are learned per circuit, which sums up to 400 random polynomials learned. We use 10 equidistant training points and the classical optimizer COBYLA. For a fair comparison the value of the cost function and the number of function evaluations is divided by the number of polynomials learned in parallel. The shaded areas indicate the respective standard deviations} \label{convergence}
    \end{minipage}
\end{figure*}

\subsection{Coupled Harmonic Oscillator}
\label{chap:coupled_HO}
In this section, we want to solve a coupled differential equation using a single QCL circuit in combination with the parameter shift rule on a simulator.
Coupled differential equations were solved with QCL circuits before, using a designated circuit for each equation \cite{Kyriienko.2021}.
We combine this problem with the finding that multiple functions can be learned with a single circuit.\\
For this purpose, a coupled harmonic oscillator with two masses $m$, two springs of spring strength $k$ and one spring of spring strength $s$ is considered.
The arrangement can be seen in Figure \ref{fig:federn}.
\begin{figure}[H]
    \centering
    \scalebox{0.5}{
\begin{tikzpicture}[x=0.75pt,y=0.75pt,yscale=-1,xscale=1]

\draw [line width=1.5]    (88,45.8) -- (88,130) ;
\draw [line width=1.5]    (569,45.8) -- (569,130) ;
\draw  [line width=1.5]  (103.95,89.89) .. controls (104.76,84.23) and (107.75,78.57) .. (113.74,78.58) .. controls (125.7,78.58) and (125.69,101.23) .. (120.17,101.22) .. controls (114.65,101.22) and (114.66,78.58) .. (126.62,78.58) .. controls (138.59,78.59) and (138.58,101.23) .. (133.06,101.23) .. controls (127.53,101.23) and (127.55,78.58) .. (139.51,78.59) .. controls (151.48,78.6) and (151.47,101.24) .. (145.95,101.24) .. controls (140.42,101.23) and (140.43,78.59) .. (152.4,78.6) .. controls (164.37,78.6) and (164.36,101.25) .. (158.83,101.24) .. controls (153.31,101.24) and (153.32,78.6) .. (165.29,78.6) .. controls (177.26,78.61) and (177.24,101.25) .. (171.72,101.25) .. controls (166.2,101.25) and (166.21,78.6) .. (178.18,78.61) .. controls (190.15,78.62) and (190.13,101.26) .. (184.61,101.26) .. controls (179.09,101.25) and (179.1,78.61) .. (191.07,78.62) .. controls (196.58,78.62) and (199.55,83.43) .. (200.61,88.62) ;
\draw  [fill={rgb, 255:red, 0; green, 0; blue, 0 }  ,fill opacity=1 ] (215.8,75.6) -- (260,75.6) -- (260,103) -- (215.8,103) -- cycle ;
\draw  [fill={rgb, 255:red, 0; green, 0; blue, 0 }  ,fill opacity=1 ] (396.3,76.1) -- (440.5,76.1) -- (440.5,103.5) -- (396.3,103.5) -- cycle ;
\draw [line width=1.5]    (88.3,89) -- (104.41,89) ;
\draw [line width=1.5]    (199.69,88.5) -- (215.8,88.5) ;
\draw  [line width=1.5]  (455.63,89.39) .. controls (456.45,83.73) and (459.48,78.07) .. (465.54,78.08) .. controls (477.64,78.08) and (477.63,100.73) .. (472.04,100.72) .. controls (466.46,100.72) and (466.47,78.08) .. (478.58,78.08) .. controls (490.68,78.09) and (490.67,100.73) .. (485.08,100.73) .. controls (479.49,100.73) and (479.51,78.08) .. (491.62,78.09) .. controls (503.72,78.1) and (503.71,100.74) .. (498.12,100.74) .. controls (492.53,100.73) and (492.55,78.09) .. (504.66,78.1) .. controls (516.76,78.1) and (516.75,100.75) .. (511.16,100.74) .. controls (505.57,100.74) and (505.59,78.1) .. (517.69,78.1) .. controls (529.8,78.11) and (529.79,100.75) .. (524.2,100.75) .. controls (518.61,100.75) and (518.63,78.1) .. (530.73,78.11) .. controls (542.84,78.12) and (542.83,100.76) .. (537.24,100.76) .. controls (531.65,100.75) and (531.67,78.11) .. (543.77,78.12) .. controls (549.36,78.12) and (552.36,82.93) .. (553.43,88.12) ;
\draw [line width=1.5]    (439.8,88.5) -- (456.1,88.5) ;
\draw [line width=1.5]    (552.5,88) -- (568.8,88) ;
\draw  [line width=1.5]  (277.06,89.38) .. controls (277.92,83.73) and (281.13,78.07) .. (287.53,78.08) .. controls (300.34,78.09) and (300.34,100.71) .. (294.42,100.7) .. controls (288.51,100.7) and (288.52,78.08) .. (301.33,78.09) .. controls (314.14,78.1) and (314.13,100.72) .. (308.22,100.71) .. controls (302.31,100.71) and (302.31,78.09) .. (315.13,78.1) .. controls (327.94,78.11) and (327.93,100.73) .. (322.02,100.72) .. controls (316.1,100.72) and (316.11,78.1) .. (328.92,78.11) .. controls (341.74,78.12) and (341.73,100.74) .. (335.81,100.73) .. controls (329.9,100.73) and (329.91,78.11) .. (342.72,78.12) .. controls (355.53,78.12) and (355.52,100.75) .. (349.61,100.74) .. controls (343.7,100.74) and (343.71,78.12) .. (356.52,78.13) .. controls (369.33,78.13) and (369.32,100.76) .. (363.41,100.75) .. controls (357.5,100.75) and (357.5,78.13) .. (370.32,78.14) .. controls (376.22,78.14) and (379.4,82.95) .. (380.54,88.13) ;
\draw [line width=1.5]    (260.3,88.5) -- (277.55,88.5) ;
\draw [line width=1.5]    (379.55,88) -- (396.8,88) ;
\draw (147,49) node [anchor=north west][inner sep=0.75pt]   [align=left] {\huge$k$};
\draw (225.5,52.5) node [anchor=north west][inner sep=0.75pt]   [align=left] {\huge$m$};
\draw (406,52.5) node [anchor=north west][inner sep=0.75pt]   [align=left] {\huge$m$};
\draw (324.5,51.5) node [anchor=north west][inner sep=0.75pt]   [align=left] {\huge$s$};
\draw (500,49) node [anchor=north west][inner sep=0.75pt]   [align=left] {\huge$k$};

\end{tikzpicture}}
    \caption{Example of the coupled harmonic oscillator with two identical masses $m$, two identical springs of spring strength $k$ and one spring of spring strength $s$.}
    \label{fig:federn}
\end{figure}

\newcommand{\mass}{m}  
\newcommand{\springConstLR}{k}  
\newcommand{\springConstM}{s}  
\newcommand{\freeVar}{x}  
\newcommand{\displaceScal}{f}  
\newcommand{\displaceVec}{\mathbf{\displaceScal}}  
\newcommand{\stiffMat}{\mathbf{S}}  
\newcommand{\freqZero}{\omega_0}
\newcommand{\freqOne}{\omega_1}
\newcommand{\displaceScalQC}[1]{\displaceScal_{\mathrm{QC} #1}}  
\newcommand{\displaceVecQC}{\mathbf{\displaceScal}_\mathrm{QC}}  

This system is described by the coupled differential equation
\begin{equation}
    \displaceVec^{\prime \prime}(\freeVar)
    = \stiffMat \displaceVec(\freeVar) \, ,
    \,\,
    \displaceVec^{\prime}(0) = \displaceVec^{\prime}_0 \, , \
    \displaceVec(0) = \displaceVec_0 \, ,
    \label{coupled_dgl}
\end{equation}
where the variable $x$ describes the time, ${\displaceVec(x) = (\displaceScal_0(x), \displaceScal_1(x))}$ collects
the displacements of the two masses, $\stiffMat$ is the stiffness matrix 
scaled by $1/\mass$,
\begin{equation}
    \stiffMat = \frac{1}{\mass}
    \pmat{
        - \springConstLR - \springConstM & \springConstM
        \\
        \springConstM & - \springConstLR - \springConstM 
    } \, ,
\end{equation}
and $\displaceVec^{\prime}_0$ and $\displaceVec_0$ are the initial 
velocity and initial displacement of the masses, respectively. For this paper we choose
the initial conditions as ${\displaceVec^{\prime}_0 = (0, 0)}$ and
${\displaceVec_0 = (1, 0})$. Then, the solution to \eqref{coupled_dgl}
is given by
\begin{equation}
    \displaceVec(\freeVar)
    = \frac{1}{2} \pmat{
        \cos (\freqZero x) - \cos (\freqOne x)
        \\
        \cos (\freqZero x) + \cos (\freqOne x)
    } \, ,
\end{equation}
with frequencies {${\freqZero^2 = k/m}$} and\break {${\freqOne^2=k/m+2s/m}$}.
The differential equation is now solved with the four qubit QCL circuit shown in Figure \ref{circ_coupled_dgl}, where the first and second qubit are measured.
\begin{figure}[H]
    \centering
    \begin{tikzpicture}
    \node[scale=0.5] {
    \begin{quantikz}
    \lstick{$\ket{0}$} & \gate{R_Y(x)}  & \ctrl{1} \gategroup[4,steps=7,style={dashed,rounded corners, inner xsep=2pt},background,label style={label position=below,anchor=north,yshift=-0.2cm}]{{$4\times$}}&  \qw &  \qw     & \targ{} & \gate{R_X(\theta_0 )}& \gate{R_Y(\theta_1 )}  & \gate{R_Z(\theta_2)}  &   \meter[]{\braket{Z}} \\[0.2cm]
    \lstick{$\ket{0}$} & \gate{R_Y(x)}  & \targ{} &  \ctrl{1} & \qw &  \qw     & \gate{R_X(\theta_3 )}& \gate{R_Y(\theta_4)}  & \gate{R_Z(\theta_5)}  &   \meter[]{\braket{Z}} \\
    \lstick{$\ket{0}$} & \gate{R_Y(x)}  & \qw  &  \targ{} & \ctrl{1} &  \qw    & \gate{R_X(\theta_6 )}& \gate{R_Y(\theta_7)}  & \gate{R_Z(\theta_8)}  &     \qw\\
    \lstick{$\ket{0}$} & \gate{R_Y(x)}  &  \qw   &  \qw   & \targ{} & \ctrl{-3} & \gate{R_X(\theta_9 )}& \gate{R_Y(\theta_{10})}  & \gate{R_Z(\theta_{11})}  &     \qw
    \end{quantikz}
    };
    \end{tikzpicture}
    \caption{Four qubit QCL circuit with $R_Y(x)$ data encoding. The first and the second qubit are measured.}
    \label{circ_coupled_dgl}
\end{figure}

\begin{figure*}[!tp]
    \centering
  \subfloat[]{%
       \scalebox{0.53}{\input{figures/coupled_HO/4_qubits_coupled_HO_d4_nu20_3para_seed2_omega0_square_1_omega1_square_16_min1to1}}}
    \hfill
  \subfloat[]{%
        \scalebox{0.53}{
\begin{tikzpicture}

\definecolor{dimgray85}{RGB}{85,85,85}
\definecolor{gainsboro229}{RGB}{229,229,229}
\definecolor{lightgray204}{RGB}{204,204,204}

\begin{axis}[
axis line style={lightgray204},
legend cell align={left},
legend style={
  fill opacity=0.8,
  draw opacity=1,
  text opacity=1,
  at={(0.03,0.97)},
  anchor=north west,
  draw=none
},
tick align=outside,
tick pos=left,
x grid style={lightgray204},
xlabel=\textcolor{dimgray85}{$x$},
xmajorgrids,
xmin=-1, xmax=1,
xtick style={color=dimgray85},
y grid style={lightgray204},
ylabel=\textcolor{dimgray85}{Error},
ymajorgrids,
ymin=-0.000370244818000021, ymax=0.00805078024392131,
ytick style={color=dimgray85}
]
\addplot [ultra thick, red]
table {%
-1 0.000656016677620296
-0.97979797979798 0.000852243423156235
-0.95959595959596 0.00107768308017986
-0.939393939393939 0.00131553371108591
-0.919191919191919 0.00155157140827938
-0.898989898989899 0.00177400823329593
-0.878787878787879 0.00197334136034871
-0.858585858585859 0.00214219495118054
-0.838383838383838 0.00227515628912897
-0.818181818181818 0.00236860769016009
-0.797979797979798 0.00242055568804259
-0.777777777777778 0.00243045895969082
-0.757575757575758 0.00239905641563523
-0.737373737373737 0.00232819682993424
-0.717171717171717 0.00222067132432144
-0.696969696969697 0.00208004995339572
-0.676767676767677 0.00191052356192412
-0.656565656565657 0.00171675200263399
-0.636363636363636 0.00150371971358832
-0.616161616161616 0.00127659955980347
-0.595959595959596 0.00104062574410458
-0.575757575757576 0.000800976489022789
-0.555555555555556 0.00056266708489372
-0.535353535353535 0.000330453790756829
-0.515151515151515 0.000108748964466332
-0.494949494949495 9.84523121971281e-05
-0.474747474747475 0.000287633957813449
-0.454545454545454 0.000455808918613021
-0.434343434343434 0.000600557559273796
-0.414141414141414 0.0007200545821478
-0.393939393939394 0.000813084740360959
-0.373737373737374 0.000879047706596969
-0.353535353535353 0.00091795255057836
-0.333333333333333 0.000930402363909955
-0.313131313131313 0.000917569650074612
-0.292929292929293 0.00088116316991782
-0.272727272727273 0.000823386997817255
-0.252525252525252 0.0007468926010612
-0.232323232323232 0.000654724803984097
-0.212121212121212 0.000550262538948521
-0.191919191919192 0.000437155318040006
-0.171717171717172 0.000319256382351862
-0.151515151515151 0.000200553499595801
-0.131313131313131 8.5098385589566e-05
-0.111111111111111 2.30642788974222e-05
-0.0909090909090908 0.000119967277562028
-0.0707070707070706 0.000201789797608098
-0.0505050505050504 0.000264923972091324
-0.0303030303030303 0.000306038157594224
-0.0101010101010101 0.000322136125657568
0.0101010101010102 0.000310611399850669
0.0303030303030305 0.000269296031067956
0.0505050505050506 0.000196503170351359
0.0707070707070707 9.10628713638451e-05
0.0909090909090911 4.76493677109557e-05
0.111111111111111 0.000219691730777294
0.131313131313131 0.000424543200762351
0.151515151515152 0.000661098899779944
0.171717171717172 0.000927673794087425
0.191919191919192 0.00122201481244655
0.212121212121212 0.00154132133322726
0.232323232323232 0.00188227390192874
0.252525252525253 0.00224107094838355
0.272727272727273 0.00261347318201532
0.292929292929293 0.00299485525502241
0.313131313131313 0.00338026419826276
0.333333333333333 0.00376448405317431
0.353535353535354 0.00414210604662368
0.373737373737374 0.00450760358412816
0.393939393939394 0.00485541127164257
0.414141414141414 0.00518000711733507
0.434343434343434 0.00547599701299106
0.454545454545455 0.00573820055074425
0.474747474747475 0.00596173719463466
0.494949494949495 0.00614211179880567
0.515151515151515 0.00627529844535468
0.535353535353535 0.00635782156460135
0.555555555555556 0.00638683329978881
0.575757575757576 0.0063601860865434
0.595959595959596 0.00627649943507536
0.616161616161616 0.00613521992991735
0.636363636363636 0.00593667349806228
0.656565656565657 0.00568210904134593
0.676767676767677 0.00537373258255498
0.696969696969697 0.00501473113698084
0.717171717171717 0.00460928559115543
0.737373737373737 0.00416257194831041
0.757575757575758 0.00368075038494986
0.777777777777778 0.00317094165410223
0.797979797979798 0.002641190468261
0.818181818181818 0.00210041559729929
0.838383838383838 0.00155834652371042
0.858585858585859 0.00102544660822904
0.878787878787879 0.000512822832570703
0.898989898989899 3.21223018964756e-05
0.919191919191919 0.000404584193314628
0.939393939393939 0.000784931138700484
0.95959595959596 0.00109639924640349
0.97979797979798 0.00132646774432542
1 0.00146277724058867
};
\addlegendentry{Error Qubit 0}
\addplot [ultra thick, olive]
table {%
-1 0.000533321437511058
-0.97979797979798 0.00054603672868514
-0.95959595959596 0.000585638604838734
-0.939393939393939 0.000642421877809829
-0.919191919191919 0.000708158277468063
-0.898989898989899 0.000775994010614567
-0.878787878787879 0.000840345571083256
-0.858585858585859 0.000896794656718658
-0.838383838383838 0.000941983017119008
-0.818181818181818 0.000973508020116087
-0.797979797979798 0.000989819684267257
-0.777777777777778 0.000990119879413665
-0.757575757575758 0.00097426434878356
-0.737373737373737 0.000942668154122583
-0.717171717171717 0.000896215090250174
-0.696969696969697 0.0008361715578864
-0.676767676767677 0.000764105324938091
-0.656565656565657 0.000681809545241996
-0.636363636363636 0.000591232342667336
-0.616161616161616 0.000494412206091033
-0.595959595959596 0.000393419379571047
-0.575757575757576 0.000290303370384937
-0.555555555555556 0.000187046636874522
-0.535353535353535 8.5524461000519e-05
-0.515151515151515 1.25290484509488e-05
-0.494949494949495 0.000105548928278854
-0.474747474747475 0.000192161468810736
-0.454545454545454 0.000271200663914062
-0.434343434343434 0.000341719256321316
-0.414141414141414 0.000402994671608981
-0.393939393939394 0.000454530171919265
-0.373737373737374 0.000496051593695479
-0.353535353535353 0.000527500061642705
-0.333333333333333 0.00054902109576227
-0.313131313131313 0.00056095054644284
-0.292929292929293 0.000563797808193822
-0.272727272727273 0.000558226770569714
-0.252525252525252 0.000545034970970387
-0.232323232323232 0.000525131411899582
-0.212121212121212 0.00049951350221733
-0.191919191919192 0.000469243570293293
-0.171717171717172 0.000435425384467905
-0.151515151515151 0.000399181096137721
-0.131313131313131 0.000361628999677746
-0.111111111111111 0.000323862475932676
-0.0909090909090908 0.000286930456514751
-0.0707070707070706 0.000251819712777341
-0.0505050505050504 0.000219439237490533
-0.0303030303030303 0.00019060694842508
-0.0101010101010101 0.000166038902140342
0.0101010101010102 0.000146341164188998
0.0303030303030305 0.000132004437759597
0.0505050505050506 0.000123401507871768
0.0707070707070707 0.000120787513685928
0.0909090909090911 0.000124303014916463
0.111111111111111 0.000133979774893697
0.131313131313131 0.000149749137228664
0.151515151515152 0.00017145283118461
0.171717171717172 0.000198855999138034
0.191919191919192 0.000231662199795729
0.212121212121212 0.000269530105410809
0.232323232323232 0.000312091575530721
0.252525252525253 0.000358970759916116
0.272727272727273 0.00040980385595224
0.292929292929293 0.000464259121334298
0.313131313131313 0.00052205672403266
0.333333333333333 0.000582987995314477
0.353535353535354 0.000646933641260006
0.373737373737374 0.000713880461021998
0.393939393939394 0.000783936118939454
0.414141414141414 0.000857341520388188
0.434343434343434 0.000934480349139011
0.454545454545455 0.00101588533671482
0.474747474747475 0.0011022408517396
0.494949494949495 0.00119438141980666
0.515151515151515 0.00129328581080634
0.535353535353535 0.00140006636246248
0.555555555555556 0.00151595324416809
0.575757575757576 0.00164227340492651
0.595959595959596 0.00178042399291523
0.616161616161616 0.00193184008076985
0.636363636363636 0.00209795658105871
0.656565656565657 0.0022801642894108
0.676767676767677 0.00247976004757278
0.696969696969697 0.00269789107698826
0.717171717171717 0.00293549359101497
0.737373737373737 0.00319322585558501
0.757575757575758 0.00347139592731482
0.777777777777778 0.00376988435958669
0.797979797979798 0.0040880622280578
0.818181818181818 0.00442470488599178
0.838383838383838 0.00477790191878114
0.858585858585859 0.00514496382346452
0.878787878787879 0.0055223259935836
0.898989898989899 0.00590545064136749
0.919191919191919 0.00628872733694219
0.939393939393939 0.00666537289033009
0.95959595959596 0.00702733134096889
0.97979797979798 0.00736517485713817
1 0.00766800637747034
};
\addlegendentry{Error Qubit 1}
\end{axis}

\end{tikzpicture}}}
    \hfill
  \subfloat[]{%
        \scalebox{0.53}{\input{figures/coupled_HO/convergence_4_qubits_coupled_HO_d4_nu20_3para_seed2_omega0_square_1_omega1_square_16_min1to1}}}
  \caption{(a) Solution of the differential equation \eqref{coupled_dgl} on a statevector simulator with $\displaceVec^{\prime}_0 = (0, 0)$,
    $\displaceVec_0 = (1, 0)$, $\omega_0^2 = 1$, $\omega_1^2 = 16$ and $\mu = 20$. The result is obtained with the QCL circuit in Figure \ref{circ_coupled_dgl} with $N = 4$, $D = 4$ and $R_Y(x)$ data encoding in combination with the parameter shift rule. The parameters are classically optimized using SLSQP. (b) The absolute values of the respective errors $|f_i(x) - f^{\text{post}}_{QC,i}(x)|$ are shown. (c) Values of the cost function versus the number of cost function evaluations.}
  \label{fig:coupled_HO_dgl} 
\end{figure*}

We chose the $R_Y(x)$ data encoding to obtain trigonometric functions, which are the appropriate choice to capture the periodic behavior expected from an undamped system of oscillators. 
To solve the differential equation with the QCL circuit in Figure \ref{circ_coupled_dgl}, the cost function
\begin{equation}
\begin{split}
    L(\boldsymbol{\theta}) =& \sum_i \Bigl(
        \| \displaceVecQC^{\prime \prime} (x_i, \boldsymbol{\theta}) - \stiffMat \displaceVecQC (x_i, \boldsymbol{\theta})\|^2\\
        &+ \mu \| \displaceVecQC^{\prime}(0, \boldsymbol{\theta}) - \displaceVec^{\prime}_0 \|^2\\
        &+ \mu  \| \displaceVecQC(0, \boldsymbol{\theta}) - \displaceVec_0 \|^2 \Bigr)
\end{split}
\end{equation}
is defined, where $\mu$ is a problem specific weight factor and
\begin{equation}
    \displaceVecQC (x, \boldsymbol{\theta}) = \Bigl(f^{\text{post}}_{QC,0}(x, \boldsymbol{\theta}), f^{\text{post}}_{QC,1}(x, \boldsymbol{\theta})\Bigr).
\end{equation}
The derivatives in the cost function are calculated with the PSR.
The result of the simulation for $\omega_0^2 = 1$, $\omega_1^2 = 16$ and 30 equidistant training points is shown in Figure \ref{fig:coupled_HO_dgl}.
It is possible to solve a coupled differential equation with just one circuit with low errors (see Figure \ref{fig:coupled_HO_dgl} (b)). 
This method could bring considerable advantages, in particular for very complicated systems with even more coupled equations, 
since only one circuit with the corresponding parameters has to be optimized, and not a set of parameters for each equation.
\section{Conclusion}
In this work, we have conducted an investigation of the QCL framework and its executability on NISQ devices.
In the beginning, simulations with different noise models and hardware experiments on ibmq\_ehningen were performed to investigate the scalability of QCL circuits on NISQ devices.
It was discovered that these circuits can be executed on NISQ hardware but only a fraction of the available qubits of ibmq\_ehningen can be effectively used before the errors become too high.
Following this, several functions were successfully learned with three-qubit QCL circuits on a simulator and subsequently also on ibmq\_ehningen.
Next, the ability to solve differential equations with QCL circuits and the parameter shift rule was investigated.
First, the parameter shift rule was tested on the IBM quantum computer to investigate the NISQ applicability.
It was possible to determine the derivative of functions. However, the resulting errors were very high.
Nevertheless, a simple differential equation could be solved to a certain extent on ibmq\_ehningen.\\
Furthermore, it was shown that multiple functions can be learned with a single QCL circuit when multiple qubits are measured which results in a faster learning in the early stages of the optimization process for some examples.
Following this idea, the differential equation of a coupled harmonic oscillator was solved with only a single circuit on a simulator.\\
Overall, it should be mentioned that the optimization process on the classical computer can be computationally intensive and takes up a large part of the computing time.\\
This problem applies not only to the QCL framework but to most variational quantum algorithms.
Therefore, a lot of research focuses on this problem, for example in the expansion and development of more suitable classical optimizers \cite{Tamiya.2022, Wiedmann.2023}.
Such methods give hope for a reduction of computation time and make the algorithm even more relevant.\\
\section*{Acknowledgments}
The authors would like to thank Vamshi Katukuri for insightful discussions and carefully proofreading this manuscript.\\
N. S. wants to thank Sungkun Hong for his advice and dedicated supervision of his Master's thesis which built the starting point of this article.\\
The authors acknowledge funding from the Ministry of Economic Affairs, Labour and Tourism Baden-Württemberg in the frame of the Competence Center Quantum Computing Baden-Württemberg (project SEQUOIA End-to-End)

\bibliography{literature}

\begin{thebibliography}{29}
\providecommand{\natexlab}[1]{#1}
\providecommand{\url}[1]{#1}
\csname url@samestyle\endcsname
\providecommand{\newblock}{\relax}
\providecommand{\bibinfo}[2]{#2}
\providecommand{\BIBentrySTDinterwordspacing}{\spaceskip=0pt\relax}
\providecommand{\BIBentryALTinterwordstretchfactor}{4}
\providecommand{\BIBentryALTinterwordspacing}{\spaceskip=\fontdimen2\font plus
\BIBentryALTinterwordstretchfactor\fontdimen3\font minus \fontdimen4\font\relax}
\providecommand{\BIBforeignlanguage}[2]{{%
\expandafter\ifx\csname l@#1\endcsname\relax
\typeout{** WARNING: IEEEtranN.bst: No hyphenation pattern has been}%
\typeout{** loaded for the language `#1'. Using the pattern for}%
\typeout{** the default language instead.}%
\else
\language=\csname l@#1\endcsname
\fi
#2}}
\providecommand{\BIBdecl}{\relax}
\BIBdecl

\bibitem[Berry(2014)]{Berry.2014}
D.~W. Berry, ``High-order quantum algorithm for solving linear differential equations,'' \emph{Journal of Physics A: Mathematical and Theoretical}, vol.~47, no.~10, p. 105301, 2014.

\bibitem[Montanaro and Pallister(2016)]{Montanaro.2016}
A.~Montanaro and S.~Pallister, ``Quantum algorithms and the finite element method,'' \emph{Physical Review A}, vol.~93, no.~3, 2016.

\bibitem[Berry et~al.(2017)Berry, Childs, Ostrander, and Wang]{Berry.2017}
D.~W. Berry, A.~M. Childs, A.~Ostrander, and G.~Wang, ``Quantum algorithm for linear differential equations with exponentially improved dependence on precision,'' \emph{Communications in Mathematical Physics}, vol. 356, no.~3, pp. 1057--1081, 2017.

\bibitem[Gaitan(2020)]{Gaitan.2020}
F.~Gaitan, ``Finding flows of a navier--stokes fluid through quantum computing,'' \emph{npj Quantum Information}, vol.~6, no.~1, 2020.

\bibitem[{Alexei Y. Kitaev}(1995)]{AlexeiY.Kitaev.1995}
{Alexei Y. Kitaev}, ``Quantum measurements and the abelian stabilizer problem,'' \emph{Electron. Colloquium Comput. Complex.}, vol. TR96, 1995.

\bibitem[Biamonte et~al.(2017)Biamonte, Wittek, Pancotti, Rebentrost, Wiebe, and Lloyd]{Biamonte.2017}
J.~Biamonte, P.~Wittek, N.~Pancotti, P.~Rebentrost, N.~Wiebe, and S.~Lloyd, ``Quantum machine learning,'' \emph{Nature}, vol. 549, no. 7671, pp. 195--202, 2017.

\bibitem[Lloyd et~al.(2020)Lloyd, Palma, Gokler, Kiani, Liu, Marvian, Tennie, and Palmer]{Lloyd.2020}
\BIBentryALTinterwordspacing
S.~Lloyd, G.~D. Palma, C.~Gokler, B.~Kiani, Z.-W. Liu, M.~Marvian, F.~Tennie, and T.~Palmer, ``Quantum algorithm for nonlinear differential equations,'' 2020. [Online]. Available: \url{https://arxiv.org/abs/2011.06571}
\BIBentrySTDinterwordspacing

\bibitem[Preskill(2018)]{Preskill.2018}
J.~Preskill, ``Quantum computing in the nisq era and beyond,'' \emph{Quantum}, vol.~2, p.~79, 2018.

\bibitem[Mitarai et~al.(2018)Mitarai, Negoro, Kitagawa, and Fujii]{Mitarai.2018}
K.~Mitarai, M.~Negoro, M.~Kitagawa, and K.~Fujii, ``Quantum circuit learning,'' \emph{Physical Review A}, vol.~98, no.~3, 2018.

\bibitem[Schuld and Killoran(2018)]{Schuld.2018}
M.~Schuld and N.~Killoran, ``Quantum machine learning in feature hilbert spaces,'' \emph{Physical Review Letters}, vol. 122, 03 2018.

\bibitem[Schuld et~al.(2019)Schuld, Bergholm, Gogolin, Izaac, and Killoran]{Schuld.2019}
M.~Schuld, V.~Bergholm, C.~Gogolin, J.~Izaac, and N.~Killoran, ``Evaluating analytic gradients on quantum hardware,'' \emph{Physical Review A}, vol.~99, 2019.

\bibitem[Hatakeyama-Sato et~al.(2023)Hatakeyama-Sato, Igarashi, Kashikawa, Kimura, and Oyaizu]{HatakeyamaSato.2022}
K.~Hatakeyama-Sato, Y.~Igarashi, T.~Kashikawa, K.~Kimura, and K.~Oyaizu, ``Quantum circuit learning as a potential algorithm to predict experimental chemical properties,'' \emph{Digital Discovery}, vol.~2, no.~1, pp. 165--176, 2023.

\bibitem[Kyriienko et~al.(2021)Kyriienko, Paine, and Elfving]{Kyriienko.2021}
O.~Kyriienko, A.~E. Paine, and V.~E. Elfving, ``Solving nonlinear differential equations with differentiable quantum circuits,'' \emph{Physical Review A}, vol. 103, no.~5, p. 052416, 2021.

\bibitem[Heim et~al.(2021)Heim, Ghosh, Kyriienko, and Elfving]{Heim.2021}
\BIBentryALTinterwordspacing
N.~Heim, A.~Ghosh, O.~Kyriienko, and V.~E. Elfving, ``Quantum model-discovery,'' 2021. [Online]. Available: \url{https://arxiv.org/abs/2111.06376}
\BIBentrySTDinterwordspacing

\bibitem[Knudsen and Mendl(2020)]{Knudsen.2020}
\BIBentryALTinterwordspacing
M.~Knudsen and C.~B. Mendl, ``Solving differential equations via continuous-variable quantum computers,'' 2020. [Online]. Available: \url{https://arxiv.org/abs/2012.12220}
\BIBentrySTDinterwordspacing

\bibitem[Paine et~al.(2023)Paine, Elfving, and Kyriienko]{Paine2.2023}
A.~E. Paine, V.~E. Elfving, and O.~Kyriienko, ``Quantum quantile mechanics: solving stochastic differential equations for generating time-series,'' \emph{Advanced Quantum Technologies}, vol.~6, no.~10, p. 2300065, 2023.

\bibitem[Schillo()]{Schillo.2023}
N.~Schillo, ``Quantum algorithms and quantum machine learning for differential equations.''

\bibitem[IBM({\natexlab{a}})]{QiskitTranspiler.2024}
\BIBentryALTinterwordspacing
IBM, ``Qiskit transpiler.'' [Online]. Available: \url{https://docs.quantum.ibm.com/api/qiskit/compiler#transpile}
\BIBentrySTDinterwordspacing

\bibitem[van~den Berg et~al.(2022)van~den Berg, Minev, and Temme]{vanBerg.2020}
E.~van~den Berg, Z.~K. Minev, and K.~Temme, ``Model-free readout-error mitigation for quantum expectation values,'' \emph{Phys. Rev. A}, vol. 105, p. 032620, Mar 2022.

\bibitem[Knill(2005)]{Knill.2005}
E.~Knill, ``Quantum computing with realistically noisy devices,'' \emph{Nature}, vol. 434, no. 7029, pp. 39--44, 2005.

\bibitem[IBM({\natexlab{b}})]{IbmAer.2024}
\BIBentryALTinterwordspacing
IBM, ``Qiskit aer noisemodel.'' [Online]. Available: \url{https://qiskit.github.io/qiskit-aer/stubs/qiskit_aer.noise.NoiseModel.html#qiskit_aer.noise.NoiseModel.from_backend_properties}
\BIBentrySTDinterwordspacing

\bibitem[Ketterer and Wellens(2023)]{Wellens.2023}
A.~Ketterer and T.~Wellens, ``Characterizing crosstalk of superconducting transmon processors,'' \emph{Physical Review Applied}, vol.~20, 09 2023.

\bibitem[Sturm et~al.(2024)Sturm, Mummaneni, and Rullkötter]{Sturm.10.04.2024}
\BIBentryALTinterwordspacing
A.~Sturm, B.~Mummaneni, and L.~Rullkötter, ``Unlocking quantum optimization: a use case study on nisq systems,'' 2024. [Online]. Available: \url{https://arxiv.org/abs/2404.07171}
\BIBentrySTDinterwordspacing

\bibitem[Spall(1998)]{Spall1998ANOO}
J.~C. Spall, ``An overview of the simultaneous perturbation method for efficient optimization,'' \emph{Johns Hopkins Apl Technical Digest}, vol.~19, pp. 482--492, 1998.

\bibitem[Kraft(1988)]{Kraft.1988}
D.~Kraft, \emph{A software package for sequential quadratic programming}, ser. Deutsche Forschungs- und Versuchsanstalt f{\"u}r Luft- und Raumfahrt K{\"o}ln: Forschungsbericht.\hskip 1em plus 0.5em minus 0.4em\relax {Wiss. Berichtswesen d. DFVLR}, 1988.

\bibitem[Powell(1994)]{Powell.1994}
M.~J.~D. Powell, ``A direct search optimization method that models the objective and constraint functions by linear interpolation,'' in \emph{Advances in Optimization and Numerical Analysis}, S.~Gomez and J.-P. Hennart, Eds.\hskip 1em plus 0.5em minus 0.4em\relax Dordrecht: {Springer Netherlands}, 1994.

\bibitem[Virtanen et~al.(2020)Virtanen, Gommers, Oliphant, Haberland, Reddy, Cournapeau, Burovski, Peterson, Weckesser, Bright, {van der Walt}, Brett, Wilson, Millman, Mayorov, Nelson, Jones, Kern, Larson, Carey, Polat, Feng, Moore, VanderPlas, Laxalde, Perktold, Cimrman, Henriksen, Quintero, Harris, Archibald, Ribeiro, Pedregosa, and {van Mulbregt}]{Virtanen.2020}
P.~Virtanen, R.~Gommers, T.~E. Oliphant, M.~Haberland, T.~Reddy, D.~Cournapeau, E.~Burovski, P.~Peterson, W.~Weckesser, J.~Bright, S.~J. {van der Walt}, M.~Brett, J.~Wilson, K.~J. Millman, N.~Mayorov, A.~R.~J. Nelson, E.~Jones, R.~Kern, E.~Larson, C.~J. Carey, {\.{I}}.~Polat, Y.~Feng, E.~W. Moore, J.~VanderPlas, D.~Laxalde, J.~Perktold, R.~Cimrman, I.~Henriksen, E.~A. Quintero, C.~R. Harris, A.~M. Archibald, A.~H. Ribeiro, F.~Pedregosa, and P.~{van Mulbregt}, ``Scipy 1.0: fundamental algorithms for scientific computing in python,'' \emph{Nature methods}, vol.~17, no.~3, pp. 261--272, 2020.

\bibitem[Tamiya and Yamasaki(2022)]{Tamiya.2022}
S.~Tamiya and H.~Yamasaki, ``Stochastic gradient line bayesian optimization for efficient noise-robust optimization of parameterized quantum circuits,'' \emph{npj Quantum Information}, vol.~8, no.~1, 2022.

\bibitem[Wiedmann et~al.(2023)Wiedmann, H{\"o}lle, Periyasamy, Meyer, Ufrecht, Scherer, Plinge, and Mutschler]{Wiedmann.2023}
\BIBentryALTinterwordspacing
M.~Wiedmann, M.~H{\"o}lle, M.~Periyasamy, N.~Meyer, C.~Ufrecht, D.~D. Scherer, A.~Plinge, and C.~Mutschler, ``An empirical comparison of optimizers for quantum machine learning with spsa-based gradients,'' 2023. [Online]. Available: \url{https://arxiv.org/abs/2305.00224}
\BIBentrySTDinterwordspacing

\end{thebibliography}
\EOD
\appendix
\begin{appendices}
    \newpage
\section{Impact of Post-processing Parameter}
\label{AppendixA}
In this part of the appendix, we demonstrate the crucial role of the post-processing parameter $\theta_{\text{post}}$ in improving the accuracy of function learning using QCL. We present a comparison of QCL performance with and without this parameter for three example functions.
We use the circuit depicted in Figure \ref{circ_simple_example_depth} with a depth of $D=3$, $R_Y(\arcsin(x))$ data encoding and a qubit number of $N=3$.
We use a statevector simulator without shot noise, evaluate the cost function on 20 equidistant training points and minimize the cost function using SLSQP with default SciPy minimizer settings.
Figure \ref{function_approx_example2} shows the results for $f_1(x) = x^3$, $f_2(x) = x^3-x^2+1$ and $f_3(x) = \sin(2x)$.
\begin{figure*}[!tp]
\centering
\subfloat[]{%
\scalebox{0.53}{\input{figures/QCL-sim/3-qubits-function-approx-d3-arcsin-tp20-xhoch3}}}
\hfill
\subfloat[]{%
\scalebox{0.53}{\input{figures/QCL-sim/3-qubits-function-approx-d3-arcsin-tp20-complex-function}}}
\hfill
\subfloat[]{%
\scalebox{0.53}{\input{figures/QCL-sim/3-qubits-function-approx-d3-arcsin-tp20-sin2x}}}
\
\subfloat[]{%
\scalebox{0.53}{
\begin{tikzpicture}

\definecolor{dimgray85}{RGB}{85,85,85}
\definecolor{gainsboro229}{RGB}{229,229,229}
\definecolor{lightgray204}{RGB}{204,204,204}

\begin{axis}[
scaled ticks=false,
ytick={0, 0.05, 0.1},
yticklabels={0,0.05,0.1},
legend cell align={left},
legend style={
  fill opacity=0.8,
  draw opacity=1,
  text opacity=1,,
  draw=none
},
axis line style={lightgray204},
tick align=outside,
tick pos=left,
x grid style={lightgray204},
xlabel=\textcolor{dimgray85}{$x$},
xmajorgrids,
xmin=-1, xmax=1,
xtick style={color=dimgray85},
y grid style={lightgray204},
ylabel=\textcolor{dimgray85}{Error},
ymajorgrids,
ymin=-0.02, ymax=0.22,
ytick={0,0.1,0.2,0.3},
yticklabels={0,0.1,0.2,0.3},
ytick style={color=dimgray85}
]
\addplot [thick, olive]
table {%
-1 0.107598878065747
-0.97979797979798 0.118942948339072
-0.95959595959596 0.110725835036104
-0.939393939393939 0.100050795952755
-0.919191919191919 0.0885639719823548
-0.898989898989899 0.0768984862179783
-0.878787878787879 0.0653719692575768
-0.858585858585859 0.054165848394773
-0.838383838383838 0.043391446903977
-0.818181818181818 0.0331197002345639
-0.797979797979798 0.0233963602359474
-0.777777777777778 0.0142505247191415
-0.757575757575758 0.00569976586931015
-0.737373737373737 0.00224661452553432
-0.717171717171717 0.00958542664061701
-0.696969696969697 0.0163181090270805
-0.676767676767677 0.0224496776439665
-0.656565656565657 0.0279879619761223
-0.636363636363636 0.0329430373673285
-0.616161616161616 0.0373267949246237
-0.595959595959596 0.0411526100479117
-0.575757575757576 0.0444350830632634
-0.555555555555556 0.0471898334939846
-0.535353535353535 0.0494333348558139
-0.515151515151515 0.0511827804970866
-0.494949494949495 0.0524559735213841
-0.474747474747475 0.0532712356039896
-0.454545454545454 0.0536473307837427
-0.434343434343434 0.0536034012350355
-0.414141414141414 0.0531589127044929
-0.393939393939394 0.0523336078037039
-0.373737373737374 0.0511474657314363
-0.353535353535353 0.0496206672898446
-0.333333333333333 0.0477735642829821
-0.313131313131313 0.0456266525595766
-0.292929292929293 0.043200548097807
-0.272727272727273 0.0405159656367718
-0.252525252525252 0.0375936994441443
-0.232323232323232 0.0344546058771333
-0.212121212121212 0.031119587448062
-0.191919191919192 0.0276095781494847
-0.171717171717172 0.0239455298289264
-0.151515151515151 0.0201483994317588
-0.131313131313131 0.0162391369536613
-0.111111111111111 0.0122386739625663
-0.0909090909090908 0.00816791256474963
-0.0707070707070706 0.00404771470134913
-0.0505050505050504 0.000101108329409725
-0.0303030303030303 0.00425780622233549
-0.0101010101010101 0.00840169998114156
0.0101010101010102 0.0125121922906949
0.0303030303030305 0.0165687780352263
0.0505050505050506 0.0205510550527024
0.0707070707070707 0.0244387352178693
0.0909090909090911 0.028211655950465
0.111111111111111 0.0318497922509204
0.131313131313131 0.0353332693736466
0.151515151515152 0.0386423762579541
0.171717171717172 0.0417575798491274
0.191919191919192 0.0446595404574813
0.212121212121212 0.0473291283219159
0.232323232323232 0.0497474415672086
0.252525252525253 0.0518958257717196
0.272727272727273 0.0537558953955466
0.292929292929293 0.0553095573594993
0.313131313131313 0.0565390371144364
0.333333333333333 0.0574269076004195
0.353535353535354 0.0579561215686146
0.373737373737374 0.058110047829345
0.393939393939394 0.0578725121017012
0.414141414141414 0.0572278432796819
0.434343434343434 0.0561609261048343
0.454545454545455 0.0546572614563588
0.474747474747475 0.0527030357510036
0.494949494949495 0.0502852013060856
0.515151515151515 0.0473915699865783
0.535353535353535 0.0440109230686251
0.555555555555556 0.0401331410597861
0.575757575757576 0.0357493582961229
0.595959595959596 0.0308521485971929
0.616161616161616 0.0254357502634334
0.636363636363636 0.0194963414880441
0.656565656565657 0.0130323811972083
0.676767676767677 0.00604503600549161
0.696969696969697 0.00146127770055082
0.717171717171717 0.00947819504774189
0.737373737373737 0.017992486628737
0.757575757575758 0.0269848340811099
0.777777777777778 0.0364281567808358
0.797979797979798 0.0462852654989354
0.818181818181818 0.0565054714012417
0.838383838383838 0.0670195055686861
0.858585858585859 0.0777315671621078
0.878787878787879 0.0885061832608938
0.898989898989899 0.0991449357378693
0.919191919191919 0.109341263092442
0.939393939393939 0.118580559981233
0.95959595959596 0.125870938814741
0.97979797979798 0.128702815081249
1 0.101747583081968
};
\addlegendentry{Error without $\theta_\text{post}$}
\addplot [thick, red]
table {%
-1 0.000701899190011845
-0.97979797979798 0.00317011058150996
-0.95959595959596 0.00332814379495938
-0.939393939393939 0.00307074280192321
-0.919191919191919 0.00267283646326577
-0.898989898989899 0.00222927959998542
-0.878787878787879 0.00178242145783836
-0.858585858585859 0.00135343338402694
-0.838383838383838 0.000953353364282483
-0.818181818181818 0.000587829198231971
-0.797979797979798 0.000259432539406745
-0.777777777777778 3.11061643989974e-05
-0.757575757575758 0.000284193291850199
-0.737373737373737 0.000500949271207252
-0.717171717171717 0.000682944479198588
-0.696969696969697 0.000832028943884233
-0.676767676767677 0.000950219720492973
-0.656565656565657 0.00103962489911569
-0.636363636363636 0.00110239152307345
-0.616161616161616 0.00114066947775446
-0.595959595959596 0.0011565862562444
-0.575757575757576 0.00115222925553857
-0.555555555555556 0.00112963335913768
-0.535353535353535 0.00109077227328497
-0.515151515151515 0.00103755255319188
-0.494949494949495 0.000971809570653792
-0.474747474747475 0.000895304889583601
-0.454545454545454 0.000809724665160438
-0.434343434343434 0.000716678787009165
-0.414141414141414 0.000617700561401729
-0.393939393939394 0.000514246780927326
-0.373737373737374 0.000407698069035081
-0.353535353535353 0.000299359415262067
-0.333333333333333 0.0001904608379854
-0.313131313131313 8.21581271673068e-05
-0.292929292929293 2.44663687861586e-05
-0.272727272727273 0.000128402939152415
-0.252525252525252 0.000228713710437339
-0.232323232323232 0.000324531864815472
-0.212121212121212 0.000415060889051938
-0.191919191919192 0.000499573862645408
-0.171717171717172 0.000577412791677282
-0.151515151515151 0.000647987993182976
-0.131313131313131 0.00071077753358394
-0.111111111111111 0.000765326723720221
-0.0909090909090908 0.000811247672330636
-0.0707070707070706 0.000848218899226105
-0.0505050505050504 0.000875985009037213
-0.0303030303030303 0.000894356426077869
-0.0101010101010101 0.000903209190661222
0.0101010101010102 0.000902484817024341
0.0303030303030305 0.000892190212902182
0.0505050505050506 0.000872397660674787
0.0707070707070707 0.000843244859929592
0.0909090909090911 0.000804935031185349
0.111111111111111 0.000757737080394272
0.131313131313131 0.000701985823720574
0.151515151515152 0.000638082271886037
0.171717171717172 0.000566493973128453
0.191919191919192 0.000487755413460671
0.212121212121212 0.000402468472482388
0.232323232323232 0.000311302932352674
0.252525252525253 0.000214997036757428
0.272727272727273 0.000114358095619192
0.292929292929293 1.02631298929748e-05
0.313131313131313 9.63404509785049e-05
0.333333333333333 0.000204434148837647
0.353535353535354 0.000312927666725846
0.373737373737374 0.00042065820600598
0.393939393939394 0.000526389793026681
0.414141414141414 0.000628812664730694
0.434343434343434 0.000726542751877821
0.454545454545455 0.00081812131081116
0.474747474747475 0.000902014771179627
0.494949494949495 0.000976614888990784
0.515151515151515 0.00104023932422387
0.535353535353535 0.00109113280279757
0.555555555555556 0.00112746907857528
0.575757575757576 0.0011473539886489
0.595959595959596 0.00114883000411803
0.616161616161616 0.00112988283341628
0.636363636363636 0.00108845085842829
0.656565656565657 0.00102243850996847
0.676767676767677 0.000929735174611868
0.696969696969697 0.00080824196030449
0.717171717171717 0.000655909786196507
0.737373737373737 0.00047079406506223
0.757575757575758 0.000251134180875801
0.777777777777778 4.52911169307457e-06
0.797979797979798 0.000297176676712474
0.818181818181818 0.000627037168327127
0.838383838383838 0.000993149642363744
0.858585858585859 0.00139263624887498
0.878787878787879 0.00181942930723533
0.898989898989899 0.00226188981975139
0.919191919191919 0.00269793297999199
0.939393939393939 0.00308369265716457
0.95959595959596 0.00332140750192644
0.97979797979798 0.00312880607205901
1 0.000867090250777802
};
\addlegendentry{Error with $\theta_\text{post}$}
\end{axis}

\end{tikzpicture}}}
\hfill
\subfloat[]{%
\scalebox{0.53}{
\begin{tikzpicture}

\definecolor{dimgray85}{RGB}{85,85,85}
\definecolor{gainsboro229}{RGB}{229,229,229}
\definecolor{lightgray204}{RGB}{204,204,204}

\begin{axis}[
scaled ticks=false,
ytick={0, 0.05, 0.1 , 0.15, 0.2, 0.25},
yticklabels={0,0.05,0.1,0.15,0.2,0.25},
axis line style={lightgray204},
legend cell align={left},
legend style={
  fill opacity=0.8,
  draw opacity=1,
  text opacity=1,,
  draw=none
},
tick align=outside,
tick pos=left,
x grid style={lightgray204},
xlabel=\textcolor{dimgray85}{$x$},
xmajorgrids,
xmin=-1, xmax=1,
xtick style={color=dimgray85},
y grid style={lightgray204},
ylabel=\textcolor{dimgray85}{Error},
ymajorgrids,
ymin=-0.02, ymax=0.22,
ytick={0,0.1,0.2,0.3},
yticklabels={0,0.1,0.2,0.3},
ytick style={color=dimgray85}
]
\addplot [thick, olive]
table {%
-1 0.200491603221555
-0.97979797979798 0.24771641377734
-0.95959595959596 0.24035179331676
-0.939393939393939 0.225676954023998
-0.919191919191919 0.208069348773188
-0.898989898989899 0.189179806031742
-0.878787878787879 0.169826329446038
-0.858585858585859 0.150474038992545
-0.838383838383838 0.131409777451369
-0.818181818181818 0.112819591605515
-0.797979797979798 0.0948278639990289
-0.777777777777778 0.0775189817087847
-0.757575757575758 0.0609501947999435
-0.737373737373737 0.0451596705509076
-0.717171717171717 0.0301717626101871
-0.696969696969697 0.0160005836104404
-0.676767676767677 0.00265250108716766
-0.656565656565657 0.00987207389897415
-0.636363636363636 0.02157737665885
-0.616161616161616 0.032471310854031
-0.595959595959596 0.0425647061087672
-0.575757575757576 0.0518707461182255
-0.555555555555556 0.0604045222801139
-0.535353535353535 0.0681826811817798
-0.515151515151515 0.0752231432471883
-0.494949494949495 0.0815448760182702
-0.474747474747475 0.0871677098636314
-0.454545454545454 0.0921121869789187
-0.434343434343434 0.0963994367596116
-0.414141414141414 0.100051072248049
-0.393939393939394 0.103089103556745
-0.373737373737374 0.105535865068799
-0.353535353535353 0.107413953896361
-0.333333333333333 0.108746177597862
-0.313131313131313 0.109555509555423
-0.292929292929293 0.109865050725637
-0.272727272727273 0.10969799672114
-0.252525252525252 0.109077609373477
-0.232323232323232 0.108027192081244
-0.212121212121212 0.106570068370486
-0.191919191919192 0.104729563193486
-0.171717171717172 0.10252898657255
-0.151515151515151 0.0999916192610926
-0.131313131313131 0.0971407001483929
-0.111111111111111 0.0939994151790132
-0.0909090909090908 0.0905908875952012
-0.0707070707070706 0.0869381693418273
-0.0505050505050504 0.0830642334999105
-0.0303030303030303 0.0789919676374782
-0.0101010101010101 0.074744167986085
0.0101010101010102 0.0703435343685764
0.0303030303030305 0.0658126658189998
0.0505050505050506 0.0611740568495704
0.0707070707070707 0.0564500943325715
0.0909090909090911 0.0516630549775019
0.111111111111111 0.0468351033959132
0.131313131313131 0.0419882907587227
0.151515151515152 0.0371445540635418
0.171717171717172 0.0323257160432716
0.191919191919192 0.0275534857622334
0.212121212121212 0.0228494599629909
0.232323232323232 0.0182351252463945
0.252525252525253 0.0137318611899581
0.272727272727273 0.00936094453640768
0.292929292929293 0.0051435546162949
0.313131313131313 0.00110078020733495
0.333333333333333 0.0027463719193721
0.353535353535354 0.00637696645761598
0.373737373737374 0.00977012664739807
0.393939393939394 0.012905018506742
0.414141414141414 0.0157608316262564
0.434343434343434 0.0183167557922925
0.454545454545455 0.0205519525167066
0.474747474747475 0.0224455203083451
0.494949494949495 0.0239764522047444
0.515151515151515 0.0251235836656422
0.535353535353535 0.02586552837606
0.555555555555556 0.0261805987628233
0.575757575757576 0.0260467070181239
0.595959595959596 0.025441241034877
0.616161616161616 0.0243409077236099
0.636363636363636 0.0227215334455083
0.656565656565657 0.0205578073680005
0.676767676767677 0.0178229478076994
0.696969696969697 0.0144882630686385
0.717171717171717 0.0105225652517499
0.737373737373737 0.00589137517543015
0.757575757575758 0.000555823928852228
0.777777777777778 0.00552889740217943
0.797979797979798 0.0124157809116568
0.818181818181818 0.020169344722121
0.838383838383838 0.0288707937357422
0.858585858585859 0.0386263377012267
0.878787878787879 0.0495813380361786
0.898989898989899 0.0619460834069838
0.919191919191919 0.0760472668099375
0.939393939393939 0.092444970553163
0.95959595959596 0.112256915436471
0.97979797979798 0.138448990203435
1 0.203602508978661
};
\addlegendentry{Error without $\theta_\text{post}$}
\addplot [thick, red]
table {%
-1 0.000367649131407299
-0.97979797979798 0.0399410827827426
-0.95959595959596 0.0404657763028385
-0.939393939393939 0.0366974439062273
-0.919191919191919 0.0315614831312921
-0.898989898989899 0.0260561243954499
-0.878787878787879 0.0206203935874226
-0.858585858585859 0.015468983087003
-0.838383838383838 0.0107099223022424
-0.818181818181818 0.00639493320144027
-0.797979797979798 0.00254390151827602
-0.777777777777778 0.000842120040115454
-0.757575757575758 0.00377353095223821
-0.737373737373737 0.00626777699227766
-0.717171717171717 0.00834661504490286
-0.696969696969697 0.0100343619872388
-0.676767676767677 0.0113567461831193
-0.656565656565657 0.0123401400522765
-0.636363636363636 0.0130110401668993
-0.616161616161616 0.0133957117807867
-0.595959595959596 0.0135199446652499
-0.575757575757576 0.0134088854796204
-0.555555555555556 0.0130869234479433
-0.535353535353535 0.0125776135461046
-0.515151515151515 0.0119036262893837
-0.494949494949495 0.0110867164815778
-0.474747474747475 0.0101477055132572
-0.454545454545454 0.00910647333431136
-0.434343434343434 0.0079819573020018
-0.414141414141414 0.0067921558675571
-0.393939393939394 0.00555413560943652
-0.373737373737374 0.00428404051518494
-0.353535353535353 0.00299710270040121
-0.333333333333333 0.00170765396384576
-0.313131313131313 0.000429137732810059
-0.292929292929293 0.000825878931660196
-0.272727272727273 0.00204569351436612
-0.252525252525252 0.00321945659064093
-0.232323232323232 0.00433716086831482
-0.212121212121212 0.00538963082287347
-0.191919191919192 0.00636851299477392
-0.171717171717172 0.00726626699736888
-0.151515151515151 0.0080761572689263
-0.131313131313131 0.00879224559093161
-0.111111111111111 0.00940938438662409
-0.0909090909090908 0.00992321080783032
-0.0707070707070706 0.0103301416138752
-0.0505050505050504 0.0106273688434988
-0.0303030303030303 0.0108128562787455
-0.0101010101010101 0.0108853366986428
0.0101010101010102 0.0108443099197921
0.0303030303030305 0.0106900416206247
0.0505050505050506 0.0104235629460619
0.0707070707070707 0.0100466708891093
0.0909090909090911 0.00956192944589196
0.111111111111111 0.00897267154027659
0.131313131313131 0.00828300171356977
0.151515151515152 0.00749779957367003
0.171717171717172 0.00662272399616615
0.191919191919192 0.00566421806724582
0.212121212121212 0.00462951475426399
0.232323232323232 0.00352664328438701
0.252525252525253 0.00236443620423166
0.272727272727273 0.00115253708326513
0.292929292929293 9.85911898031544e-05
0.313131313131313 0.00137765758636266
0.333333333333333 0.00267253368558285
0.353535353535354 0.00397024411437985
0.373737373737374 0.00525695684514194
0.393939393939394 0.00651797357145123
0.414141414141414 0.00773772045458421
0.434343434343434 0.00889973962979895
0.454545454545455 0.00998668198987551
0.474747474747475 0.0109803019353216
0.494949494949495 0.0118614550125383
0.515151515151515 0.0126100996758439
0.535353535353535 0.0132053048394838
0.555555555555556 0.0136252654792506
0.575757575757576 0.0138473293697637
0.595959595959596 0.0138480392072686
0.616161616161616 0.0136031960256688
0.636363636363636 0.0130879522076316
0.656565656565657 0.0122769459012415
0.676767676767677 0.0111444938805361
0.696969696969697 0.00966486782377363
0.717171717171717 0.00781269128236384
0.737373737373737 0.00556351412961753
0.757575757575758 0.00289465308689851
0.777777777777778 0.000213559619034909
0.797979797979798 0.00377588433949272
0.818181818181818 0.0077992348066469
0.838383838383838 0.0122779658006409
0.858585858585859 0.0171860114438267
0.878787878787879 0.0224630853557617
0.898989898989899 0.0279887949178514
0.919191919191919 0.0335295316892374
0.939393939393939 0.0386146982750055
0.95959595959596 0.0421853482333706
0.97979797979798 0.0411623321650476
1 0.00150624739737848
};
\addlegendentry{Error with $\theta_\text{post}$}
\end{axis}

\end{tikzpicture}}}
\hfill
\subfloat[]{%
\scalebox{0.53}{
\begin{tikzpicture}

\definecolor{dimgray85}{RGB}{85,85,85}
\definecolor{gainsboro229}{RGB}{229,229,229}
\definecolor{lightgray204}{RGB}{204,204,204}

\begin{axis}[
scaled ticks=false,
ytick={0, 0.05, 0.1 , 0.15},
yticklabels={0, 0.05, 0.1 , 0.15},
legend cell align={left},
legend style={
  fill opacity=0.8,
  draw opacity=1,
  text opacity=1,,
  draw=none
},
axis line style={lightgray204},
tick align=outside,
tick pos=left,
x grid style={lightgray204},
xlabel=\textcolor{dimgray85}{$x$},
xmajorgrids,
xmin=-1, xmax=1,
xtick style={color=dimgray85},
y grid style={lightgray204},
ylabel=\textcolor{dimgray85}{Error},
ymajorgrids,
ymin=-0.02, ymax=0.22,
ytick={0,0.1,0.2,0.3},
yticklabels={0,0.1,0.2,0.3},
ytick style={color=dimgray85}
]
\addplot [thick, olive]
table {%
-1 0.188505642197756
-0.97979797979798 0.0985992395489381
-0.95959595959596 0.0756481975331258
-0.939393939393939 0.0640220350879855
-0.919191919191919 0.0582260817589874
-0.898989898989899 0.0560614384248931
-0.878787878787879 0.0563594480603737
-0.858585858585859 0.0583987284036549
-0.838383838383838 0.0616900012384799
-0.818181818181818 0.0658792410552552
-0.797979797979798 0.0706980566447791
-0.777777777777778 0.0759358152600714
-0.757575757575758 0.0814228549256913
-0.737373737373737 0.0870198081014308
-0.717171717171717 0.0926105059414426
-0.696969696969697 0.0980970866376313
-0.676767676767677 0.103396516976499
-0.656565656565657 0.108438051537411
-0.636363636363636 0.113161332328571
-0.616161616161616 0.117514936892839
-0.595959595959596 0.121455247289543
-0.575757575757576 0.124945552995136
-0.555555555555556 0.127955327136839
-0.535353535353535 0.130459633011594
-0.515151515151515 0.132438629764118
-0.494949494949495 0.133877154362439
-0.474747474747475 0.134764362841363
-0.454545454545454 0.135093417966854
-0.434343434343434 0.134861213518566
-0.414141414141414 0.134068127633594
-0.393939393939394 0.132717799332235
-0.373737373737374 0.130816923614362
-0.353535353535353 0.128375061483385
-0.333333333333333 0.125404462001854
-0.313131313131313 0.121919894064464
-0.292929292929293 0.117938486031135
-0.272727272727273 0.113479571724605
-0.252525252525252 0.108564541585659
-0.232323232323232 0.103216698011009
-0.212121212121212 0.097461114086458
-0.191919191919192 0.0913244950805686
-0.171717171717172 0.0848350421889477
-0.151515151515151 0.0780223181219321
-0.131313131313131 0.0709171142132136
-0.111111111111111 0.0635513187972067
-0.0909090909090908 0.0559577866613233
-0.0707070707070706 0.0481702094278856
-0.0505050505050504 0.0402229867608413
-0.0303030303030303 0.032151098326046
-0.0101010101010101 0.0239899764616963
0.0101010101010102 0.0157753795383634
0.0303030303030305 0.00754326600664187
0.0505050505050506 0.000670330854765963
0.0707070707070707 0.00882942746632012
0.0909090909090911 0.0168982137177272
0.111111111111111 0.0248411759428489
0.131313131313131 0.0326232190587765
0.151515151515152 0.0402097878283946
0.171717171717172 0.0475669872085148
0.191919191919192 0.0546617017512158
0.212121212121212 0.0614617140345889
0.232323232323232 0.0679358221109865
0.252525252525253 0.0740539559764526
0.272727272727273 0.0797872930848164
0.292929292929293 0.085108372954752
0.313131313131313 0.0899912109487414
0.333333333333333 0.0944114113406674
0.353535353535354 0.0983462798352086
0.373737373737374 0.101774935759358
0.393939393939394 0.104678424216991
0.414141414141414 0.107039828584912
0.434343434343434 0.108844383837976
0.454545454545455 0.11007959132792
0.474747474747475 0.110735335813921
0.494949494949495 0.110804005764072
0.515151515151515 0.11028061823176
0.535353535353535 0.109162949981289
0.555555555555556 0.107451677023946
0.575757575757576 0.105150525373254
0.595959595959596 0.10226643670031
0.616161616161616 0.0988097537609388
0.636363636363636 0.0947944321168451
0.656565656565657 0.0902382869975615
0.676767676767677 0.0851632874835443
0.696969696969697 0.0795959150657616
0.717171717171717 0.073567610924466
0.737373737373737 0.0671153474330662
0.757575757575758 0.0602823769725319
0.777777777777778 0.0531192397086784
0.797979797979798 0.0456851601120482
0.818181818181818 0.0380500465544326
0.838383838383838 0.0302974643937084
0.858585858585859 0.0225292586289685
0.878787878787879 0.0148731457847823
0.898989898989899 0.00749607861410384
0.919191919191919 0.000630038220491014
0.939393939393939 0.00537134103579917
0.95959595959596 0.00988128639713703
0.97979797979798 0.0114194404827727
1 0.00404444851076569
};
\addlegendentry{Error without $\theta_\text{post}$}
\addplot [thick, red]
table {%
-1 0.00135953280244727
-0.97979797979798 0.0118225654404824
-0.95959595959596 0.0117960796202617
-0.939393939393939 0.0101012134176807
-0.919191919191919 0.00787950635475065
-0.898989898989899 0.00555979864267753
-0.878787878787879 0.00334402205294559
-0.858585858585859 0.00133480588887747
-0.838383838383838 0.000416520413524135
-0.818181818181818 0.00188792142011818
-0.797979797979798 0.00307517911763744
-0.777777777777778 0.00398533310429183
-0.757575757575758 0.0046326646393684
-0.737373737373737 0.00503610370160024
-0.717171717171717 0.00521748281857426
-0.696969696969697 0.00520032084728561
-0.676767676767677 0.00500895272646884
-0.656565656565657 0.00466789339058338
-0.636363636363636 0.00420136522713432
-0.616161616161616 0.00363294297500782
-0.595959595959596 0.00298528508727613
-0.575757575757576 0.00227993021045136
-0.555555555555556 0.00153714373174851
-0.535353535353535 0.000775803568571098
-0.515151515151515 1.33172666517689e-05
-0.494949494949495 0.000734435507326103
-0.474747474747475 0.00145313955801751
-0.454545454545454 0.00213006328267651
-0.434343434343434 0.00275406300762171
-0.414141414141414 0.00331557466441335
-0.393939393939394 0.00380659450880017
-0.373737373737374 0.00422065026621321
-0.353535353535353 0.00455276384306824
-0.333333333333333 0.00479940655453637
-0.313131313131313 0.00495844767252496
-0.292929292929293 0.00502909698265219
-0.272727272727273 0.00501184194815119
-0.252525252525252 0.00490838000673388
-0.232323232323232 0.00472154646898676
-0.212121212121212 0.00445523844096868
-0.191919191919192 0.00411433515669396
-0.171717171717172 0.00370461507642716
-0.151515151515151 0.00323267008264638
-0.131313131313131 0.00270581708607892
-0.111111111111111 0.00213200733843913
-0.0909090909090908 0.00151973373581904
-0.0707070707070706 0.000877936386453226
-0.0505050505050504 0.000215906708381605
-0.0303030303030303 0.000456809683835935
-0.0101010101010101 0.00113051104916429
0.0101010101010102 0.0017954382963891
0.0303030303030305 0.00244187421859304
0.0505050505050506 0.00306024293925519
0.0707070707070707 0.00364120932375797
0.0909090909090911 0.00417577810954661
0.111111111111111 0.00465539250647293
0.131313131313131 0.00507203201614032
0.151515151515152 0.00541830921492553
0.171717171717172 0.00568756523989206
0.191919191919192 0.00587396370959503
0.212121212121212 0.00597258280250473
0.232323232323232 0.00597950520411633
0.252525252525253 0.00589190561907177
0.272727272727273 0.00570813552633576
0.292929292929293 0.00542780483238048
0.313131313131313 0.00505186004882185
0.333333333333333 0.00458265858501916
0.353535353535354 0.00402403870128665
0.373737373737374 0.00338138461195314
0.393939393939394 0.00266168615633366
0.414141414141414 0.00187359236549167
0.434343434343434 0.00102745813777505
0.454545454545455 0.000135383088997409
0.474747474747475 0.000788758546671708
0.494949494949495 0.00172929833556412
0.515151515151515 0.00266876607633237
0.535353535353535 0.00358789636442802
0.555555555555556 0.00446565197783388
0.575757575757576 0.00527926753975905
0.595959595959596 0.00600431796068179
0.616161616161616 0.00661481760873017
0.636363636363636 0.00708335818569494
0.656565656565657 0.00738129617459038
0.676767676767677 0.00747900491148112
0.696969696969697 0.0073462125207524
0.717171717171717 0.00695245629505548
0.737373737373737 0.00626769855502674
0.757575757575758 0.00526317201091231
0.777777777777778 0.0039125603703567
0.797979797979798 0.00219368411290299
0.818181818181818 9.09752141344811e-05
0.838383838383838 0.00240076313672277
0.858585858585859 0.00527039591853351
0.878787878787879 0.00848048588519823
0.898989898989899 0.0119494243242018
0.919191919191919 0.0155163184483691
0.939393939393939 0.018862462295687
0.95959595959596 0.0212971237217713
0.97979797979798 0.0209190094821469
1 0.00293491442376881
};
\addlegendentry{Error with $\theta_\text{post}$}
\end{axis}

\end{tikzpicture}}}
\
\subfloat[]{%
\scalebox{0.53}{\input{figures/QCL-sim/convergence-3-qubits-function-approx-d3-arcsin-tp20-xhoch3.tex}}}
\hfill
\subfloat[]{%
\scalebox{0.53}{\input{figures/QCL-sim/convergence-3-qubits-function-approx-d3-arcsin-tp20-complex-function.tex}}}
\hfill
\subfloat[]{%
\scalebox{0.53}{\input{figures/QCL-sim/convergence-3-qubits-function-approx-d3-arcsin-tp20-sin2x.tex}}}
\caption{Comparative analysis of QCL performance with (red lines) and without (green lines) the post-processing parameter $\theta_{\text{post}}$ for three functions: (a) $f_1(x) = x^3$, (b) $f_2(x) = x^3-x^2+1$, and (c) $f_3(x) = \sin(2x)$. The dashed black lines show the initial functions with randomly chosen starting parameters and $\theta_\text{post} = 1$. (d)-(f) show the absolute errors on a fine grid. (g)-(i) display the cost function values versus the number of cost function evaluations during optimization.}
\label{function_approx_example2}
\end{figure*}
The results clearly demonstrate the limitations of QCL without the post-processing parameter.
While the qualitative behavior of the target functions is captured, significant deviations are observed without $\theta_{\text{post}}$ (green lines in Figures \ref{function_approx_example2} (d)-(f)).
The absolute errors $|f(x) - f_{QC}(x)|$ are substantially higher without $\theta_{\text{post}}$ (green lines in Figures \ref{function_approx_example2} (a)-(c)).
Without $\theta_{\text{post}}$, the optimization process progresses more slowly and stagnates earlier.\\
In conclusion, the inclusion of $\theta_{\text{post}}$ leads to more accurate function approximations, significantly reduced errors, faster convergence and lower final cost values.
It also extends the value range to $f^{\text{post}}_{QC}(x, \boldsymbol{\theta}) \in \mathbb{R}$ for $\theta_{\text{post}} \in \mathbb{R}$.

    \section{Necessity of circular entanglement}
\label{AppendixC}
In this part of the appendix, we demonstrate the importance of circular entanglement in our QCL approach, despite its potential drawbacks in terms of hardware efficiency and error resilience as discussed in Section \ref{sec_circ_ent}.
Figure \ref{circular_comp} compares the performance of QCL circuits with circular and linear entanglement on a simulator that incorporates hardware noise (noise model 3 from Section \ref{chap:hardware_noise}). The function $f(x) = x^3$ is learned using both entanglement strategies with $R_Y(\arcsin(x))$ data encoding and a qubit number of $N=3$. The cost function is evaluated on 10 equidistant training points and optimized using the COBYLA algorithm.
\begin{figure*}[!tp]
\centering
\subfloat[]{%
\scalebox{0.77}{\input{figures/circ_ent_comp/function_approximation}}}
\hfill
\subfloat[]{%
\scalebox{0.77}{
\begin{tikzpicture}

\definecolor{darkgray176}{RGB}{176,176,176}
\definecolor{lightgray204}{RGB}{204,204,204}

\begin{axis}[
nodes={scale=0.8, transform shape},
axis line style={lightgray204},
legend cell align={left},
legend style={fill opacity=0.8, draw opacity=1, text opacity=1, draw=none},
log basis y={10},
tick align=outside,
tick pos=left,
x grid style={darkgray176},
xlabel={Cost Function Evaluations},
xmajorgrids,
xmin=-6.1, xmax=150.1,
xtick style={color=black},
y grid style={darkgray176},
ylabel={Cost Function Value},
ymajorgrids,
ymin=0.00537747770975809, ymax=10.5172423985914,
ymode=log,
ytick style={color=black},
ytick={0.0001,0.001,0.01,0.1,1,10,100,1000},
yticklabels={
  \(\displaystyle {10^{-4}}\),
  \(\displaystyle {10^{-3}}\),
  \(\displaystyle {10^{-2}}\),
  \(\displaystyle {10^{-1}}\),
  \(\displaystyle {10^{0}}\),
  \(\displaystyle {10^{1}}\),
  \(\displaystyle {10^{2}}\),
  \(\displaystyle {10^{3}}\)
}
]
\addplot [semithick, blue]
table {%
1 4.56262288900179
2 3.2760724953118
3 4.05006137871372
4 4.03085771204706
5 3.37162820175899
6 2.92718409339136
7 2.81470975868629
8 2.69041665992086
9 2.89204861876859
10 2.73821012219795
11 3.58856224976997
12 2.47183044567772
13 2.51194854227712
14 2.44610877437078
15 2.54339712457376
16 2.48767494925262
17 2.69688837372303
18 2.56111170765673
19 2.37619681599209
20 2.43970919432139
21 2.51509959889918
22 2.37830478987125
23 2.28931010813885
24 1.85597883469739
25 1.70964262175279
26 1.98318752800484
27 2.20357880342383
28 2.12861348378365
29 1.40730775821352
30 1.37021920954724
31 1.45186708614148
32 2.60898914315155
33 1.24272881367607
34 1.70653805857785
35 1.59792302806143
36 2.49650866533276
37 1.4455036942454
38 1.40263097059541
39 1.20267390415886
40 1.28173803367416
41 1.14589636081873
42 1.35002730945604
43 0.99361461432959
44 0.871303175768573
45 0.955752344080105
46 0.926111502984607
47 1.04727232866633
48 0.908201099359757
49 0.989462989753955
50 0.814121871052698
51 1.01649442743233
52 0.886349859866701
53 0.893918043793073
54 0.935808819534431
55 0.941152806059737
56 0.913163231104851
57 0.783469375664759
58 0.747605653104021
59 0.764328258928552
60 0.841626092520535
61 0.930675348679705
62 0.651317468976428
63 0.568112917782355
64 0.519197243371722
65 0.550595401141476
66 0.486646316955693
67 0.43624373272872
68 0.393586617795217
69 0.319774186258207
70 0.461709280376028
71 0.391817488946271
72 0.358989465515315
73 0.364473555856275
74 0.365264063630197
75 0.367049651137049
76 0.328649956400742
77 0.330053584564135
78 0.396882261036845
79 0.253676633106158
80 0.321785810163891
81 0.295718625042295
82 0.288051903331361
83 0.324609776688505
84 0.294306164183456
85 0.305185011544751
86 0.301611960236729
87 0.340034946164795
88 0.318492157057338
89 0.327513552296289
90 0.326541751759469
91 0.292932153052273
92 0.285249188075551
93 0.327177002646821
94 0.310096208109308
95 0.312021063596467
96 0.336402637068483
97 0.253938001674331
98 0.322984447032273
99 0.341708805280332
100 0.273794571981897
101 0.283100679375518
102 0.274650182681442
103 0.309629790635643
104 0.334314252727964
105 0.306059371690945
106 0.337123638089926
107 0.29847028420363
108 0.303295140119337
109 0.337430604476401
110 0.252175823956891
111 0.277663789368547
112 0.327194118294217
113 0.312359706286671
114 0.254762038257125
115 0.343949010509653
116 0.282557517059801
117 0.34602021139644
118 0.374734954580929
119 0.346307934387105
120 0.342100343477981
121 0.345000872948581
122 0.344205349546879
123 0.360006353791992
124 0.266352632514586
125 0.320068970166201
126 0.321042417750777
127 0.303706414508843
128 0.278596606222241
129 0.29379791216115
130 0.275715860422898
131 0.28683249709994
132 0.312617259360284
133 0.302822867362101
134 0.320453807668281
135 0.292898502279391
136 0.288018020686987
137 0.24739269997083
138 0.254666365807607
139 0.288449467938961
140 0.301147003888541
141 0.298240558048848
142 0.294134029478182
143 0.341846412054876
};
\addlegendentry{Linear Entanglement}
\addplot [semithick, red]
table {%
1 7.45240211808272
2 3.50603377240371
3 3.42148928406352
4 1.63102394524322
5 2.99496718392634
6 3.70271197679328
7 2.94209469695789
8 1.8513206091664
9 3.510823286807
10 1.75801789997573
11 1.48258280944075
12 1.43942058684765
13 0.708560243688286
14 0.558551959117183
15 0.331776567516116
16 1.86638003835807
17 0.100874592723674
18 0.174074148117424
19 0.189023971856006
20 1.3121551116508
21 0.241697080097964
22 1.20426577559637
23 0.192512973870328
24 0.71324860270947
25 0.214128034831924
26 1.64563381749641
27 0.266364946164585
28 0.102288423984367
29 0.146414148828362
30 0.0938493500371146
31 0.140877844005056
32 0.147535745883343
33 0.063542635385978
34 0.0565894331386343
35 0.0880177870132081
36 0.0371262000928356
37 0.0744694712562549
38 0.078242710771149
39 0.0244868881979918
40 0.103719123334493
41 0.0931609928211812
42 0.0448189613426998
43 0.0372099258069087
44 0.0800473443181739
45 0.0479114545743994
46 0.0376807765129301
47 0.0286814447182436
48 0.0243460816532921
49 0.0419982256702302
50 0.010133760879826
51 0.0466703849285165
52 0.0292407517014634
53 0.0282319982054335
54 0.00759795079857572
55 0.0192077425552512
56 0.0377077272346681
57 0.031636070989906
58 0.0291118832069037
59 0.0309022691458192
60 0.0261581409610416
61 0.0485722739299665
62 0.0258833208858864
63 0.029970491714092
64 0.0465213755162154
65 0.0180041252313087
66 0.0309275398259659
67 0.0350978668298692
68 0.00939450795316005
69 0.0207949125724697
70 0.0313111339503309
71 0.0536493606271129
72 0.0172411053921217
73 0.0188277027284085
74 0.0398972260187272
75 0.0229217823644715
76 0.023681671765346
77 0.0432383245488913
78 0.0240703036678319
79 0.0455455991422163
80 0.0198268404595315
81 0.0253261150658949
82 0.0148197605794942
83 0.0322044531140173
84 0.0358326049052821
85 0.0230371249635843
86 0.0308580979153499
87 0.0133259650547118
88 0.0246206917883322
89 0.0284335336248178
90 0.020027456525438
91 0.0075889942156125
92 0.0383565342092782
93 0.034013940175953
94 0.0272920385139298
95 0.0234697873504425
96 0.0347188358472308
97 0.0121815545549824
98 0.0281937300417636
99 0.0473605092410834
100 0.0123108967229784
101 0.0139462346981538
102 0.0452041306365675
103 0.0273621969878912
104 0.028362877915913
105 0.0275481949951216
106 0.0279852904980911
107 0.0532849394781644
108 0.0129339267682047
109 0.0427119961450093
110 0.0199982730236957
111 0.0236614273823786
};
\addlegendentry{Circular Entanglement}
\end{axis}

\end{tikzpicture}}}
\caption{$f(x) = x^3$ learned using QCL circuits on a simulator with noise model 3 with 2000 shots, $R_Y(\arcsin(x))$ data encoding and a qubit number of $N=3$. The cost function is evaluated on 10 equidistant training points and is classically minimized using COBYLA. The results for QCL with circular entanglement and linear entanglement are compared (a). The initial function (dashed black line) shows $f^{\text{post}}_{QC}(x)$ with the randomly chosen starting parameters before the optimization process and $\theta_{\text{post}} = 1$. In (b) the respective values of the cost functions versus the number of cost function evaluations during the optimization process are plotted.}
\label{circular_comp}
\end{figure*}
Figure \ref{circular_comp} (a) clearly shows that the circuit with circular entanglement (red line) achieves a significantly better approximation of the target function compared to the circuit with linear entanglement (blue line). Circular entanglement allows the QCL circuit to learn the cubic nature of the function more accurately.
The convergence plot in Figure \ref{circular_comp} (b) further supports this observation. The linear entanglement circuit (blue line) struggles to reduce the cost function beyond a certain point, suggesting a limitation in its expressivity.
These results demonstrate that, despite the advantages of linear entanglement in terms of hardware efficiency and error resilience (see Section \ref{sec_circ_ent}), circular entanglement is necessary to achieve the required expressivity in our experiments.

    \section{Calibration data of ibmq\_ehningen}
\label{AppendixB}
This part of the appendix presents the calibration data of ibmq\_ehningen from January 30, 2024, 8:42:06+01:00, which forms the basis for the noise models used in our simulations. The individual X, CNOT, and readout errors are given in Tables~\ref{table-x-errors-ibmq-ehningen}, \ref{table-cx-errors-ibmq-ehningen}, and \ref{table-readout-errors-ibmq-ehningen}.
The median T1 and T2 times are 113 µs and 96 µs, respectively.
It is noteworthy that there is significant variation in error rates across different qubits and qubit pairs. For example, the X gate errors range from 0.00014 to 0.00422, while the CNOT gate errors range from 0.00377 to 0.06896. This high variability underscores the importance of using a detailed noise model for accurate simulations.
\begin{table}[!tp]
    \centering
    \begin{tabular}{c|c}
        Qubit & X error
        \\
        \hline
        19& 0.00013591302669571659
        \\
        16& 0.00014569737631580065
        \\
        24& 0.0001651366443341561
        \\
        6& 0.00016680730700449124
        \\
        3& 0.00019475108276179376
        \\
        25& 0.00019511898972169597
        \\
        12& 0.0001977410765303121
        \\
        23& 0.0002005349723274833
        \\
        13& 0.00020272128722526137
        \\
        9& 0.0002079151356677378
        \\
        26& 0.0002107794975931981
        \\
        0& 0.00021184286817880178
        \\
        1& 0.0002263078174650456
        \\
        15& 0.00023837868227682494
        \\
        11& 0.0002477045424814508
        \\
        22& 0.0002549839225068226
        \\
        18& 0.00027892545750661713
        \\
        21& 0.0002979112936644023
        \\
        5& 0.00031537329438427003
        \\
        2& 0.00033257814929062255
        \\
        4& 0.0003574360500279698
        \\
        20& 0.0003953408174393563
        \\
        14& 0.0004337410173584769
        \\
        8& 0.0004944174096551563
        \\
        17& 0.0005776839080296207
        \\
        10& 0.001741461094155414
        \\
        7& 0.004221992895832456
        \\
        median& 0.00023837868227682494
    \end{tabular}
    \caption{X errors of ibmq\_ehningen from January 30, 2024, 8:42:06+01:00 in descending order.}
    \label{table-x-errors-ibmq-ehningen}
\end{table}
\begin{table}[!tp]
    \centering
    \begin{tabular}{c|c}
        Qubit pair & CNOT error
        \\
        \hline
        (23, 24) & 0.003766111620421869
        \\
        (22, 25) & 0.005100573458733021
        \\
        (1, 4) & 0.005725284943773834
        \\
        (18, 21) & 0.0059052091403848095
        \\
        (19, 22) & 0.006018110357534884
        \\
        (0, 1) & 0.006244572705932261
        \\
        (14, 16) & 0.006699047143990389
        \\
        (16, 19) & 0.006816483127679324
        \\
        (2, 3) & 0.00686035596001125
        \\
        (19, 20) & 0.006865771449287628
        \\
        (8, 9) & 0.007161408838511407
        \\
        (21, 23) & 0.007217990657473805
        \\
        (5, 8) & 0.007345102646509005
        \\
        (1, 2) & 0.007464918811527332
        \\
        (3, 5) & 0.0074783630145963675
        \\
        (12, 15)& 0.007548664291686519
        \\
        (24, 25)& 0.008734138864550017
        \\
        (15, 18)& 0.009328357231174117
        \\
        (13, 14)& 0.009752602982268627
        \\
        (11, 14)& 0.010083426759830982
        \\
        (25, 26)& 0.01018025088966229
        \\
        (8, 11)& 0.01066070025821142
        \\
        (17, 18)& 0.013317040648282985
        \\
        (6, 7)& 0.01607606156008279
        \\
        (10, 12)& 0.017207769362116293
        \\
        (12, 13)& 0.021712709764238475
        \\
        (7, 10)& 0.043634415742329125
        \\
        (4, 7)& 0.06896314743392529
        \\
        median& 0.00747164091306185
    \end{tabular}
    \caption{CNOT errors of ibmq\_ehningen from January 30, 2024, 8:42:06+01:00 in descending order.}
    \label{table-cx-errors-ibmq-ehningen}
\end{table}
\begin{table}[!tp]
    \centering
    \begin{tabular}{c|c}
        Qubit & readout error
        \\
        \hline
        21& 0.006699999999999928 
        \\
        13& 0.007300000000000084 
        \\
        14& 0.007399999999999962
        \\
        16& 0.00770000000000004
        \\
        23& 0.00770000000000004
        \\
        24& 0.00869999999999993
        \\
        15& 0.009000000000000008
        \\
        4& 0.009400000000000075
        \\
        9& 0.009400000000000075
        \\
        25& 0.009700000000000042
        \\
        19& 0.010199999999999987
        \\
        1& 0.010499999999999954
        \\
        2& 0.011199999999999988
        \\
        0& 0.011500000000000066
        \\
        12& 0.012599999999999945
        \\
        18& 0.012800000000000034
        \\
        26& 0.013399999999999967
        \\
        6& 0.013800000000000034
        \\
        8& 0.014000000000000012
        \\
        10& 0.015300000000000091
        \\
        5& 0.016699999999999937
        \\
        22& 0.01760000000000006
        \\
        20& 0.018199999999999994
        \\
        7& 0.020000000000000018
        \\
        3& 0.020499999999999963
        \\
        11& 0.0232
        \\
        17& 0.02859999999999996
        \\
        median& 0.011500000000000066
    \end{tabular}
    \caption{Readout errors of ibmq\_ehningen from January 30, 2024, 8:42:06+01:00 in descending order.}
    \label{table-readout-errors-ibmq-ehningen}
\end{table}
\end{appendices}

\end{document}